\documentclass[%
 aip,
 amsmath,amssymb,
 reprint,%
]{revtex4-1}

\usepackage{graphicx}
\usepackage{dcolumn}
\usepackage{bm}

\usepackage[utf8]{inputenc}
\usepackage[T1]{fontenc}
\usepackage{mathptmx}
\usepackage{graphicx}  
\usepackage{dcolumn}   
\usepackage{bm}        
\usepackage{amssymb}   
\usepackage{subfigure}  
\usepackage{hyperref}
\usepackage{ragged2e}
\usepackage{mathtools}
\usepackage{xcolor}
\usepackage{nccmath}

\usepackage{CJKutf8}

\begin{document}

\preprint{AIP/123-QED}

\title{Probing the pressure dependence of sound speed and attenuation in bubbly media: Experimental observations, {a theoretical model} and numerical calculations}
\author{AJ. Sojahrood}
\email[Corresponding author email: ]{amin.jafarisojahrood@ryerson.ca}

\affiliation{Department of Physics, Ryerson University, Toronto, Ontario, Canada}
\affiliation{Institute for Biomedical Engineering, Science and Technology (IBEST) a partnership between Ryerson University and St. Mike's Hospital, Toronto, Ontario, Canada}
\author{Q. Li \begin{CJK*}{UTF8}{gbsn}
		{(李倩)}
\end{CJK*}}
\affiliation{Department of Biomedical Engineering, Boston University, Boston, MA, USA}
\author{H. Haghi}
\affiliation{Department of Physics, Ryerson University, Toronto, Ontario, Canada}
\affiliation{Institute for Biomedical Engineering, Science and Technology (IBEST) a partnership between Ryerson University and St. Mike's Hospital, Toronto, Ontario, Canada}
\author{R. Karshafian}
\affiliation{Department of Physics, Ryerson University, Toronto, Ontario, Canada}
\affiliation{Institute for Biomedical Engineering, Science and Technology (IBEST) a partnership between Ryerson University and St. Mike's Hospital, Toronto, Ontario, Canada}
\author{T.M. Porter}
\affiliation{Department of Biomedical Engineering, Boston University, Boston, MA, USA}
\affiliation{Department of Mechanical Engineering, Boston University, Boston, MA, USA}
\author{M.C. Kolios}
\affiliation{Department of Physics, Ryerson University, Toronto, Ontario, Canada}
\affiliation{Institute for Biomedical Engineering, Science and Technology (IBEST) a partnership between Ryerson University and St. Mike's Hospital, Toronto, Ontario, Canada}
\begin{abstract}
The problem of attenuation and sound speed of bubbly media has remained partially unsolved. Comprehensive data regarding pressure-dependent changes of the attenuation and sound speed of a bubbly medium are not available. Our theoretical understanding of the problem is limited to linear or semi-linear theoretical models, which are not accurate in the regime of large amplitude bubble oscillations. Here, by controlling the size of the lipid coated bubbles (mean diameter of $\approx$5.4$\mu$m), we report the first time observation and characterization of the simultaneous pressure dependence of sound speed and attenuation in bubbly water below, at and above MBs resonance (frequency range between 1-3MHz). With increasing acoustic pressure (between 12.5-100kPa), the frequency of the attenuation and sound speed peaks decreases while maximum and minimum amplitudes of the sound speed increase. We propose a nonlinear model for the estimation of the pressure dependent sound speed and attenuation with good agreement with the experiments. The model calculations are validated by comparing with the linear and semi-linear models predictions. One of the major challenges of the previously developed models is the significant overestimation of the attenuation at the bubble resonance at higher void fractions (e.g. 0.005).   We addressed this problem by incorporating bubble-bubble interactions and comparing the results to experiments. Influence of the bubble-bubble interactions increases with increasing pressure.  Within the examined exposure parameters, we numerically show that, even for low void fractions (e.g. 5.1$\times$10$^{-6}$) with increasing pressure the sound speed may become 4 times higher than the sound speed in the non-bubbly medium.
\end{abstract}

\maketitle
\section{Introduction}
Acoustically excited microbubbles (MBs) are present in a wide range of phenomena; they have applications in sonochemistry \cite{1}; oceanography and underwater acoustics \cite{2,3,Leighton1}; material science \cite{4}, sonoluminescence \cite{5} and in medicine \cite{6,7,8,9,10,11,12}. Due to their broad and exciting biomedical applications, {MBs have many emerging applications in diagnostic and therapeutic ultrasound}\cite{12}. MBs are used in ultrasound molecular imaging \cite{6,7} and recently have been used for the non-invasive imaging of the brain microvasculature \cite{7,tanter}. MBs are being investigated for site-specific enhanced drug delivery \cite{8,9,10,11} and for the non-invasive treatment of brain pathologies (by transiently opening the impermeable blood-brain barrier (BBB) to deliver macromolecules \cite{9}; with the first in human clinical BBB opening reported in 2016 \cite{8}).\\ However, several factors limit our understanding of MB dynamics which consequently hinder our ability to optimally employ MBs in these applications. The MB dynamics are nonlinear and chaotic \cite{13,14,15}; furthermore, the typical lipid shell coating adds to the complexity of the MBs dynamics due to the nonlinear behavior of the shell (e.g., buckling and rupture \cite{16}). Importantly, the presence of MBs changes the sound speed and attenuation of the medium \cite{17,18,19,20,21}. These changes are highly nonlinear and depend on the MB nonlinear oscillations which in turn depend on the ultrasound pressure and frequency, MB size and shell characteristics \cite{17,18,19,20}.
The increased attenuation due to the presence of MBs in the beam path may limit the pressure at the target location. This phenomenon is called pre-focal shielding (shadowing) \cite{20,21}. Additionally, changes in the sound speed can change the position and dimensions of the focal region; thus, reducing the accuracy of focal placement (e.g., for targeted drug delivery). In imaging applications, MBs can limit imaging in depth due to the shadowing caused by pre-focal MBs \cite{14,20,21,23,24}. In sonochemistry, changes in the attenuation and the sound speed impact the pressure distribution inside the reactors and reduce the procedure efficacy \cite{18,19}.\\
An accurate estimation of the pressure dependent attenuation and sound speed in bubbly media remains one of the unsolved problems in acoustics \cite{25}. Most current models are based on linear approximations which are only valid for small amplitude MB oscillations \cite{17,Hoff}.  Nonlinear propagation of pressure waves in bubbly media containing coated and uncoated bubbles is theoretically studied using the Korteweg–de Vries–Burgers (KdVB) equation in \cite{kanagawa1,kanagawa2,kanagawa3,kanagawa4} and the nonlinear coefficients of wave propagation were derived through linearization on the effective equations. Linear approximations, however, are not valid for the typical exposure conditions encountered in the majority of ultrasound MB applications.\\ In an effort to incorporate the nonlinear MB oscillations in the attenuation estimation of bubbly media, a pressure-dependent MB scattering cross-section has been introduced \cite{2,26}. While the models introduce a degree of pressure dependency (e.g. only the pressure dependence of the scattering cross section were considered while the damping factors were estimated using the linear model), they still incorporate linear approximations for the calculation of the other damping factors (e.g. liquid viscous damping, shell viscous damping and thermal damping). Additionally, they neglect the nonlinear changes of the sound speed in their approximations. We have shown in\cite{27,28,pof1}, that the changes in liquid and shell viscous damping and thermal dissipation are pressure dependent and significantly deviate from linear predictions even at moderate pressures (e.g. 40 kPa).\\  
Louisnard \cite{18} and Holt and Roy \cite{29} have derived models based on employing the energy conservation principle. In Louisnard\'{}s approach \cite{18} the pressure dependent imaginary part of the wave number is calculated by computing the total nonlinear energy loss during bubble oscillations. However, this method still uses the linear approximations to calculate the real part of the wave number; thus, it is unable to predict the changes of the sound speed with pressure. Holt and Roy calculated the energy loss due to MB nonlinear oscillations and then calculated the attenuation by determining the extinction cross-section \cite{29}. Both approaches in \cite{18} and \cite{29} use the analytical form of the energy dissipation terms.  In the case of coated MBs with nonlinear shell behavior, such calculations are complex and can result in inaccuracies. The existing approaches for sound speed computations based on the Woods model \cite{29,30} are either limited to bubbles whose expansion is essentially in phase with the rarefaction phase of the local acoustic pressure, or {require tedious calculations} in nonlinear regimes of oscillations {from the spine of the $\frac{dP}{dV}$ loops} (e.g. \cite{31}) where {$P$} is pressure and {$V$} is the MB volume.\\ {Sojahrood et al. \cite{34,35} introduced a nonlinear model to calculate the pressure dependent attenuation and sound speed of the bubbly media considering full nonlinear bubble oscillations. Later, Trujillo \cite{trujio} used a similar approach to derive the pressure dependent terms for attenuation and sound speed. However, the models were only validated against the linear model \cite{17}. Pressure dependent predictions of the models were not tested against experiments.}\\
Experimental investigation of the pressure and frequency dependence of the attenuation of bubbly media has been limited to few studies of coated MBs suspensions \cite{23,26,32}. Although pressure dependent attenuation measurements have been performed on mono-disperse bubble populations \cite{26,32}, to our best knowledge there is no study that investigated the pressure dependent sound speed in the bubbly media. Application of mono-disperse or narrow sized bubble populations are critical in observing the influence of pressure on the sound speed. In case of poly-disperse solutions, due to the contributions from different resonant sub-populations at each pressure, inference of the sound speed as a function of acoustic pressure and relating it to the bubble behavior is a near impossible task. Moreover, in the absence of a comprehensive model to calculate the pressure dependent sound speed and attenuation, the relationship between the changes in the acoustic pressure and variations in the sound speed and attenuation are not fully understood.\\ The objective of this work is to gain {fundamental} insight on the simultaneous dependence of sound speed and attenuation on the excitation pressure. To achieve this, we carried out attenuation and sound speed measurements of bubbly water samples at acoustic pressure amplitudes of (12.5 kPa-100 kPa) using monodispersions of stabilized lipid coated MBs. We then derive a simple model describing the relationship between the acoustic pressure and the sound speed and attenuation in bubbly media at resonant low Mach number {(maximum bubble wall velocity $\approx$40m/s)} regimes of oscillation which treats the MB oscillations with their full nonlinearity. Here, we report the first time controlled observations of the pressure dependence of the sound speed of a bubbly medium. The predictions of the model are in good agreement with experiments.\\ 
{ In the appendix we extended the theoretical analysis. First, the model predictions are verified against the linear and semi-linear models for free bubbles, bubbles encapsulated with elastic shells and bubbles immersed in elastic materials. Then, we extended the numerical simulations to higher void fractions and pressure amplitudes. One of the advantages of the introduced model is the ability to take into account bubble-bubble interaction effects. We show that at higher void fractions the interaction reduces the attention and the frequency of the attenuation peak. Thus, one of the well known problems of the linear models \cite{17} which is the significant overestimation of the attenuation near the resonance frequency of bubbles is potentially addressed. We numerically show that with increasing acoustic pressure the attenuation and sound speed can increase significantly (e.g. 4 times the medium sound speed) even for small void fractions (e.g. 5.1$\times$10$^{-6}$). {Finally, noting that the main contribution of this work is laying out the theoretical foundations}, the {potential future} applications of the model for the  accurate shell characterization of lipid coated MBs and the sound propagation in bubbly media are discussed.}\\ 
\section{Methods}
\subsection{Experiments}
\begin{figure}
	\includegraphics[scale=0.3]{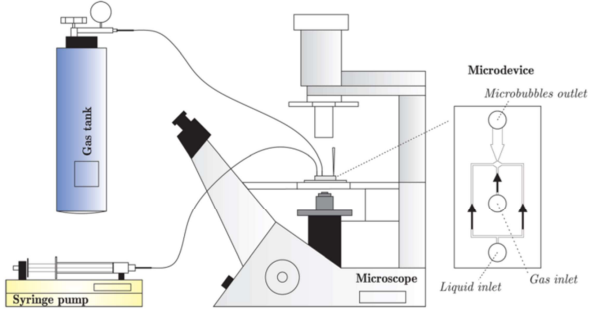}
	\caption{Schematic of the flow-focusing microfluidic procedure for the production of monodisperse lipid coated microbubbles \cite{parrales}.}
\end{figure}
To experimentally explore the pressure-dependent changes of the sound speed and
attenuation for coated MBs first we need to make monodisperse MBs sizes. Commercially available MBs are polydisperse, and at each pressure a subpoulation can be resonant and thus the unambiguous identification of pressure dependent effects will be challenging. Thus, we aim to first produce monodisperse bubbles. Monodisperse lipid shell MBs were produced using flow-focusing in a microfluidic device as previously described \cite{32,parrales}. {Figure} 1 shows the schematic of the procedure. Figure 2 shows the size distribution of the MBs in our experiments. The setup for the attenuation and sound speed measurements is the same as the one used in \cite{parrales}. Figure 3 shows the setup for the measurements of the attenuation and sound speed.  A pair of single-element 2.25 MHz unfocused transducers (Olympus, Center Valley,$\approx$ PA; bandwidth 1-3.0 MHz) were aligned coaxially in a tank of deionized water and oriented facing each other. Monodisperse MBs were injected into a sample chamber that was made with a plastic frame covered with an acoustically transparent thin
film. The dimensions of sample chamber were 1.4 x 3.5 x 3.5 cm (1.4 cm acoustic
path length), and a stir bar was used to keep the MBs dispersed. The
transmit transducer was excited with a pulse generated by a pulser/receiver
(5072PR, Panametrics, Waltham,~MA) at a pulse repetition frequency (PRF) of 100
Hz. An attenuator controlled the pressure output of the transmit transducer
(50BR-008, JFW, Indianapolis, IN), which was calibrated with a 0.2-mm broadband
needle hydrophone (Precision Acoustics, Dorset, UK). Electric signals generated
by pulses acquired by the receive transducer were sent to the Gagescope
(Lockport, IL) and digitized at a sampling frequency of 50 {MHz}. All received
signals were recorded on a desktop computer (Dell, Round Rock, TX) and processed
using Matlab software (The MathWorks, Natick, MA).~ The peak negative pressures
of the acoustic pulses that are used in experiments were 12.5, 25, {50, and 100}
kPa.\\ Attenuation and sound speed were then calculated by comparing the power and phase spectra of the received signals before and after injection of the MBs in to the chamber:
{\begin{equation}
\alpha(\omega)=\frac{20}{d}\log _{10}\left(\frac{P_l}{P_{MBs}}\right)
\label{eq:1}
\end{equation}}
and
\begin{equation}
C(\omega)=\frac{\omega d C_l}{\omega d +C_l(\phi_{MBs}-\phi_l)}
\label{eq:2}
\end{equation}
 where $\alpha(\omega)$ is the frequency dependent attenuation of the bubbly medium, $d$ is acoustic path length, $P_l$ and $P_{MBs}$ are the power spectrum of the received signals in the absence and in the presence of the MBs respectively. $C(\omega)$ is the frequency dependent sound speed of the bubbly medium, $C_l$ is the sound speed in the liquid in the absence of the MBs, and $\phi_{MBs}$ and $\phi_l$ are the phase of the received signal in the presence and absence of the MBs respectively.    
\begin{figure}
	\includegraphics[scale=0.5]{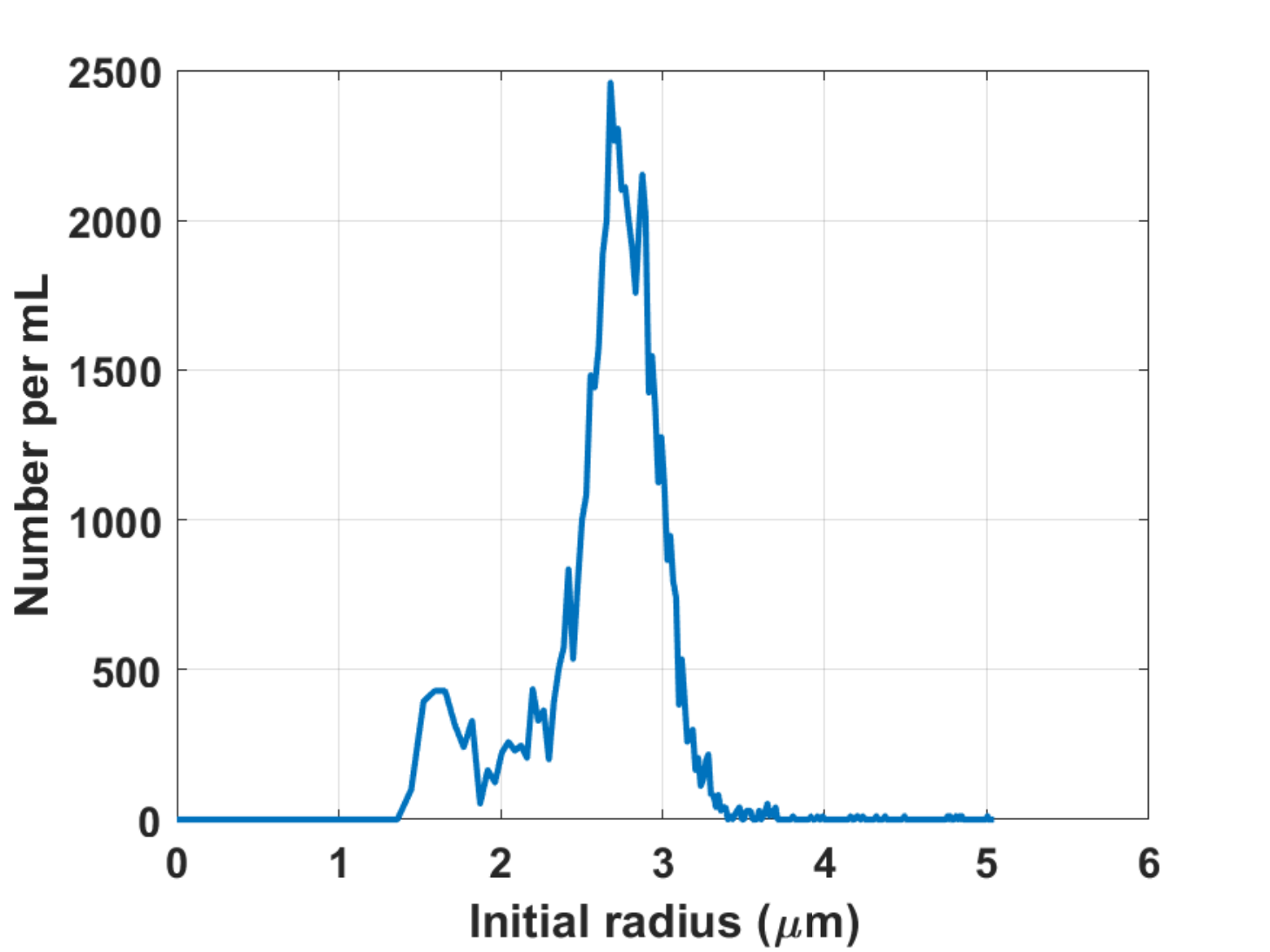}
	\caption{ Size distribution of the MBs in the experiments measured by Coulter-counter. The volume fraction {$\beta_0$} can directly be calculated from the size distribution.}
\end{figure}

\begin{figure}
	\includegraphics[scale=1.2]{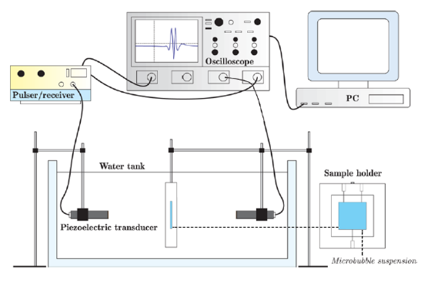}
	\caption{The schematic of the setup for the measurements. A broadband pulse with 2.25 MHz center frequency is transmitted by the transducer on the right hand. After propagation through the chamber, the pulse will be received by the transducer on the left hand side.}
\end{figure}

\subsection{Theoretical and numerical approach}
\subsubsection{Caflisch model}
To drive the terms for pressure dependent attenuation and sound speed in bubbly media, we start with the Caflisch equation \cite{37} for the propagation of the acoustic waves in a bubbly medium.
The model treats the bubbly media as as a continuum. This implies that the radial
oscillations of all the bubbles inside a small volume of the mixture at point $r$, can be described by a continuous spatio-temporal radius function $R(r,t)$.  Using the mass and momentum conservation in the mixture we have:
\begin{equation}
\frac{1}{\rho_l {C_l}^2}\frac{{\partial{}}P}{\partial{}t}+\nabla{}.v=\sum_{i=1}^N\frac{\partial{}\beta_i}{\partial{}{t}}
\label{eq:3}
\end{equation}
\begin{equation}
\rho_l \frac{\partial{}v}{\partial{}t}+\nabla{}p=0
\label{eq:4}
\end{equation}
In these equations, {$P(r,t)$} is pressure, {$v(r,t)$ is the velocity field, . is the dot product}, $\ C_l$ is the speed of sound in the liquid in the absence of bubbles, ${\rho{}}_l$ is the liquid density and  ${\beta{}}_i$ is the local volume fraction occupied by the gas at time $t$ of  the  \textit{ith} microbubbles (MBs). ${\beta{}}_i$ is given by ${\beta{}}_i(t)=\frac{4}{3}\pi{}{R_i(t)}^3N_{i}$  where {$R_i$\textit{(t)}} is the
instantaneous radius of the MBs with initial radius of $R{_{0i}}$ and $N_{i}$ is the number of the corresponding MBs per unit volume in the medium. The summation is performed over the whole population of the MBs inside the volume.\\ 
The velocity field can be eliminated between {Eqs.} \ref{eq:3} and \ref{eq:4} to yield an equation involving only
the pressure field \cite{17,18,19}:
\begin{equation}
{\nabla{}}^2\left(P\right)=\frac{1}{C_l^2}\frac{{\partial{}}^2P}{\partial{}t^2}-{\sum_{i=1}^N\
	\ \rho_l{}}\frac{{\partial{}}^2{\beta{}}_i}{\partial{}t^2}
\label{eq:5}
\end{equation}
{To calculate the attenuation and sound speed we need to determine the wave number and transfer the time averaged form of Eq. \ref{eq:5} in the form of the Helmholtz equation given by:}
\begin{equation}
{\nabla{}}^2\left(P\right)+k^2(P)=0
\label{eq:6}
\end{equation}
 {where $k$ is the wave number (\textit{k=k$_{r}$-i$\alpha{}$}). The sound speed can be calculated from \textit{k$_{r }$} which is the real part of the wave number, and the attenuation \textit{$\alpha{}$} from the imaginary part of the wave number.}\\ 
A general solution to Eq. \ref{eq:6} is given as:
\begin{equation}
P(r,t)=\frac{1}{2}p(r)e^{(-i\omega t)}+\frac{1}{2}\bar{p}(r)e^{(i\omega t)}
\end{equation}
\label{eq:7}
where $\omega$ is the angular frequency and $\bar{p}$ is the complex conjugate of $p$. Each term on the right side is a particular solution to Eq. \ref{eq:6}. Mathematically, the Helmholtz equation (Eq. \ref{eq:6}) is a homogeneous partial differential equation, thus each particular solution is also a solution to the 
Eq. \ref{eq:6}. Thus, if we input the two solutions in Eq. \ref{eq:6} we will have:
\begin{equation}
{\nabla{}}^2\left({p}\right)=-\frac{\omega^2}{C_l^2}{p}-2{\sum_{i=1}^N\
	\ \rho_l{}}\frac{{\partial{}}^2{\beta{}}_i}{\partial{}t^2}
\label{eq:8}
\end{equation}
and
\begin{equation}
{\nabla{}}^2\left(\bar{p}\right)=-\frac{\omega^2}{C_l^2}\bar{p}-2{\sum_{i=1}^N\
	\ \rho_l{}}\frac{{\partial{}}^2{\beta{}}_i}{\partial{}t^2}
\label{eq:9}
\end{equation}
where $\bar{p}$ is the complex conjugate of $p$.\\ 
Next step is to calculate the wave number in the presence of the bubbles: $k^2=-\frac{{\nabla{}}^2\left(P\right)}{P}$.\\ To achieve this, Eq. \ref{eq:8}  was multiplied by  $\frac{\ \bar{p}}{p\bar{p}}\ $ and  Eq. \ref{eq:9} was multiplied by   $\frac{p}{p\bar{p}}$.  The pressure dependent real and imaginary parts of $k^2$ were derived using the time average of the results of the addition and subtraction of the new equations:
\begin{equation}
\langle\Re(k^2)\rangle=\frac{{\omega{}}^2}{C_l^2}+\frac{2{\rho_l{}}}{T{\left\vert{}p\right\vert{}}^2}\sum_{i=1}^N\int_0^T{\Re(p)}\frac{{\partial{}}^2{\beta{}}_i}{\partial{}t^2}dt
\label{eq:10} 
\end{equation}
\begin{equation}
\langle\Im(k^2)\rangle =\frac{2{\rho_l{}}}{T{\left\vert{}p\right\vert{}}^2}\sum_{i=1}^N\int_0^T {\Im(p)}\frac{{\partial{}}^2{\beta{}}_i}{\partial{}t^2}dt
\label{eq:11} 
\end{equation}
\raggedbottom \raggedbottom 
where $\Re$ and $\Im$ denote the real and imaginary parts respectively,
\textit{$<$$>$} denotes the time average, and \textit{T} is the time averaging interval. The
contribution of each MB with ${\beta{}}_i$ is summed. Using Eqs.
\ref{eq:10} and \ref{eq:11}, we can now calculate the pressure-dependent sound speed and
attenuation in a bubbly medium. To do this, the radial oscillations of the MBs in
response to an acoustic wave need to be calculated first. {Equations} \ref{eq:10} and \ref{eq:11} need to be
solved by integrating over the  ${\beta{}}_i$ of each of the MBs in the population. The advantage of this technique is the simultaneous calculation of the pressure dependent sound speed and attenuation in the bubbly medium.\\ This approach is verified in Appendix A against the linear model and in Appendix B against the semi-linear model.
\raggedbottom \raggedbottom 
\begin{table*}
	\begin{tabular}{ |p{2cm}||p{3.5cm}|p{2cm}|p{2cm}|p{2cm}|}
		\hline
		\multicolumn{5}{|c|}{Thermal parameters of the air and {C$_3$F$_8$} at 1 atm} \\
		\hline
		Gas type  & {L(W/mK)} &{$c_p$(kJ/kgK)} &{$c_v$ (kJ/kgK)}& {$\rho_g$ (kg/m$^3$)}\\
		\hline
		Air \cite{49}   & $0.01165+C\times T^2$ &1.0049&   0.7187&1.025\\
		C3F8 \cite{50} &   0.012728  & 0.79   &0.7407&8.17\\
		\hline
	\end{tabular}
	\caption{Thermal properties of the gases used in simulations({C=5.528$\times$10$^{25}$ W/mK$^2$)}}.
	\label{table:1}
\end{table*}
\subsubsection{Lipid coated bubble model including bubble-bubble interaction}
\raggedbottom \raggedbottom 
To numerically simulate the attenuation and sound speed, radial oscillations of the lipid coated MBs with a size distribution given in Fig. 2 should be simulated first. Moreover, effect of MB-MB interaction must be included as the pulsation of each MBs generates a pressure at the location of each MBs in its vicinity, which consequently may influence each MB behavior. To model the MB oscillations the Marmottant model \cite{16}, which accounts for radial-dependent shell properties of lipid coated MBs, was modified to include MB multiple scattering using the approach introduced in \cite{58}: 
\raggedbottom \raggedbottom 
\begin{widetext}
\begin{equation}
\begin{gathered}
R_i\ddot{R_i}+\frac{3}{2}\dot{R_i}^2=\frac{1}{\rho}\left([P_0+\frac{2\sigma(R_{0i})}{R_{0i}}](\frac{R_i}{R_{i0}})^{3k}(1-\frac{3k}{C_l}\dot{R_i})-\frac{2\sigma(R_{i})}{R_{i}}-\frac{4\mu\dot{R_i}}{R_i}-\frac{4\kappa_s\dot{R_i}}{R_i^2}-P_0-P_{ac}(t)-{\rho\sum_{j=1,j\neq i}^{N}\frac{R_j}{d_{ij}}(R_j\ddot{R_j}+2\dot{R_j}^2)}\right) 
\end{gathered}
\label{eq:12}
\end{equation}
\end{widetext}
\raggedbottom
In this equation $R_{i0}$ is the initial radius of the ith bubble, $P_0$ is the atmospheric pressure $C_l$ is the sound speed, $k$ is the polytropic exponent, $\mu$ is the viscosity of the liquid. {The last term in the right hand side of Eq. \ref{eq:12} defines the pressure radiated by neighboring MBs with $R_j$ at the location of ith MB and $d_{ij}$ represents the distance between centers of ith and jth bubble} {$\sigma(R_{i})$} is the surface tension {acting on the ith MB} which is a function of bubble radius and is given by :  
{\begin{equation}
\sigma(R_{i})=
\begin{dcases}
0    \hspace{1cm} if \hspace{0.5cm} R_i<=R_{bi}\\
\chi \left ({(\frac{R_i}{R_{bi}})}^2-1\right)  \hspace{1cm}  if \hspace{0.5cm} R_{bi}<R_i<R_{ri}\\
\sigma_{water}   \hspace{1cm}           if \hspace{0.5cm} R_i>=R_{ri}
\end{dcases}
\label{eq:13}
\end{equation}}
where $\sigma_{water}$ is the water surface tension, $R_{bi}={R_{0i}}/{\sqrt{1+\frac{\sigma(R_{0i})}{\chi}}}$ is the buckling radius, $\sigma(R_{0i})$ is the initial surface tension, $\chi$ is the shell elasticity. $\kappa_s$ in Eq. \ref{eq:12}, is the surface diltational viscosity. {$R_{ri}$ is the rupture radius (break up radius in this paper similar to \cite{helr2,sojahroodr2}) and is given by $R_r=R_{bi}\sqrt{1+{\sigma_{ri}}/{\chi}}$ where $\sigma_{ri}$ is the rupture surface tension of the ith MB.}\\
Using the approach in \cite{Man1,55,59}, Eq.\ref{eq:12} can be written in a matrix format as:
\begin{figure*}
	\includegraphics[scale=0.55]{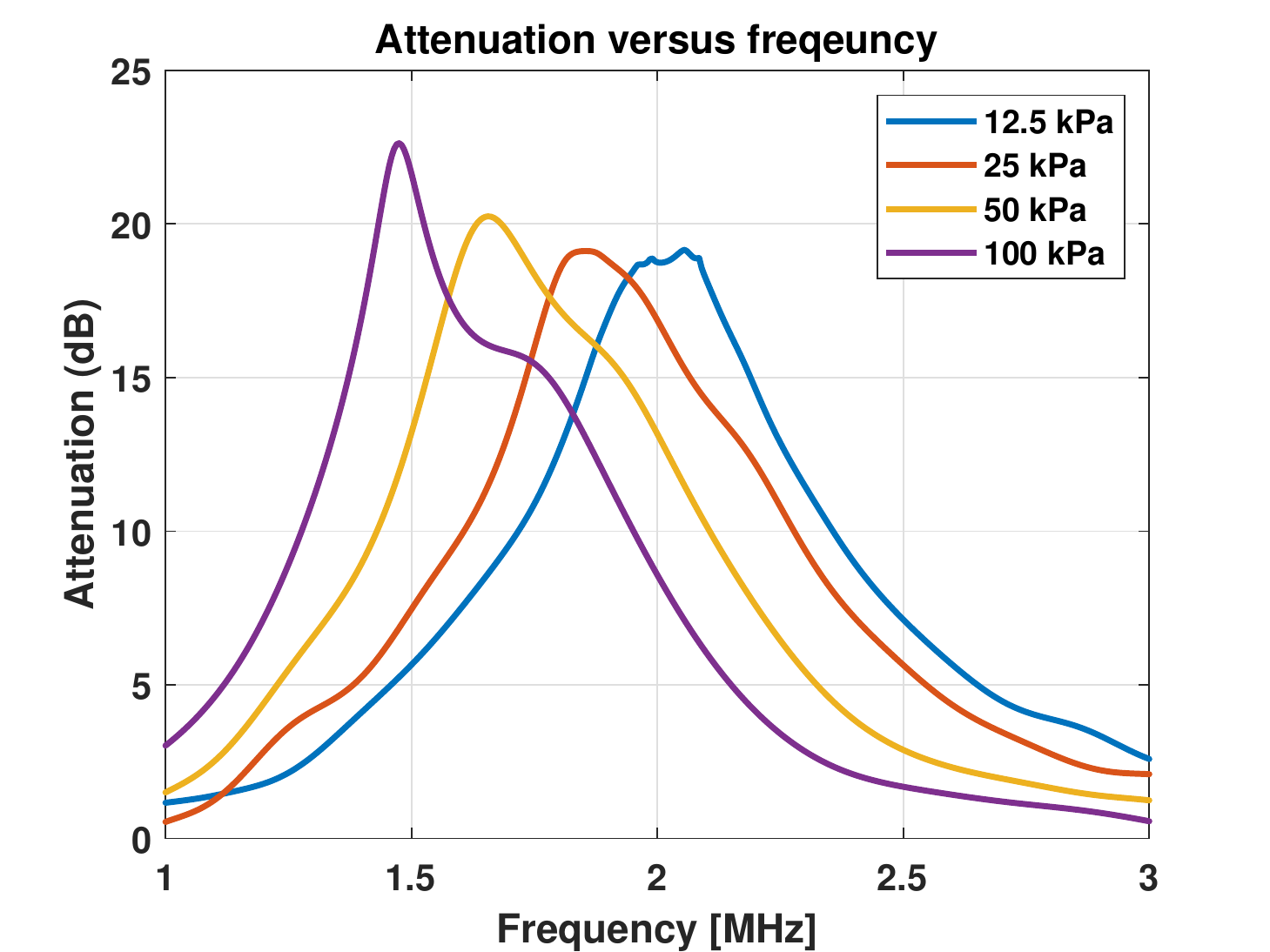}  \includegraphics[scale=0.55]{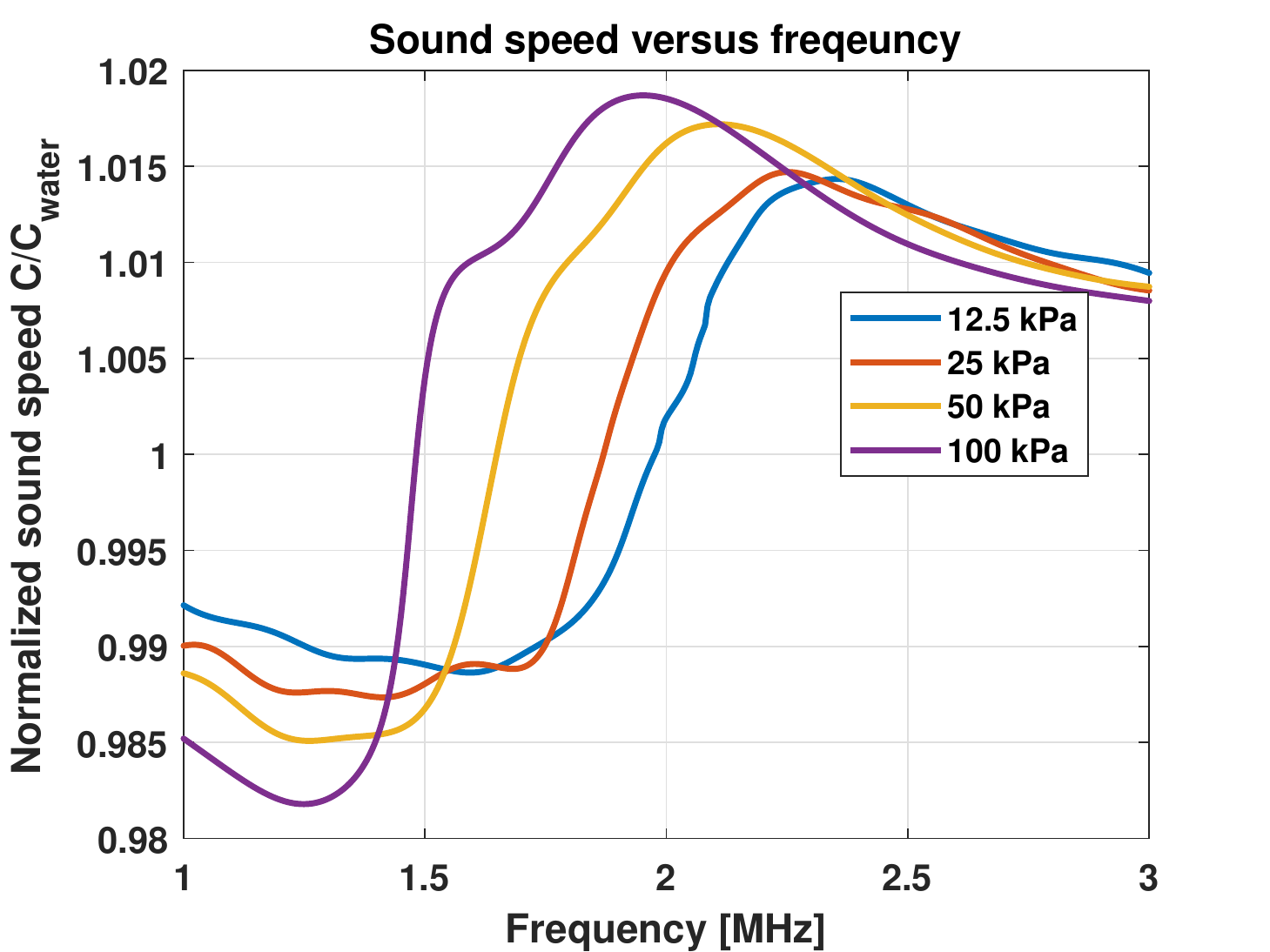}\\
(a) \hspace{8cm} (b)
	\caption{ Experimentally measured a) attenuation  and b) sound speed of the bubbly medium for four different pressures.}
\end{figure*}
\begin{equation}
\begin{pmatrix} 
\ddot{R}_1\\ \\
\ddot{R}_2\\ \\
...\\ \\
\ddot{R}_N\\ \\
\end{pmatrix}=\begin{pmatrix} 
R_1 & \frac{R_2^2}{d_{12}} & ... & \frac{R_N^2}{d_{1N}}\\ \\
\frac{R_1^2}{d_{21}} & R_1 & ... & \frac{R_N^2}{d_{2N}}\\ \\
... & ... & ... & ... \\ \\
\frac{R_1^2}{d_{N1}} & \frac{R_2^2}{d_{N2}} & ... & R_N\\ \\
\end{pmatrix}^{-1}\begin{pmatrix} 
A_1\\ \\
A_2 \\ \\
...\\ \\
A_N
\end{pmatrix}
\label{eq:14}
\end{equation}
where
\begin{equation}
\begin{gathered}
A_i=\frac{1}{\rho}\left(\left[P_0+\frac{2\sigma(R_{0i})}{R_{0i}}\right]\left(\frac{R_i}{R_{i0}}\right)^{3k}\left(1-\frac{3k}{C_l}\dot{R_i}\right)-\frac{2\sigma(R_{i})}{R_{i}}\right.\\
-\left.\frac{4\mu\dot{R_i}}{R_i}-\frac{4\kappa\dot{R_i}}{R_i^2}-P_{ac}(t)-\frac{3\rho}{2}\dot{R_i}^2\right)-\sum_{j\neq i}^{}\frac{2R_j\dot{R_j}^2}{d_{ij}}
\end{gathered}
\label{eq:15}
\end{equation}
The gas inside the bubble is C$_3$F$_8$ and the properties are given in {Table 1}. We have neglected the thermal effects in the numerical simulations. In \cite{28} we have shown that in case of the coated bubbles that enclose C$_3$F$_8$ gas cores, thermal dissipation has negligible contribution to the overall dissipated power in the ultrasound exposure range that was used here. Thus, thermal effects are neglected to reduce the complexity of the multiple scattering equation that is used in the simulations  (Eq.\ref{eq:12}).
\subsubsection{Simulation procedure}
 {For the numerical simulations, the experimentally measured size of the MBs were used as the input for initial MB sizes. Size and concentration of the MBs in the experiments are given in Fig. 2. There are 63 size bins in the range between 1.44$\mu$m$<R_0<$3.3$\mu$m. Each $R_{0i}$ has a number density of $N_i$. These 63 MBs were distributed randomly in a cube of diameter 0.1{cm} to simulate the total concentration of {6.3$\times$10$^{4}$ MBs/mL} in the experiments. { The total gas volume in the experiments was $\approx$5.1$\times$10$^{-12}$m$^3$/mL. When written in m$^3$/m$^3$, this results in $\beta_0$ of 5.1$\times$10$^{-6}$. The minimum distance between the MBs was set to be 15$\mu$m to avoid MBs collisions \cite{wang1,zhang1,ouch}}. At each case (12.5, 25, 50 and 100kPa), the numerical simulations were repeated for 40 frequency values between 1-3MHz. At each of those frequencies, the corresponding pressure was extracted from the driving experimental pressure spectrum. To calculate the attenuation at a given frequency and pressure, then, Eq. \ref{eq:12} was solved and the radial oscillations of each MB was recorded for the duration of the experimental pulse.  Next the radial oscillations were used as an input to calculate the attenuation and sound speed in Eq. \ref{eq:10} and \ref{eq:11}. Since the acoustic pressure that excites the MBs in Eq. \ref{eq:12} is given by $P$=$P_a$$\sin(2\pi ft)$, thus $\Re(p)$=$P_a$$\sin(2\pi ft)$ and $\Im(p)$=-$P_a$$\cos(2\pi ft)$. Similar to \cite{trujio} this is simply achieved by an angle shift ($P_a$e$^{(i\omega t-\pi/2)}$=$P_a$$\left(\sin(2\pi ft)-i\cos(2\pi ft)\right)$. Thus, real and imaginary parts of $k^2$ can be calculated from:
	\begin{equation}
	\langle\Re(k^2)\rangle=\frac{{\omega{}}^2}{C_l^2}+\frac{2{\rho_l{}}}{T{P_a}}\sum_{i=1}^{63}N_i\int_0^T{\sin(2\pi ft)}\frac{{\partial{}}^2{\beta{}}_i}{\partial{}t^2}dt
	\label{eq:16} 
	\end{equation}
	\begin{equation}
	\langle\Im(k^2)\rangle =-\frac{2{\rho_l{}}}{T{P_a}}\sum_{i=1}^{63}N_i\int_0^T {\cos(2\pi ft)}\frac{{\partial{}}^2{\beta{}}_i}{\partial{}t^2}dt
	\label{eq:17} 
	\end{equation}
where T is the pulse length in the experiments.\\
Attenuation and sound speed can be calculated from:
\begin{equation}
\alpha(p,f)=\Im \left(\sqrt{\Re(k^2)+i\Im(k^2)}\right)
	\label{eq:18} 
\end{equation}
    \begin{equation}
  C(p,f)=\frac{{\omega}}{\Re \left(\sqrt{\Re(k^2)+i\Im(k^2)}\right)}
  	\label{eq:18} 
  \end{equation}
 Dimension of the $\alpha(p,f)$ is in Np/m. To calculate the attenuation in dB/m Eq. \ref{eq:18} is multiplied by 8.6858.\\
This procedure was done for different values of the shell elasticity (0.1 N/m$<$\textit{$\chi{}$} $<$2.5 N/m), initial surface tension (0 N/m $<$ \textit{$\sigma{}$}($R_0)\ $$<$ 0.072 N/m),  shell viscosity (1$\times$10$^{-9}$ kg/s $<$${\mu{}}_s$$<$1$\times$10$^{-7}$ kg/s), and rupture radius {$(R_{0i}<R_{ri}$ $<$ 2$R_{0i})$}. The best fit results are presented. These values are within the parameter ranges that were previously reported for lipid coated MBs \cite{26,42,43,44,45,Hel1,Hel2,Seg2}.}
\section{Results}
\begin{figure*}
	\includegraphics[scale=0.19]{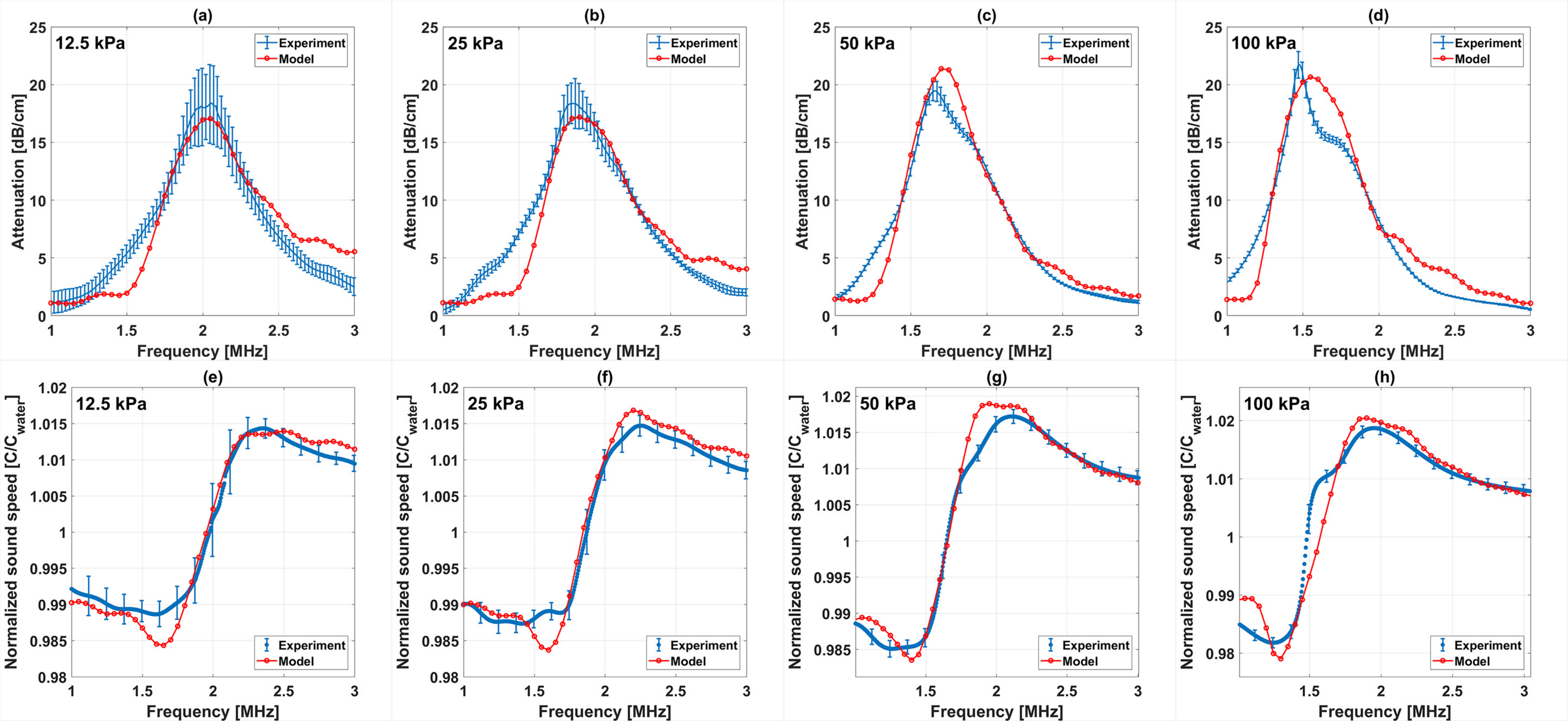}
	\caption{ Experimentally measured (blue) and simulated (red) attenuation of the
		sample for a) 12.5 kPa, b) 25 kPa, c) 50 kPa and d) 100 kPa . Sound speed of the
		sample for e) 12.5 kPa, f) 25 kPa, g) 50 kPa and h) 100 kPa . Errors bars represent the standard deviation.}
\end{figure*}
{Figures 4a-b} show representative samples of the experimentally measured attenuation and sound speed of the mixture respectively. The attenuation of the bubbly medium increases as the pressure increases from 12.5 kPa to 100 kPa and the frequency of the maximum attenuation decreases from ~2.045 MHz to ~1.475 MHz. The maximum sound speed of the medium increases with pressure and the corresponding frequency of the maximum sound speed decreases by pressure increases. To our best knowledge this is the \textit{first} experimental demonstration of the pressure dependence of the sound speed.\\ It is also interesting to note that at the pressure dependent resonance frequencies (measured attenuation peaks in Fig. 4a) the sound speed is \textit{equal} to the sound speed of the water in the absence of bubbles.\\ 
To compare the predictions of the model with experiments, Figs. 5a-h illustrate the results of the experimentally measured (blue) (with standard deviations of the 100 data points at each condition) and numerically simulated (red) sound speed of the medium as a function of frequency, for 4 different pressure exposures of (12.5, 25, 50 and 100 kPa).
The shell parameters that were used to compare with the experimental results are \textit{$\chi{}$}=1.1 N/m, \textit{$\sigma{}$}($R_0)$= 0.016 N/m,\ ${\mu{}}_s$=7$\times$10$^{-9}$ {kg/s} and
$R_{breakup}$=1.1$R_0$. These values are within the estimated values for lipid coated MBs \cite{26,42,43,44,45,Hel1,Hel2,Seg2}.\\ As the pressure increases, the frequency at which the maximum attenuation occurs (which indicates the resonance frequency) decreases (from ~2.02 MHz at 12.5 kPa to 1.475 MHz at 100 kPa) and the magnitude of the attenuation peak increases (from 16.5 dB/cm at 2 MHz to 21.8 dB/cm at 1.475 MHz).
At 12.5 kPa and for frequencies below $\sim$ 2
MHz, the speed of sound in the bubbly medium is smaller than the sound speed of
water. Above 2 MHz, speed of sound increases and reaches a maximum at 2.25 MHz
with a magnitude of $\sim$ 1.015$C_l$.  At $\sim$
{2 MHz} the sound speed is equal to $C_l$. This is also the frequency where the attenuation is maximum. According to the linear theory \cite{56} at the resonance frequency the sound speed of the bubbly medium is equal to the sound speed of the medium without the bubbles. As the pressure increases to 25, 50 and 100 kPa the frequency at which the speed of sound in the bubbly sample is
equal to $C_l$ decreases to 1.87, 1.65 and 1.48 MHz respectively. The
frequency at which the maximum sound speed occurs decreases as the pressure increases and the magnitude of the maximum sound speed increases to $\sim$ 1.019$C_l$
at 100 kPa. The minimum sound speed decreases from $\sim$ 0.989$C_l$ at 12.5 kPa (peak at1.6 MHz) to $\sim$ 0.981$C_l$ at 100 kPa and (1.25 MHz). 
At each pressure, the frequency at which attenuation is maximized (the pressure dependent resonance) is approximately equal to the frequency where the sound speed becomes equal to $C_l$. Thus, it can be observed that, even at the pressure dependent resonance frequency ($PDf_r$) which is a nonlinear effect \cite{14}, the velocity is in phase with the driving force and $\frac{C}{C_l}=1$ (page 290) \cite{56}. This observation is also consistent with the numerical results of the uncoated bubble in Appendix D in Figs. 6a-f and Figs. 7a-f.\\
\section{Discussion}
{In ultrasound applications, the bubble activity in the target heavily depends on the acoustic pressure that can reach the target. However, there is no comprehensive model that can be used to estimate attenuation and sound speed changes for large amplitude bubble oscillations. The majority of published works employed linear models that are derived using the Comander and {Prosperetti} approach \cite{17}. Since these models are developed under the assumption of very small amplitude bubble oscillations ($R=R_0(1+x)$ where $x\ll 1$), they are not valid for most ultrasound MB applications \cite{18,louisnard1,louisnard2}. On the other hand  recent semi-linear models \cite{2,18,26} still have some inherent linear assumptions.}\\ {Sojahrood et al, \cite{pof1} classified the pressure dependent dissipation regimes of uncoated and coated bubbles over a wide range of pressure and for frequencies below, at, and above resonance in tandem with bifurcation analysis of the bubble radial oscillations. They showed that dissipation due to shell viscosity, radiation, thermal and liquid viscosity exhibit strong pressure dependency and linear estimations are inaccurate even at pressures {\textit{as low as 25kPa}\cite{27,28,pof1}.\\ The models that include both the pressure dependent real and imaginary part of $k^2$ \cite{34,35,trujio} have not been compared to experiments in pressure dependent regimes. Trujillo\cite{trujio} validated their model against only the linear model and experiments at very low excitation pressures ($\approx$ 10Pa). Despite good agreement with the linear model, the model still suffers from significant overestimation of the attenuation near the bubble resonance.}  Pressure dependent behavior of the attenuation and sound speed thus remained limited to numerical studies \cite{34,trujio2} and qualitative comparisons \cite{36}. Additionally, simultaneous pressure dependent sound speed and attenuation has not been reported experimentally.\\ Due to the polydisersity of MBs in the majority of applications and the absence of a comprehensive model, {thus}, there is no detailed study on the simultaneous changes of the attenuation and sound speed as a function of acoustic pressure. {In case of the polydisperse conventional contrast agents (1-10$\mu$m size range), at each exposure parameter only a small sub-population exhibit resonant behavior. Thus, the overall response of the MBs may mask the intricate changes in MB dynamics \cite{23,acs} and the corresponding changes in the attenuation and sound speed, thus observation of such changes may not be possible. Moreover, due to MB-MB interactions and mode locking between different size MBs\cite{59}, interpreting the the overall response of the MBs becomes more difficult. The monodisperse agents used in this work, exhibit unifrom acoustic behavior \cite{26} which enhances the nonlinear effects. Due to this higher potency \cite{23,26,acs}, observation of pressure dependent changes are simpler.}\\ Here, by confining the MB sizes (using monodisperse MB solutions), we report the first time, {controlled, simultaneous} experimental observations of the pressure dependent sound speed {and attenuation} in a bubbly medium. {The complex influence of the bubble-bubble interactions \cite{36,pof2,pof3,guedra} were minimized by using lower void fractions to allow {for an unambiguous} visualization of the simultaneous relationship of sound speed and attenuation with acoustic pressure}. We {then presented} a pressure dependent attenuation-sound speed model for the low mach number regimes of oscillations. Predictions of the model are in good agreement with experimentally measured attenuation and sound speed of mono-dispersions of lipid coated bubbles. 
{\subsection{Justification for the choice of acoustic exposure parameters to achieve strong and controllable nonlinear oscillations during experiments}
   The choice of the acoustic exposure parameters ($P_a$ between 12.5-100kPa) would initially \textit{appear} to result in linear oscillations; however the choice of resonant lipid coated MBs results in strong nonlinear oscillations within the exposure parameter ranges that are tested. The main reasons for the chosen pressure ranges are discussed.
\subsubsection{Enhanced nonlinearity due to the Lipid coating}
Lipid coated MBs are interesting highly nonlinear oscillators. The nonlinear behavior of the shell, including buckling and rupture \cite{10,over} is intermixed with the nonlinearity of the MB oscillations itself. This results in the generation of the \textit{nonlinear oscillations at pressures as low as 10kPa} \cite{pof2,sojahroodr2,Hel2,over,frink}.  The presence of the shell increases the viscosity at the bubble wall, however, despite this, the pressure threshold for nonlinear oscillations of the lipid coated bubbles is significantly lower than the pressure required for the 1/2 order subharmonic (SH) generations for the uncoated bubbles \cite{frink}. Sojahrood et al \cite{pof2,sojahroodr2} compared the nonlinear oscillations of the uncoated and lipid coated bubbles using bifurcation analysis \textit{over a wide range of frequency and pressures}. They showed that even at pressures below 30kPa, the lipid coated MB exhibits strong nonlinearity including the generation of 1/2 and 1/3 order SHs and higher order superharmonic resonances \cite{pof2,sojahroodr2,Hel2,over,frink}. Attenuation measurements by Segers et al \cite{26} reveals strong nonlinear behavior (significant 62.5$\%$ reduction of the frequency of the attenuation peak together with the generation of new peaks at frequencies below and above main resonance) in the 10-50kPa range.   
\subsubsection{Interrogation of the bubble oscillations near resonance}
In our study, experiments were performed using frequencies in the near vicinity of the bubble resonance frequency. We have previously shown that (for uncoated and viscoelastic shell MBs \cite{27,28,pof1}), in the vicinity of linear resonance frequency (0.5$f_r$-2$f_r$) even an increase in pressure from 1kPa to 30-50kPa will result in significant changes in the dissipated powers and 50-100$\%$ expansion ratio near the bubble resonance. Additionally, the frequency of the main resonance peak shifts and new sub-resonant peaks in the power dissipation curves appear. This behavior is also studied in detail in \cite{pros1}. These studies predict expansion ratio of up to 280$\%$ for pressures below the Blake threshold and in the vicinity of resonance.\\
This non-linearity is enhanced for lipid coated MBs. Near the linear resonance frequency of the lipid coated MBs \cite{26,Hel2,over,sojahroodr2}, even changes as low as 20kPa can result in $\approx$67$\%$ shift in the resonance frequency of the MBs. The shift in the resonance frequency is highly nonlinear and can not be described by linear bubble models (pure viscoelastic shell models \cite{Hoff} or Rayleigh-Plesset type models for the uncoated bubble\cite{22}). In our experiments we went far above the onset of nonlinear oscillations. The mean initial radius of the MBs was approximately 2.7$\mu$m. A lipid coated MB with the shell parameters obtained in our study, experiences 4.5, 9, 25 and 61$\%$ expansion at 12.5, 25, 50 and 100 kPa respectively. Thus, at pressures even as low as 12.5kPa the oscillations are beyond the linear regime (expansion ratio $<$1$\%$). The same MB attained a maximum wall velocity of 30 m/s at 100kPa.\\ For subresonant regimes (well below the resonance frequency of typical clinical MBs of diameters 1-2$\mu$m) that are used in drug delivery studies (500kHz and pressures $\approx$ 500kPa \cite{rafi}) and for the case of stable and non-destructive oscillations ($R/R_0<2$\cite{flynn}), a coated MB with $R_0$=1$\mu$m reaches velocities of $\approx$ 30-40 m/s before possible MB destruction \cite{pof1}. Thus, both radial expansion and wall velocities in our experiments were in the nonlinear regime and of practical relevance.\\ In sonochemistry, frequencies well below the MB resonance are used. Sonochemical procedures typically use a frequency of 20kHz with mean diameters of the MBs of approximately 10$\mu$m\cite{jin,size1,size2}. For a an uncoated bubble this results in a non-damped resonance frequency of $\approx$ 720kHz. Below the Blake threshold of $\approx$103kPa, the oscillation amplitude grows slowly with increasing pressure ($\approx$ 45$\%$ expansion ratio and 0.3m/s wall velocity at 93kPa). In the vicinity of the Blake threshold and higher, the maximum bubble oscillation increases exponentially with pressure increases. At the Blake threshold the bubble expansion ratio is $\approx$74$\%$ with maximum wall velocity of 1.6m/s. As soon as the pressure increases above the Blake threshold the oscillation amplitude and bubble wall velocity increases rapidly. At 113kPa, the expansion ratio is 300$\%$ and wall velocity reaches 40m/s. In such oscillation regimes at frequencies well below the MB resonance, the non-linearity sets in more strongly near and above the Blake threshold\\  Since in our study we interrogated the oscillations near resonance, and additionally used lipid coated MBs, the onset of nonlinearity was at much lower pressures than have been predicted for uncoated microbubbles (e.g. 12.5-25kPa). 
\subsubsection{Possible MB destruction at higher pressures}
Increasing the pressure beyond the 100kPa may result in MB destruction. In our work, 81$\%$ of the MB population was between 2.5-3.1$\mu$m. At 100kPa, the MBs with initial radius of 3.1$\mu$m are predicted to have expansion ratio of 76$\%$ with wall velocities near 40m/s. This is close to the minimum MB destruction threshold of 100$\%$ expansion ratio ($R/R_0$=2\cite{flynn} and for a review of the minimum threshold please see discussion in \cite{14,54}). Since we interrogated the MBs in the vicinity of their resonance, at slightly higher pressures (e.g. for an uncoated MB with $R_0$=2$\mu$m, the MB is likely to be destroyed at 125kPa and 80kPa at $f=f_r$ and $f=0.9f_r$\cite{pof1}), a sub-population of the MBs may undergo destruction, thus interpretation of the attenuation and sound speed results may be difficult. Since, the goal of this study was reporting the pressure dependence of sound speed and attenuation and verifying the predictions of the model we choose the pressure in a range that would \textbf{minimize} the unwanted bubble destruction. Nevertheless, based on numerical simulations, we believe that we achieved wall velocities of 30-40 m/s which are in the range of the maximum predicted wall velocities during stable oscillations for MBs with initial radii between 1-10$\mu$m\cite{pof1,54}.   
\subsubsection{Inaccuracies due to the increased attenuation at higher pressures}
Another limitation of using higher pressures is the potential increased attenuation from MBs. This creates a steeper pressure gradient within the experimental chamber and thus complicates the interpretation of the results. \\
Therefore, we conclude that the choice of the acoustic parameters and MB sizes were suitable for strong non-linearity while minimizing possible factors attributing to the discrepancy between the model and the experiments.}  	
{\subsection{Advantages of the model, new insights and applications}}
The model predictions were in good agreement with experiments.  In Appendix D, the predictions of the model were also validated against the linear models at a very small pressure excitations ($P_a$=1 kPa) and for three different cases of the uncoated bubble, the coated bubble and the bubble in viscoelastic media. At higher pressures, the predictions of the imaginary part of $k^2$ were validated by comparing the predictions of the model with the Louisnard model \cite{18} and for the three different cases mentioned above. {The proposed approach provides several advantages and generates new insights.}
{\subsubsection{Pressure dependence of the $\Re \left(k^2\right)$:}} Advantages of the developed model in this paper are its simplicity and its ability to calculate the pressure dependent real part and imaginary part of the $k^2$. We showed that pressure dependent effects of the real part of the $k^2$ that are neglected in semi-linear models are important. Attenuation and sound speed estimations using the semi-linear models that does not take into account the pressure dependent effects of the real part of the $k^2$ deviate significantly from the predictions of the approach presented in this paper {(section 3 of Appendix D)}; and these deviations increase with increasing pressure. {These deviations may lead to overestimation of the attenuation by a factor of 2 even at moderate acoustic pressures (Fig. 8 in section 3 of Appendix D).} Moreover, the pressure dependence of the sound speed can be calculated more precisely since the derived model in this paper can calculate the real and imaginary part of the $k^2$. To our best knowledge, unlike current sound speed models, this model does not have a $\frac{dP}{dV}$ term (e.g.  \cite{28}); thus it does not encounter difficulties addressing the low amplitude nonlinear oscillations. 
{\subsubsection{The requirement of the radial oscillations and pressure as inputs only and possible applications:}}
 Another advantage of the model is that it uses as input only the radial oscillations of the MBs. There is no need to calculate the energy loss terms, and thus our approach is simpler and faster. {This provides important advantages to applications in underwater acoustics and acoustic characterization of contrast agents. In the measurements related to the gassy intertidal marine sediments, pressures in the range of 15-20kPa are used \cite{Leighton1}. The bubble size range is between 80$\mu$m to 1mm \cite{2,Leighton1,51}. The behavior of such big bubbles is very sensitive to the pressure amplitude changes with thermal effects significantly influencing the bubble oscillations \cite{pros1}. The energy dissipation as calculated by the linear model deviates significantly as the pressure increases due to two reasons: 1- The pressure dependent change in the resonance frequency and 2- Inaccuracy of the linear thermal approximation. These are discussed in detail in \cite{pros2}. Thus accurate modeling and estimation of the bubble populations requires a model that encompass the full non-linear bubble oscillations.
 Since our nonlinear model \textit{only} uses the radial oscillations and the acoustic pressure as input, it can robustly be used in combination with many different models of MB oscillation. Moreover, attenuation at multiple pressures provides more data that may be used in more accurate estimation of important parameters that characterize bubble populations in these applications. Moreover, in cases of  polymer or highly compressible shells(such as in \cite{15,marm2} and \cite{comp}) the appropriate models can be used. When nonlinear shell behavior occurs (e.g. \cite{16,41})} it may provide more accurate estimates since there is no need for simplified analytical expressions.\\
When fitting the shell parameters of ultrasound contrast agents \cite{43,44,45,46}, {pressure dependent attenuation, and} sound speed values provide more information for accurate characterizations of the shell parameters, especially in case of microbubbles with more complex rheology \cite{16,41}. There may exist multiple combination of values of the initial surface tension, shell elasticity, and shell viscosity that fits well with measured attenuation curves; however, only one combination provides good fit to \textbf{both} the measured sound speed and attenuation values, {and over all the excitation pressure values. A numerical example of a lipid coated MB is shown in Appendix G.} Sound speed curves provide more accurate information on the effect of shell parameters on the bulk modulus of the medium (e.g. shell elasticity) while attenuation graphs are more affected by damping parameters; thus the attenuation and sound speed curves can be used in parallel in order to achieve a more accurate characterization of the shell parameters.\\
{Several studies used linear models to estimate the shell properties of lipid coated MBs \cite{parrales,32,43,44,45,46,faez,xia} using attenuation measurements. These studies reported shell elasticity of 0.28-1N/m and shell viscosity of 3-60$\times$10$^{-9}$kg/s for MBs with 0.7-12.6$\mu m$ over a frequency range between 0.5-29MHz (for a comprehensive review of these studies please refer to \cite{helfield2}). These studies were not able to report the initial surface tension of the MBs. Segers et al \cite{23} estimated a shell elasticity of 2.5N/m, shell viscosity of 2-5$\times$10$^{-9}$kg/s and initial surface tension of 0.02N/m for monodisperse MBs of $\approx$5$\mu$m over a frequency range between 1-6MHz. Attenuation measurements were performed at multiple pressures between 10-100kPa and a semi-linear model was used to fit the shell parameters. By measuring acoustic scattering from single Definity$\textregistered$ MBs and an "acoustic spectroscopy" approach, Helfield et al \cite{Hel2} estimated shell elasticity values between 0.5-2.5N/m and shell viscosity of 0.1-6$\times$10$^{-9}$kg/s for MBs with sizes between 1.7-4$\mu$m over 4-14MHz. They studied the scattered pressure from single MBs at pressures $<$25kPa and used a linear model for fitting the shell parameters. Using optical imaging and flow cytometry, Tu et al\cite{Tu1} directly fit the radial oscillations of the Definity$\textregistered$ MBs to the Marmottant model using 1MHz sonications with acoustic pressures between 95-333kPa. They estimated shell elasticity values of 0.46-0.7 and shell viscosity of 0.1-10$\times$10$^{-9}$kg/s for MBs with sizes between 1.5-6$\mu$m. However, the initial surface tension was not reported in their study. To our best knowledge the latter 3 studies \cite{23,Hel2,Tu1} are the only studies that fit the shell parameters of the MBs taking into account the acoustic pressure amplitude. In our study we estimated the shell elasticity of 1.1N/m, shell viscosity of 7$\times$10$^{-9}$kg/s and initial surface tension of 0.016N/m for MBs with a mean size of 5.6$\mu$m over a frequency range of 1-3MHz and acoustic pressure of 12.5-100kPa. Our findings are in good agreement with the reported values in \cite{23,Hel2,Tu1}.\\ {The study by Helfield et al \cite{Hel2} used linear assumptions and was limited to very small pressures and bubble oscillation amplitude. The study by Tue et al \cite{Tu1} has sensitivity to the MB oscillation amplitude, and thus, requires higher pressures for large enough MB oscillations (which is limited by optical resolution). Both of the methods are challenging experiments and time consuming. In the first method the initial surface tension and rupture radius of the MB is not reported due to the linear assumptions in the model. In the second method, due to the higher pressures used, MBs spend more of their oscillation time in the ruptured state and behave similar to free bubbles \cite{sojahroodr2}. At such high pressures results become less sensitive to the initial surface tension and are affected more by the other parameters due to expansion dominated behavior \cite{2,4}. Thus, estimation of the initial surface tension that is critical in MB behavior at lower pressures (e.g. 50kPa) is not possible and was not reported in \cite{Tu1}. In addition, during the collapse where significant changes in radius occurs in a short nano-second interval, the optical methods are less accurate in resolving instantaneous bubble behavior. The methods presented in our work and in the semi-linear work of Segers et al \cite{26} are simpler. Additionally the methods have the advantages of providing means to extract the values for the initial surface tension and rupture radius. Taking into account the pressure dependent sound speed changes was shown to significantly influence the estimated attenuation values. This becomes even more important in case of the lipid coated MB that exhibits large variations in resonance with small pressure amplitude changes. Thus, another advantage of our nonlinear method is that in addition to accounting for nonlinear MB radial oscillations, it also provides more accurate predictions of the sound speed in the bubbly medium.} To our best knowledge, our study is the first study that uses the pressure dependent sound speed and attenuation curves in tandem to fit the shell parameters of the MBs.}
{\subsubsection{Inclusion of the MB-MB interactions:}}
{MB-MB interactions have been shown to significantly influence the resonance frequency \cite{guedra,sojahrood3,hamaguchi} and the nonlinear behavior of the MBs \cite{guedra,sojahrood3,58,59,Man1}. As the void fraction increases, multiple scatterings by bubbles become stronger\cite{leroy2}. However, even at higher void fractions (e.g. 5$\times$10$^{-3}$) studies based on linear models \cite{17} were not able to incorporate this effect. As such, one of the well-known challenges of these models was the significant overestimation of the attenuation in the vicinity of the bubble resonance frequency even during the linear regime of the oscillations. Investigations at higher acoustic pressure amplitudes and void fractions, also neglected the MB-MB interactions \cite{18,20,trujio,trujio2,jin,brenner,Louisnard3}. In this study we numerically examined the interaction effects over a wide range of void fractions $\beta_0=10^{-7}-10^{-4}$. We show that even in the relatively low void fractions (e.g. $\beta_0=5.1\times 10^{-6}$ used in experiments), bubble-bubble interactions are important (Fig. 9). The influence of the bubble-bubble interactions on the attenuation and sound speed increases with void fraction.\\ The reason for strong interaction in our experiments can be explained as follows. The $\beta$ in experiments was estimated to be 5.1$\times$10$^{-6}$ with 63$\times$10$^{10}$ MBs/$m^3$. The average distance between the MBs can be calculated as $\approx$1/n$^{(1/3)}$ where n is the number density of the MBs. Thus, we can assume that the mean distance between two neighboring bubbles is $\approx$251$\mu$m. This results in a relatively strong interaction between MBs. For example at 100kPa, the resonant pressure radiated by a lipid coated MB with $R_0$ of 2.7$\mu$m a distance of 251$\mu$m is 12.75kPa. When in a solution, each MB receives a sum of the pressures of the multiple neighboring bubbles, plus the incident acoustic pressure field. Since the average scattered pressure by each MB is not negligible, the interaction effects can not be neglected. Moreover, since we interrogated the MBs behavior near their resonance frequency, this effect became more significant.\\
At a high void fraction $\beta_0=5.1\times10^{-3}$, we compared the predictions of our nonlinear model with the experimental results of Silberman \cite{silberman}. It was shown that by including the bubble-bubble interaction we can significantly improve the fit to experiments and potentially address the near resonance attenuation overestimation by the linear models (Fig. 10 in Appendix E).\\
{A wide range of pressure amplitudes are employed in diagnostic and therapeutic ultrasound. As examples, in diagnostic ultrasound peak negative pressures of $\approx$ 100kPa at 2.5MHz \cite{t1} and 127kPa at 5MHz\cite{t2} have been applied. In therapeutic ultrasound the focal pressure amplitudes of 180kPa-300kPa have been used in blood brain barrier opening \cite{t3,t4}. In MB enhanced hyperthermia, sonothrombolysis and drug delivery peak negative pressures between 110-975kPa \cite{rafi,t5,t6,t7} have been applied. In MB enhanced high intensity focused ultrasound heating pressures in the range of 1.5MPa were applied \cite{29}. In all of the therapeutic applications, focused transducers employ large pressure gradients in the focal region. Near and at the focus due to the sharp pressure gradients, the pressure increases by more than 10 fold. Thus, it is estimated that the pressure at the pre-focal region is usually below 200kPa for drug delivery and blood brain barrier opening and below 500kPa for thermal ablation applications. The fundamental knowledge of the pressure dependent attenuation in the pre-focal region is of great importance in optimizing the exposure parameters (e.g. shielding reduction).} 
We numerically investigated the pressure dependent attenuation and sound speed at pressures higher than those in the experiments (200kPa-1MPa). We showed that changes in the sound speed and attenuation are strongly non-linear and bubble-bubble interactions become stronger as pressure increases. An interesting numerical observation is that at higher pressures the sound speed of the bubbly medium (assuming monodispersity) becomes 2.5-4 times higher than the sound speed in water, even for the low void fractions used in clinical ultrasound (Fig. 10).\\
In sonochemistry and sonoluminescence applications, due to the large increase in the attenuation above the Blake threshold the wave can get attenuated significantly so that the standing wave patterns are lost in sonoreactor geometries\cite{Louisnard3,jin}. Thus, accurate prediction of the attenuation is necessary in the reactor geometry design and optimization of the acoustic exposure parameters \cite{jin,brenner,Louisnard3}  Despite the improvement in the estimation of the pressure distribution using the semi-linear Helmholtz model when compared to experiments \cite{jin,louisnard2}, these methods still suffer from insufficient accuracy in the estimation of the attenuation or pressure dependent wavelengths\cite{jin}. By taking into account the simultaneous pressured dependent variations of the real and imaginary part of $k^2$  in the presence of bubble-bubble interactions, a more realistic interpretation of the changes of attenuation and sound speed in bubbly media is expected.}
 {\subsection{Limitations of this study}}
{The experiments and the introduced nonlinear model have inherent limitations. Despite the good agreement between theory and experiment, there were also small discrepancies between the model predictions and experimental measurements. Some of the limitations of the study are discussed here.}
\subsubsection{{Neglecting the pressure variation within the chamber and uniform shell properties}}
{In modeling the MB behavior, we have not considered pressure variations within the MB chamber due to MB attenuation. The importance of considering the pressure changes within the chamber is discussed in \cite{versluis}}. Additionally, we assumed that all the MBs have the identical shell composition and properties, and effects like strain softening or shear thinning of the shell \cite{41}, and possible MB destruction, were neglected.
As the purpose of the current work was to investigate the simultaneous pressure dependence of the sound speed and attenuation and to develop a model to accurately predict these effects, we have used the simplest model for lipid coated MBs. Investigation of models that incorporate more complex rheological behaviors of the shell which takes into account the effect of shell properties on sound speed and attenuation is the subject of future work.
{\subsubsection{Low void fraction of $\approx$5.1$\times$10$^{-6}$ of micron sized MBs in experiments}}
The changes of the magnitudes of the sound speed {peaks} in our experiments were {small}; this is due to the {small size} and low concentration of the MBs in our experiments. {However, it is not only the peak of the sound speed that changed with pressure but also the frequency of the sound speed peaks (frequency of sound speed peak decreased by $\approx$30$\%$ between 12.5kPa and 100kPa).} Bigger bubbles (mm sized \cite{17,trujio}) have stronger effects on the {magnitude of the sound speed} changes of the medium because of their higher compressibility. Moreover, resonance frequency is predicted to change more rapidly with increasing acoustic pressure \cite{14}. {When volume matched, however, although the sound speed changes of the medium are larger in case of bigger bubbles, however, the attenuation changes of the medium \cite{17,trujio,silberman} are quite lower than the case of smaller MBs. As an instance maximum attenuation of water with bubbles of $R_0$=2.07mm and {$\beta_0$}=5.3$\times$10$^{-2}$ was less than 10dB/cm in Silberman et. al\cite{silberman} while the maximum sound speed reached $\approx$2$\times$10$^4$m/s.\\The focus of this paper is on the dynamics of the micron sized ultrasound contrast agents (1-10$\mu$m), thus we did not test the cases of mm sized bubbles. Although the changes of the sound speed amplitude are small, we were able to measure these changes in experiments consistent with model predictions indicating the sensitivity of the measurements and the accuracy of the theoretical approach. The void fraction in our study is also relevant to clinical applications of ultrasound. One of the clinically used UCAs is Definity$^{\textregistered}$ which is used at lower concentrations for nonlinear imaging and at higher concentrations for clinical therapeutic applications \cite{dannold,riley,def}. For therapeutic applications, the the dosage of Definity$^{\textregistered}$ MBs is $10\mu$L per kg weight of the human body (with maximum allowable dose of $20\mu$L per kg weight \cite{def}).  In novel ultrasound imaging applications, a small total dose of $0.1-0.2 mL$ is used for diagnostic applications \cite{def}. Thus, for a 100kg person with approximately 7500mL total blood volume, 1-2 mL of MBs should be injected for clinical therapeutic applications. Each mL of Definity$^{\textregistered}$ has $\approx$1.2$\times$10$^{10}$ MBs  \cite{dannold,riley,def} with gas volume of $\approx$27$\times$10$^{9}$ $\mu$m$^3$/mL. {For a 100kg patient, thus, a total gas volume of $\approx$27-54$\times$$10^{9}$$\mu$m$^3$ is injected. Lower and upper limit of $\beta_0$ (void fraction in m$^3$/m$^3$) thus can be calculated as 27$\times$$10^{-9}$m$^3$/7.5$\times$$10^{-3}$m$^3$ and 54$\times$$10^{-9}$m$^3$/7.5$\times$$10^{-3}$m$^3$ respectively.} This results in {$\beta_0$} of 3.6-7.2$\times$10$^{-6}$ for clinical therapeutic applications. For imaging applications {$\beta_0$} would be 3.6-7.2$\times 10^{-7}$ inside blood. In our study the concentration of MBs was {6.3$\times$10$^4$MBs/mL} with {$\beta_0$} of $\approx$ 5.1 $\times$10$^{-6}$. Thus, the void fraction in our work is relevant to clinical therapeutic applications of the UCAs in blood. When the whole body is considered as the medium, the void fraction would be even lower. $7\%$ of the body is blood, and assuming homogeneous distribution of blood the actual physical values for {$\beta_0$} in the human body are between 2.5 $\times 10^{-8}$-5 $\times 10^{-7}$.}\\ In many {pre-clinical} applications (e.g. drug delivery and {vascular disruption}) high concentrations {(e.g 100-200$\mu$L per kg)} of microbubbles are employed, thus greater changes in the sound speed amplitude are expected (e.g. please refer to Fig. 6 and 9 in Appendix D with $\beta =10^{-5}$). {However, experimental measurements of the higher void fractions in case of micron sized bubbles (1-10$\mu$m) are challenging. As it was discussed and measured in the current work, the attenuation of smaller MBs are very high even when low void fractions are used (e.g. ~20dB/cm at 100kPa in Fig. 5d). Increasing the {$\beta_0$} further increases the attenuation and thus a very large pressure gradient will occur within the sample \cite{versluis}. For example for an attenuation of 60dB/cm just within the first 1mm of the sample holder, the ultrasound amplitude will reduce in half. This causes difficulty in interpretation of the results or numerical implementation of the model. In order to accurately measure the pressure dependent sound speed and attenuation in high void fractions, technically complex experiments are needed where ultrasound only propagates through a very thin sheet of MBs \cite{versluis,leroy}. Moreover, bubble-bubble interactions becomes significant (Please refer to Appendix E {and F}) which further adds to the difficulty of the interpretation of the results. At higher void fractions the steep pressure gradients changes inside the chamber and the bubble-bubble interaction causes large resonance shifts \cite{sojahrood3,guedra} resulting in significant widening of the attenuation curves even in case of monodisperse MBs \cite{36}. Thus, experimental detection of a clear relationship between pressure and sound speed and attenuation becomes very difficult. For more accurate detection of this relationship thus, using lower void fractions are crucial. In Appendix E, we presented numerical simulations on the relationship between void fraction and bubble-bubble interaction on the pressure dependent attenuation and sound speed. The experimental detection of the attenuation and sound speed changes at higher void fractions is outside of the scope of this paper and can be the subject of future studies.}  
\subsubsection{{Assumption of single frequency wave propagation in the model}}
{Similar to the approach by Louisnard \cite{18}, we only considered the case of mono-frequency propagation in the model.}
Since the nonlinear ultrasound propagation effects were neglected, the presented model in this work may loose accuracy at higher pressures where the higher order harmonics are generated along the propagating path of the ultrasound waves. Moreover, the simple Helmholz equation can not model the nonlinear propagation of the waves and dissipation terms should take into account the nonlinear waves effects \cite{kanagawa1,kanagawa2,kanagawa3,kanagawa4,kanagawa5,kanagawa6}. Derivation of the pressure dependent attenuation and sound speed at {higher pressures, and} Mach numbers, thus, requires changes to the bubble oscillation model in tandem with considering the nonlinear wave propagation effects and is outside of the scope of the current study.
\subsubsection{{Low Mach numbers in experiments}}
 {To minimize MBs destruction we applied pressures below the Blake threshold of the MBs and we estimate that the maximum MB velocity reached 40m/s in our experiments. We did not experimentally examine the exposure parameters that result in high Mach number regimes. In dealing with high pressure applications, especially at lower frequencies (e.g. 20kHz) used in sonochemistry, care must be taken when pressure exceeds that of the Blake threshold. Above the Blake threshold, with increasing pressure, the maximum bubble wall velocity increases very fast. The Keller-Miksis equation takes into account the compressibility effects by only keeping the first order terms. Thus, it is only valid for when $\frac{|\dot{R}|}{c}<1$ \cite{yasui,pros2}. However, in numerical simulations using the Keller-Miksis equation, this condition is often violated \cite{20,jin,brenner} during the violent bubble collapse above the Blake threshold. At higher wall velocities, energy dissipation is dominated by the compressibility effects. Thus a proper representation of the compressibility effects is required for more accurate estimation of the attenuation. When the bubble wall velocity exceeds that of the water sound speed, either corrections to the Keller-Miksis equation should be applied \cite{yasui,pros2} or Glimore equation should be used as it has a  higher validity range up to $\frac{|\dot{R}|}{c}<2.2$ \cite{zee}. Zheng et al \cite{Zhang} included the second-order compressibility terms in the simulation of the bubble oscillations. They showed that above approximately the Mach number of 0.59, the predicted energy dissipation deviates from the Keller-Miksis predictions and thus the influence of the liquid compressibility should be fully considered for proper modeling. It is estimated that, at higher Mach numbers ($>$0.59-1\cite{Zhang,pros2,yasui}), higher order liquid compressibility effects dampen further the bubble oscillations, thus the attenuation will be smaller than the value as predicted by the Keller-Miksis equation. When using the nonlinear model, for more accurate predictions, above Mach numbers between 0.59-1, radial oscillations should be calculated using the Glimore equation or models that include the higher order compressibility terms.
 \subsubsection{MB size distribution}
 In our experiments despite using a microfluidic technique, the bubble size distribution was still not fully monodisperse (1.1$\mu m<R_0<$3.3$\mu m$). Thus, the behavior of the MBs were not quite uniform. This contributed to broadening the attenuation width and sound speed changes and reduction in the attenuation and sound speed peaks. Nevertheless, we still were able to observe the pressure dependent effects and the contribution of each population of MBs at a particular size were considered in the modeling.  However, using a narrower size range or using MBs that are generated with more complex acoustic size sorting techniques \cite{seg3} will allow for a more uniform behavior of the MBs and more clear separation of attenuation peaks at each pressure. Additionally, due to uniform and more potent behavior \cite{26,acs,seg3}, for the same volume fraction a higher attenuation or sound speed peak is expected.\\
{Due to the narrow size distribution of the MBs and near resonance sonication, in our experiments we were able to detect strong nonlinearity at relatively low pressure amplitudes. However, it should be noted that commercial contrast agents are polysiperse and subharmonic emission thresholds and onset of nonlinearity could easily extend to few hundred kPa.}} 
 \section{Conclusion}
In summary, we report the first controlled observation of the pressure dependence of sound speed in bubbly media. The relationship between the sound speed and attenuation with pressure was established both theoretically and experimentally. We have presented a model for the calculation of the pressure-dependent attenuation and sound speed in a bubbly medium. The model is
free from any linearization in the MB dynamics. The accuracy of the model
was verified by comparing it to the linear model \cite{17} at low pressures and the
semi-linear Lousinard model \cite{18} at higher pressure amplitudes that cause nonlinear oscillations.  The predictions of the model are in good agreement with experimental observations.  We showed that to accurately model the changes of the attenuation and sound propagation in a bubbly medium we also need to take into account how the sound speed changes with pressure and frequency.\\ {There is a  the limited range of experiments in the near resonance regime and with monodisperse bubbles. Future experiments employing polydisperse commercial MBs, sonication of MBs below their resonance and application of higher acoustic pressure are needed to gain better insight on the attenuation and sound speed phenomena as well as testing the model predictions for the entire range of biomedical ultrasound applications. Nevertheless, application of this model will help to shed light on the effect of the nonlinear oscillations on the acoustical properties of bubbly media. This might help to discover new potential exposure parameters to further optimize and improve the current applications.}\\
{\textbf{Acknowledgments}}\\
{We like to thank Prof. Yuan Xun, Prof. Pedro Goldman and Prof. David E. Goertz for helpful discussions.} The work is supported by the Natural Sciences and Engineering Research Council of Canada (Discovery Grant RGPIN-2017-06496), NSERC and the Canadian Institutes of Health Research ( Collaborative Health Research Projects ) and the Terry Fox New Frontiers Program Project Grant in Ultrasound and MRI for Cancer Therapy (project $\#$1034). A. J. Sojahrood is supported by a CIHR Vanier Scholarship and Qian Li is supported by NSF CBET grant $\#$1134420.\\
{\textbf{Data availability}}\\
The data that support the findings of this study are available
from the corresponding author upon reasonable request.

\appendix

\section{Verification of the predictions of the Eqs. \ref{eq:10} and \ref{eq:11} in linear oscillation regimes}
The appendices structure is as follows:\\
1- We first recall {three} models for uncoated bubbles (Eq. \ref{eq:A1}), bubbles coated with linear viscoealstic shells (Eq. \ref{eq:A2}) and bubbles immersed in elastic materials (Eq. \ref{eq:A3}). The linear terms for the attenuation and sound speed of the bubbly media are then presented for each model (Appendix \ref{section:B}). The attenuation and sound speed calculated by Eqs. \ref{eq:10} and \ref{eq:11} are compared with the linear models for each case at acoustic pressure amplitude of 1 kPa (linear regime of oscillations) in Appendix \ref{subsection:D1}. We show that the predictions of the model are in good agreement with the linear model in the linear oscillation regimes.\\
2- At higher pressures, predictions of the linear models are no longer valid.  Thus predictions of the model are compared with the Louisnard model \cite{18}. First, nonlinear dissipation power terms are presented for each model in Appendix \ref{section:C}.  The maginary part of wave number squared ($\langle\Im(k^2)\rangle$) is then calculated at different pressures and model predictions are compared with the Louisnard model \cite{18} at different pressure amplitudes (Appendix \ref{subsection:D2}). We show that that  ($\langle\Im(k^2)\rangle$) as calculated by our model and the Louisnard model \cite{18} are in good agreement. However, one advantage of our model is the capability to calculate the ($\langle\Re(k^2)\rangle$) which results in correction of the overestimation of the attenuation in the semi-linear model.\\
3- At higher void fractions bubble-bubble interactions become significant. We numerically investigate the influence of the increasing void fraction on the changes of the attenuation and sound speed. We compare the cases with and without bubble-bubble interaction. Finally, we compare the predictions of the introduced model with the one of the experimental results of \cite{silberman}. We show that inclusion of the bubble-bubble interaction can partially improve the significant overestimation of the attenuation at linear resonance \cite{17,trujio}.\\\\
\textbf{The bubble models}\\\\
The volume fraction occupied by a bubble with $\beta_i$ (Eq. \ref{eq:5}) depends on the $R(t)$ of each bubble. The $R(t)$ curve of each bubble is determined by solving the model that describes the bubble oscillations (Eqs. \ref{eq:A1} (uncoated bubble), \ref{eq:A2} (coated bubble) and \ref{eq:A3} (bubble immersed in elastic materials)). The predictions of Eq. \ref{eq:10} and \ref{eq:11}, will be numerically verified in case of an uncoated free bubble model, a coated bubble model and a model of a bubble immersed in sediment or tissue. In each case the model predictions are validated against the linear regime of oscillations at very low pressures ($1 kPa$) by comparing the predictions with the linear models. Linear models are derived using the Commander $\&$ Porspereti approach in \cite{17}.
\subsection{Uncoated free bubble}
Radial oscillations of an uncoated bubble can be modeled to the first order of Mach number by solving the Keller-Miksis \cite{22} (KM) model:
\begin{equation}
{\rho\left(1-\frac{\dot{R}}{C_l}\right)R\ddot{R}+\rho\frac{3}{2}\dot{R}\left(1-\frac{R}{3C_l}\right)=\left(1+\frac{\dot{R}}{C_l}\right)G_g+\frac{R}{C_l}\frac{d}{dt}G_g}
	\label{eq:A1}
\end{equation}
where $G_g=P_g-\frac{4\mu_L\dot{R}}{R}-\frac{2\sigma}{R}-P_0-P_a \sin(2 \pi f t)$.\\
In this equation, $R$ is radius at time t, $R_0$ is the initial bubble radius, $\dot{R}$ is the wall velocity of the bubble, $\ddot{R}$ is the wall acceleration, $\rho{}$ is the liquid density (998 kg/m$^3$), $C_l$ is the sound speed (1481 m/s), $P_g$ is the gas pressure, $\sigma{}$ is the surface tension (0.0725 N/m), $\mu{}$ is the liquid viscosity (0.001 Pa s), $P_0$ is the atmospheric pressure (101.325 kPa), and $P_a$ and \textit{f} are the amplitude and frequency of the applied acoustic pressure. The values in the parentheses are for pure water at 293 K. In this paper the gas inside the bubble is either air or C$_3$F$_8$ and water is the host media.\\
\subsection{Coated bubble}
The dynamics of the coated bubble can be modeled using the Keller-Miksis-Church-Hoff model (KMCH) \cite{28}. We have derived this model by adding the compressibility effects to the first order of Mach number in \cite{28}. The model is presented in Eq. 6:
\begin{widetext}
	\begin{equation}
	\begin{gathered}
	\rho \left(\left(1-\frac{\dot{R}}{C_l}\right)R\ddot{R}+\frac{3}{2}\dot{R}^2\left(1-\frac{\dot{R}}{3C_l}\right)\right)=\\
	\left(1+\frac{\dot{R}}{C_l}+\frac{R}{C_l}\frac{d}{dt}\right)\left(P_g-\frac{4\mu_L\dot{R}}{R}-\frac{12\mu_{sh}\epsilon R_0^2\dot{R}}{R^4}-12G_s\epsilon R_0^2 \left(\frac{1}{R^3}-\frac{R_0}{R^4}\right)-P_0-P\right)
	\label{eq:A2}
	\end{gathered}
	\end{equation}
\end{widetext}
in this equation $\mu_{sh}$ is the viscosity of the  shell (coating), $\epsilon$ is the thickness of the coating, $G_s$ is the shell shear modulus, $P_g$ is the gas pressure inside the bubble, $P_0$ is the atmospheric pressure (101.325 kPa) and P is the acoustic pressure given by $P=P_a\sin(2\pi ft)$ with $P_a$ and $f$ are respectively the excitation pressure and frequency. In this paper for all of the simulations related to coated bubbles $G_s$ {is} 45 MPa and 
$\mu_{sh}$ {is given by} 1.49$(R_0$($\mu m)$-0.86)$/\theta$ (nm) \cite{38} ($sh$ stands for shell (coating)) with $\theta$ {is} 4 nm unless otherwise stated.
\subsection{Bubble in sediment or tissue}
The Yang and Church model \cite{39} describes the radial oscillations of an uncoated bubble in a viscoelastic medium (e.g. marine sediments or tissue):
\begin{widetext}
	\begin{equation}
	\rho\left(1-\frac{\dot{R}}{C_l}\right)R\ddot{R}+\frac{3}{2}\rho\dot{R}^2\left(1-\frac{\dot{R}}{3C_l}\right)=\\\left(1+\frac{\dot{R}}{C_l}+\frac{R}{C_l}\frac{d}{dt}\right)\left(P_g-\frac{2\sigma}{R}-\frac{4\mu_s\dot{R}}{R}-\frac{4G}{3R^3}\left(R^3-R_0^3\right)-P_0-P_a\sin(2\pi ft)\right)
	\label{eq:A3}
	\end{equation}
\end{widetext}
This equation, similar to Eqs. \ref{eq:A1} and \ref{eq:A2}  accounts for compressibility effects to the first order of Mach number, thus inherits the acoustic radiation losses. Several approaches for the incorporation of such losses into a Rayleigh-Plesset type equation were outlined in \cite{26}. The introduced new constant $G$ describes the shear modulus and $\mu_s$ describes the shear viscosity of the sediment or tissue. In this paper we considered a tissue with  $G$ {is} 0.5 MPa, $\mu_s$=0.00287 Pa.s and $\sigma$ is 0.056 N/m (blood surface tension)\cite{39}.
\subsection{Gas pressure and thermal effects}
\textbf{4a-Linear thermal model}\\\\
The linear thermal model \cite{17,40} is a popular model that has been widely used in studies related to oceanography \cite{4,5} and the modeling and charecterization of coated bubble oscillations\cite{41,42,43,44,45}. In this model through linearization, thermal damping is approximated by adding an artificial viscosity term to the liquid viscosity. Furthermore, a variable isoentropic index is used instead of the polytropic exponent of the gas.\\ In this model $P_g$ is given by:
\begin{equation}
P_g=P_{g0}\left(\frac{R_0}{R}\right)^{3k_i}
\label{eq:A4}
\end{equation} 
Where the polytropic exponent $\gamma$ is replaced by isoentropic indice ($k_i$):
\begin{equation}
k_i=\frac{1}{3}\Re(\phi)
\label{eq:A5}
\end{equation}
The liquid viscosity is artificially increased by adding a thermal viscosity ($\mu_{th}$) to the liquid viscosity. This thermal viscosity ($\mu_{th}$) is given by:
\begin{equation}
\mu_{th}=\frac{P_{g0} \Im(\phi)}{4\omega}
\label{eq:A6}
\end{equation}
In the above equations the complex term $\phi$ is calculated from
\begin{equation}
\phi=\frac{3\gamma}{1-3\left(\gamma-1\right)i\chi\left[\left(\frac{i}{\chi}\right)^{\frac{1}{2}}{\coth}\left(\frac{i}{\chi}\right)^{\frac{1}{2}}-1\right]}
\label{eq:A7}
\end{equation}
where $\gamma$ is the polytropic exponent and $\chi=D/\omega R_0^2$ represents the thermal diffusion length where $D$ is the thermal diffusivity of the gas. $D= L/C_p\rho_g$ where $C_p$, $\rho_g$, and $L$ are specific heat in constant pressure, density and thermal conductivity of the gas inside the bubble.\\
To calculate the radial oscillations of the coated bubble and uncoated bubble while including the linear thermal effects Eqs. \ref{eq:A1} , \ref{eq:A2}  $\&$ \ref{eq:A3}  are coupled with Eq. \ref{eq:A4}  and the liquid viscosity is increased by $\mu_{th}$ (Eq.\ref{eq:A6}). The linear thermal model is used to derive the attenuation and sound speed terms in the regime of linear oscillations. In case of the uncoated bubble and the uncoated bubble in viscoelastic medium (Eq.\ref{eq:A1} $\&$\ref{eq:A3} ) $P_g=P_0${$+$}2$\sigma$/$R_0$. In the case of the coated bubble (Eq.\ref{eq:A2} ) $P_g=P_0$ \cite{Hoff}.\\\\
\textbf{4b- Full thermal model}\\\\
If temperature dependent thermal effects are considered, $P_g$ is given by Eq. 12 \cite{47}:
\begin{equation}
P_g=\frac{N_gKT}{\frac{4}{3}\pi R(t)^3-N_g B}
\label{eq:A8}
\end{equation}
here $N_g$ is the total number of the gas molecules, $K$ is the Boltzman constant and B is the molecular co-volume. The average temperature inside the gas can be calculated using Eq. 13 \cite{47}:
\begin{equation}
\dot{T}=\frac{4\pi R(t)^2}{C_v} \left(\frac{L\left(T_0-T\right)}{L_{th}}-\dot{R}P_g\right)
\label{eq:A9}
\end{equation}
here $C_v$ is the heat capacity at constant volume, $T_0$= {293 K} is the initial gas temperature, $L_{th}$ is the thickness of the thermal boundary layer. $L_{th}$ is given by $L_{th}={\min}(\sqrt{\frac{DR(t)}{|\dot{R(t)}|}},\frac{R(t)}{\pi})$ where $D$ is the thermal diffusivity of the gas. $D$ can be calculated using $D=\frac{L}{c_p \rho_g}$ where {$L$} is the gas thermal conductivity, $c_p$ is specific heat at constant pressure and $\rho_g$ is the gas density. \\
Predictions of the full thermal model have been shown to be in good agreement with predictions of the models that incorporate the thermal effects using the PDEs \cite{48}. To calculate the radial oscillations of the coated bubble and uncoated bubble while including the full thermal effects Eqs.\ref{eq:A1} (coated bubble) or Eq.\ref{eq:A2}  (uncoated bubble) or Eq. \ref{eq:A3} (bubble in sediment or tissue) are coupled with Eq. \ref{eq:A8} and Eq. \ref{eq:A9} and then solved using the ode45 solver in Matlab. The relative and absolute tolerances were 1$\times$10$^{-13}$ and 1$\times$10$^{-14}$.
\section{Attenuation and sound speed equations for linear regime of oscillations} 
\label{section:B}
The linear thermal equations (Eqs.\ref{eq:A4},\ref{eq:A5}, \ref{eq:A6} $\&$ \ref{eq:A7}) were coupled to the bubble models (Eqs.\ref{eq:A1},\ref{eq:A2} $\&$ \ref{eq:A3}) to derive the attenuation and sound speed terms.\\ For the linear regime, radial oscillations can be considered as $R=R_0(1+x)$ where {$x$} is a small displacement amplitude \cite{2,17,Hoff,51}.  Thus, $\dot{R}=R_0\dot{x}$ and $R^{-n}=R_0(1-nx)$ (higher order small terms are neglected in the Taylor series expansion of $R^{-n}$). A function $g(t)=e^{i\omega t}$ is also defined \cite{2,17,Hoff,51}. The incident pressure can be linearized as \cite{2,51}:
\begin{equation}
P_ag(t)=\frac{\rho \ddot{R}R_0}{\left(1-\frac{i\omega R_0}{C_l}\right)}
\label{eq:B1}
\end{equation}
Using these linear approximations a linear analytical solution to Eqs. \ref{eq:A1}, \ref{eq:A2} and \ref{eq:A3} can be provided. These solutions can be written in the following general form of forced damped oscillations:
\begin{equation}
\begin{gathered}
\alpha \ddot{x}+2\beta\dot{x}+\gamma x=-P_Ae^{i\omega t}
\end{gathered}
\label{eq:B2}
\end{equation}
where constants $\alpha$, $\beta$ and $\gamma$ can be defined by solving the appropriate equations. We can transfer Eq.\ref{eq:B2} to the frequency domain by setting $x(t)=x(\omega)e^{i\omega t}$:\\ 
\begin{equation}
\begin{gathered}
-\alpha\omega^2x(\omega)e^{i\omega t}+2i\beta\omega x(\omega)e^{i\omega t}+\gamma x(\omega)e^{i\omega t}=-P_Ae^{i\omega t}
\end{gathered}
\label{eq:B3}
\end{equation} 
Eq.\ref{eq:B3} can be simplified by dividing both sides by $\alpha e^{i\omega t}$ and inputting $\omega_0=\frac{\gamma}{\alpha}$ ($\omega_0$ is resonance angular frequency). Thus: 
\begin{equation}
x(\omega)\left[\omega_0^2-\omega^2+\frac{2i\beta}{\alpha}\omega\right]=-\frac{P_A}{\alpha}
\label{eq:B4}
\end{equation} 
\subsection{Free uncoated bubble (KM model) constants} 
In case of the uncoated bubble model Eq. \ref{eq:A1}, the constants $\alpha$, $\beta$ and $\gamma$ of Eq.\ref{eq:B3} can be derived using an approach similar to \cite{2,51}:
\begin{equation}
\begin{dcases}
\alpha=\rho R_0^2+\frac{4\mu R_0}{C_l}\\
\beta=2\mu_{th}-\frac{\sigma}{C_l}+2\mu+\frac{\left(\frac{\omega R_0}{C_l}\right)}{1+\left(\frac{\omega R_0}{C_l}\right)^2}\frac{\omega}{2}\left(\rho R_0^2\right)\\
\gamma=P_{g0}\Re(\phi)-\frac{2\sigma}{R_0}+\frac{\omega^2 \rho R_0^2 }{1+\left(\frac{\omega R_0}{C_l}\right)^2}
\end{dcases}
\label{eq:B5}
\end{equation} 
and the angular resonance frequency $\omega_0=2\pi f_r$ ($f_r$ is the linear resonance frequency) is given by:
\begin{equation}
\omega_0=\sqrt{\frac{P_{g0}\Re(\phi)-\frac{2\sigma}{R_0}+\frac{\omega^2 \rho R_0^2 }{1+\left(\frac{\omega R_0}{C_l}\right)^2}}{\rho R_0^2+\frac{4\mu\ R_0}{C_l}}}
\label{eq:B6}
\end{equation} 
The constant $\delta_{total}$ is the total damping and is defined as $\delta_{total}=\delta_{liquid}+\delta_{radiation}+\delta_{thermal}=\frac{\beta}{\alpha}$ (in left hand side of equation \ref{eq:B4}). Thus:
\begin{equation}
\begin{gathered}
\delta_{total}=\frac{\beta}{\alpha}=\frac{2\mu_{th}-\frac{\sigma}{C_l}+2\mu+\frac{\left(\frac{\omega R_0}{C_l}\right)}{1+\left(\frac{\omega R_0}{C_l}\right)^2}\frac{\omega}{2}\left(\rho R_0^2\right)}{\rho R_0^2+\frac{4\mu R_0}{C_l}}
\end{gathered}
\label{eq:B7}
\end{equation} 
where $\delta_{Vis}$, $\delta_{th}$, $\delta_{Rad}$ $\&$ $\delta_{int}$ are damping constants due to liquid viscosity, thermal loss, re-radiation and interfacial effects.
\begin{equation}
\begin{dcases}
\delta_{liquid}=\frac{2\mu}{\rho R_0^2+\frac{4\mu R_0}{C_l}}\\
\\
\delta_{thermal}=\frac{2\mu_{th}}{\rho R_0^2+\frac{4\mu R_0}{C_l}}\\
\\
\delta_{radiation}=\frac{\frac{\left(\frac{\omega R_0}{C_l}\right)}{1+\left(\frac{\omega R_0}{C_l}\right)^2}\frac{\omega}{2}\left(\rho R_0^2\right)}{\rho R_0^2+\frac{4\mu R_0}{C_l}}\\
\\
\delta_{int}=\frac{-\sigma}{\rho R_0^2+\frac{4\mu R_0}{C_l}}
\end{dcases}
\label{eq:B8} 
\end{equation} 
\subsection{Coated bubble (KMCH) model }
Linearizing Eq. \ref{eq:A2} for coated bubbles we can arrive in an analytical solution in the form of Eq. \ref{eq:B4} where the angular linear resonance frequency is given by
\begin{equation}
\begin{gathered}
\omega_0=\sqrt{\frac{P_{g0}\Re(\phi)+\frac{12G_s\epsilon}{R_0}+\frac{\omega^2 \rho R_0^2}{1+\left(\frac{\omega R_0}{C_l}\right)^2}}{\rho R_0^2 +\frac{4\mu R_0}{C_l}+\frac{12\mu_{sh}\epsilon}{C_l}}}
\end{gathered}
\label{eq:B9} 
\end{equation} 
and 
\begin{equation}
\begin{dcases}
\alpha=\rho R_0^2+\frac{4\mu R_0}{C_l}\\
\beta=2\mu_{th}+2\mu+\frac{\left(\frac{\omega R_0}{C_l}\right)}{1+\left(\frac{\omega R_0}{C_l}\right)^2}\frac{\omega}{2}\left(\rho R_0^2\right)-2GR_0\\
\gamma=P_{g0}\Re(\phi)-\frac{2\sigma}{R_0}+\frac{\omega^2 \rho R_0^2 }{1+\left(\frac{\omega R_0}{C_l}\right)^2}+4G
\end{dcases}
\label{eq:B10} 
\end{equation} 
In this case, existence of the shell introduces an extra term for damping due to shell viscosity $\delta_{shell}$. Thus, $\delta_{total}=\delta_{liquid}+\delta_{radiation}+\delta_{shell}+\delta_{thermal}$ where:
\begin{equation}
\begin{gathered}
\begin{dcases}
\delta_{liquid}=\frac{2\mu}{\rho R_0^2 +\frac{4\mu R_0}{C_l}+\frac{12\mu_{sh}\epsilon}{C_l}}\\
\\
\delta_{radiation}=\frac{\frac{\left(\frac{\omega R_0}{C_l}\right)}{1+\left(\frac{\omega R_0}{C_l}\right)^2}\frac{\omega}{2} \left(\rho R_0^2\right)}{\rho R_0^2 +\frac{4\mu R_0}{C_l}+\frac{12\mu_{sh}\epsilon}{C_l}}\\
\\
\delta_{shell}=\frac{\frac{6\mu_{sh}\epsilon}{R_0}+\frac{6G_s\epsilon}{C_l}}{\rho R_0^2 +\frac{4\mu R_0}{C_l}+\frac{12\mu_{sh}\epsilon}{C_l}}\\
\\
\delta_{thermal}=\frac{2\mu_{th}}{\rho R_0^2 +\frac{4\mu R_0}{C_l}+\frac{12\mu_{sh}\epsilon}{C_l}}\\
\end{dcases}
\end{gathered}
\label{eq:B11} 
\end{equation} 
\subsection{Bubble immersed in tissue or sediment}
The linear analytical solution to Eq.\ref{eq:A3}  for bubbles immersed in tissue or sediments can be written again in the form of Eq. \ref{eq:B4}. The constants of the equation can be written as follows \cite{2,51}:
\begin{equation}
\omega_0=\sqrt{\frac{P_{g0}\Re(\phi)-\frac{2\sigma}{R_0}+\frac{\omega^2 \rho R_0^2 }{1+\left(\frac{\omega R_0}{C_l}\right)^2}+4G}{\rho R_0^2+\frac{4\mu R_0}{C_l}}}
\label{eq:B12} 
\end{equation} 
and 
\begin{equation}
\begin{dcases}
\alpha=\rho R_0^2+\frac{4\mu R_0}{C_l}\\
\beta=2\mu_{th}-\frac{\sigma}{C_l}+2\mu+\frac{\left(\frac{\omega R_0}{C_l}\right)}{1+\left(\frac{\omega R_0}{C_l}\right)^2}\frac{\omega}{2}\left(\rho R_0^2\right)-2GR_0\\
\gamma=P_{g0}\Re(\phi)-\frac{2\sigma}{R_0}+\frac{\omega^2 \rho R_0^2 }{1+\left(\frac{\omega R_0}{C_l}\right)^2}+4G
\end{dcases}
\label{eq:B13} 
\end{equation} 
The total damping this case has a term related to the elasticity of the sediment or the tissue ($\delta_{Sed}$). Thus, $\delta_{total}=\delta_{liquid}+\delta_{radiation}+\delta_{Sed}+\delta_{thermal}$ where:
\begin{equation}
\begin{dcases}
\delta_{liquid}=\frac{2\mu}{\rho R_0^2+\frac{4\mu R_0}{C_l}}\\
\\
\delta_{thernal}=\frac{2\mu_{th}}{\rho R_0^2+\frac{4\mu R_0}{C_l}}\\
\\
\delta_{radiation}=\frac{\frac{\left(\frac{\omega R_0}{C_l}\right)}{1+\left(\frac{\omega R_0}{C_l}\right)^2}\frac{\omega}{2}\left(\rho R_0^2\right)}{\rho R_0^2+\frac{4\mu R_0}{C_l}}\\
\\
\delta_{int}=\frac{-\sigma}{\rho R_0^2+\frac{4\mu R_0}{C_l}}\\
\\
\delta_{Sed}=\frac{2GR_0}{\rho R_0^2+\frac{4\mu R_0}{C_l}}\\
\end{dcases}
\label{eq:B14} 
\end{equation}
\subsection{Derivation of the linear equations of attenuation and sound speed}
Using the linear formulations for $R$, ${\partial{}^2\beta{}}/{\partial{}t^2}=4\pi {R_0}^3\ddot{x}$. Moreover $x(\omega)$ can be calculated from Eq. \ref{eq:B4} :
\begin{equation}
x(\omega)=\frac{-\frac{P_A}{\alpha}}{\left[\omega_0^2-\omega^2+2i \delta_{total} \omega\right]}
\label{eq:B15} 
\end{equation}
and $\dot{x}(\omega)=-\omega x(\omega)$ $\&$ $\ddot{x}(\omega)=\omega^2 x(\omega)$. Inputting these into Eq. \ref{eq:5} and eliminating $e^{i\omega t}$ yields:
\begin{equation}
{\nabla{}}^2\left(P\right)+k^2P=0
\label{eq:B16} 
\end{equation}
where $k$ is the complex wave number ($k={\omega}/{C_l}-i\alpha$):
\begin{equation}
k^2=(\frac{\omega}{C_l})^2+4\pi \omega^2 \sum_{j=1}^N \frac{R_{0j}}{\omega_{0}^2-\omega^2+2i\delta_{total} \omega}
\label{eq:B17} 
\end{equation}
where $R_{0j}$ is the initial radius of the bubble number j. Attenuation and sound speed can easily be obtained from equation \ref{eq:B17}.
\section{Validation of Eqs. \ref{eq:10} \ref{eq:11} at nonlinear regime of oscillations}
\label{section:C}
Louisnard \cite{18} derived the pressure dependent term for the imaginary part of the wave number. Thus, at higher pressures, predictions of the imaginary part of the wave number as calculated by Eqs. \ref{eq:10} and \ref{eq:11} are verified numerically with predictions of the Louisnard model \cite{18} and for different pressure amplitudes.\\ The Louisnard model was modified by  Jamshidi $\&$ Brenner \cite{19} to include the compressibility effects to the first order of Mach number. Using this approach, they were able to present the nonlinear terms that describe the power loss due to radiation, thermal and viscous effects. In \cite{27} we provided critical corrections to the derived terms in \cite{19} for uncoated bubbles. Using our approach in \cite{27}, we have derived the terms describing the nonlinear power loss in the case of the coated bubbles in \cite{28}. Here, we will also derive the terms describing the nonlinear power loss for bubbles that are immersed in sediments or tissues using our approach in \cite{27}.\\ 
 In Louisnard's approach, firstly, the terms for the nonlinear energy dissipation are derived accounting for large bubble oscillation amplitude. One starts with the mass and momentum conservation equations in a bubbly medium \cite{37}:
\begin{equation}
\begin{dcases}
\frac{1}{\rho C_l^2}\frac{\partial P}{\partial t}+\nabla.v=\frac{\partial \beta}{\partial t}
\\
\rho\frac{\partial v}{\partial t}=-\nabla P       
\end{dcases}
\label{eq:C1} 
\end{equation} 
here $P(r,t)$ $\&$ $v(r,t)$ are the pressure and velocity field. The above equation can be written as:
\begin{equation}
\frac{\partial}{\partial t}\left(\frac{1}{2}\frac{P^2}{\rho C_l^2}+\frac{1}{2}\rho v^2\right)=NP\frac{\partial V}{\partial t}
\label{eq:C2} 
\end{equation}
Where $V(r,t)$ is the instantaneous volume of the bubbles at time $t$ and $N$ is the number of bubbles per unit volume. $V$ can be calculated by solving the related bubble model (Eqs.\ref{eq:A1}, \ref{eq:A2} and \ref{eq:A3}). In order to get an energetic interpretation of the bubble radial motion, both sides of the bubble model (e.g. Eq. \ref{eq:A1} or \ref{eq:A2} or \ref{eq:A3}) can be multiplied by the time derivative of the bubble volume ${\partial V}/{\partial t}$ and using the equation of the  kinetic energy of the liquid \cite{18} $K_l=2\pi \rho R^3 \dot{R}^2 $ and Eq.\ref{eq:C2}, one can arrive at:
\begin{equation}
\begin{gathered}
\frac{\partial}{\partial t}\left(\frac{1}{2}\frac{P^2}{\rho C_l^2}+\frac{1}{2}\rho v^2+NK_l+4N\pi R^2 \sigma\right)+\nabla.(Pv)=\\-N\left(\pi_{total}\right)
\end{gathered}
\label{eq:C3}
\end{equation}
where $\pi_{total}$ is total dissipated energy  term. Because Louisnard used the plain Rayleigh-Plesset equation that does not incorporate the compressibility effects of the liquid; he was not able to derive the terms that describe the nonlinear loss due to radiation effects. Jamshidi $\&$ Brenner \cite{19} used the K-M equation (Eq. \ref{eq:A1}) that incorporates the compressibility effects up to the first order of Mach number. Thus, they were able to derive the nonlinear radiation loss terms. In \cite{27} we provided critical corrections to the terms derived in \cite{19}.\\ 
By taking a time average of both sides of Eq.\ref{eq:C3} and eliminating terms that are zero:
\begin{equation}
\nabla.<Pv>=-N\left(\Pi_{Total}\right)
\label{eq:C4}
\end{equation} 
Where $\Pi_{Total}$ is the total dissipated power. Eq. \ref{eq:C4} expresses the conservation of mechanical energy averaged over one or many periods of oscillations. The physical interpretation of this equation is that the the acoustic energy leaving a volume of bubbly liquid is always smaller than the one incident on it \cite{18}.  This is due to the losses during the bubble oscillations.
Each bubble therefore acts as a dissipator of acoustic energy. The physical origin of wave attenuation is thus self-contained in the Caflish model, even for nonlinear oscillations, provided that a correct model is used to describe the losses in the bubble oscillation. In \cite{37}, Caflish et. al proposed a conservation equation similar to Eq. 35; however, since they disregarded viscosity and assumed isothermal oscillations, mechanical energy was conserved. Eq. \ref{eq:C4} as derived by Louisnard \cite{18} reverts the same equation solved in 1D by Rozenberg \cite{52} in the case of purely traveling waves, but in the latter work, the dissipated power was fitted from experimental data, rather than being calculated from single bubble dynamics as done by Louisnard \cite{18}.\\
\subsection{Nonlinear dissipation terms of the uncoated free bubble}
In case of the uncoated bubble model Eq.\ref{eq:A1}, $\Pi_{Total}$ is the sum of the following dissipated powers \cite{27}:
\begin{widetext}
	\begin{equation}
	\begin{dcases}
	\Pi_{Thermal}=\frac{-4\pi}{T}\int_{0}^{T}R^2\dot{R}P_g dt\\ \\
	\Pi_{Liquid}=\frac{16\pi\mu_L}{T}\int_{0}^{T}R\dot{R}^2dt\\ \\
	\begin{gathered}
	\Pi_{Radiation}=\frac{1}{T}\int_{0}^{T} \left[\frac{4\pi}{C_l}\left(R^2\dot{R}\left(\dot{R}P+R\dot{P}-\frac{1}{2}\rho \dot{R}^3-\rho R\dot{R}\ddot{R}\right)\right)\right.\\ 
	\left.-\left(\frac{\dot{R}}{C_l}P_g+\frac{R}{C_l}\dot{P}_g\right)\frac{\partial V}{\partial t}+\frac{16\pi\mu_LR^2\dot{R}\ddot{R}}{C_l}\right]dt
	\end{gathered}
	\end{dcases}
	\label{eq:C5}
	\end{equation} 
\end{widetext}
where $T$ is the integration time interval.\\
\subsection{Nonlinear dissipation terms of the coated bubble}
For the coated bubble Eq.\ref{eq:A2},  $\Pi_{Total}$ is the sum of the following dissipated powers \cite{28}:
\begin{widetext}
	\begin{equation}
	\begin{dcases}
	\Pi_{Thermal}=\frac{-4\pi}{T}\int_{0}^{T}R^2\dot{R}P_g dt\\ \\
	\Pi_{Liquid}=\frac{16\pi\mu_L}{T}\int_{0}^{T}R\dot{R}^2dt\\ \\
	\Pi_{Shell}=\frac{48\pi\mu_{sh}\varepsilon R_0^2}{T}\int_{0}^{T}\frac{\dot{R}^2}{R^2}dt\\ \\
	\Pi_{Gs}=\frac{48\pi G_s\varepsilon R_0^2}{T}\int_{0}^{T}\left(\frac{\dot{R}}{R}-\frac{R_0\dot{R}}{R^2}\right)dt\\ \\
	\begin{gathered}
	\Pi_{Radiation}=\frac{1}{T}\int_{0}^{t}\left(4\pi\left[\frac{R^2\dot{R}^2}{C_l}\left(P-P_g\right)+\frac{R^3\dot{R}}{C_l}\left(\dot{P}-\dot{P}_g\right)+\frac{4\mu_LR^2\dot{R}\ddot{R}}{C_l}\right.\right.\\
	\left.+12\mu_{sh}\varepsilon R0^2 \left(\frac{\dot{R}\ddot{R}}{C_lR}-\frac{3\dot{R}^3}{C_l R^2}\right)+12G_s\varepsilon R0^2\left(\frac{-2\dot{R}^2}{cR}+\frac{3R_0\dot{R}^2}{C_lR^2}\right) \right]\\
	\left. -\frac{\rho R^2\dot{R}^4}{2C_l}-\frac{\rho R^3 \dot{R}^2\ddot{R}}{C_l}\right)dt
	\end{gathered}
	\label{eq:C6}
	\end{dcases}
	\end{equation}
\end{widetext}
\setlength{\belowdisplayskip}{-10pt}
\subsection{Nonlinear dissipation terms of the bubble in sediment or tissue}
For the bubbles immersed in sediments or tissue Eq.\ref{eq:A3},  $\Pi_{Total}$ is derived using the same approach in \cite{27,28}:
\setlength{\belowdisplayskip}{0pt}
\begin{widetext}
	\begin{equation}
	\begin{dcases}
	\Pi_{Thermal}=\frac{-4\pi}{T}\int_{0}^{T}R^2\dot{R}P_g dt\\ \\
	\Pi_{Liquid}=\frac{16\pi\mu_L}{T}\int_{0}^{T}R\dot{R}^2dt\\ \\
	\begin{gathered}
	\Pi_{Radiation}=\frac{1}{T}\int_{0}^{T} \left[\frac{4\pi}{C_l}\left(R^2\dot{R}\left(\dot{R}P+R\dot{P}-\frac{1}{2}\rho \dot{R}^3-\rho R\dot{R}\ddot{R}+\frac{4G\dot{R}}{3C_lR^3}\left(R^3-R_0^3\right)\right)\right)\right.\\ 
	\left.-\left(\frac{\dot{R}}{C_l}P_g+\frac{R}{C_l}\dot{P}_g\right)\frac{\partial V}{\partial t}+\frac{16\pi\mu_LR^2\dot{R}\ddot{R}}{C_l}\right]dt
	\end{gathered}\\ \\
	\Pi_{Sediment}=\frac{-1}{T}\int_{0}^{T}\frac{16G\pi\dot{R}}{3R}\left(R^3-R_0^3\right)dt
	\end{dcases}
	\label{eq:C7}
	\end{equation}
\end{widetext}
\setlength{\belowdisplayskip}{0pt}
\subsection{Pressure dependent attenuation and sound speed in Louisnard model}
Louisnard \cite{18} used the equations of energy dissipation and obtained the imaginary part of the $k^2$. In this model, the imaginary part of the $k^2$ is pressure dependent and is given by: 
\begin{equation}
\Im(k^2)=2 \rho \omega \sum_{j=1}^N \frac{\Pi(R_{0j})_{Total}}{|P|^2}
\label{eq:C8}
\end{equation}
where $\Pi(R_{0j})_{Total}$ is the total dissipated power due to the oscillations of the $jth$ bubble with initial radius of $R_{0j}$.\\
The real part of the $k^2$ is still calculated by the linear model (Eq. \ref{eq:B15}) and is given by:
\begin{equation}
\Re(k^2)=(\frac{\omega}{C_l})^2+4\pi \omega^2 \sum_{j=1}^N \frac{R_{0j}(\omega_{0j}^2-\omega^2)}{(\omega_{0j}^2-\omega^2)^2-4\delta_{totalj}^2 \omega^2}
\label{eq:C9}
\end{equation}
The Louisnard model \cite{18} (Eq. \ref{eq:C8}) incorporates the pressure effects in the imaginary part of the wave number. However, because the real part of the wave number is still estimated from the linear approximations it loses accuracy in predicting the sound speed and attenuation, especially in oscillation regimes where the sound speed changes are significant.
\section{Validation results}
\begin{figure*}
	\begin{center}
		\includegraphics[scale=0.55]{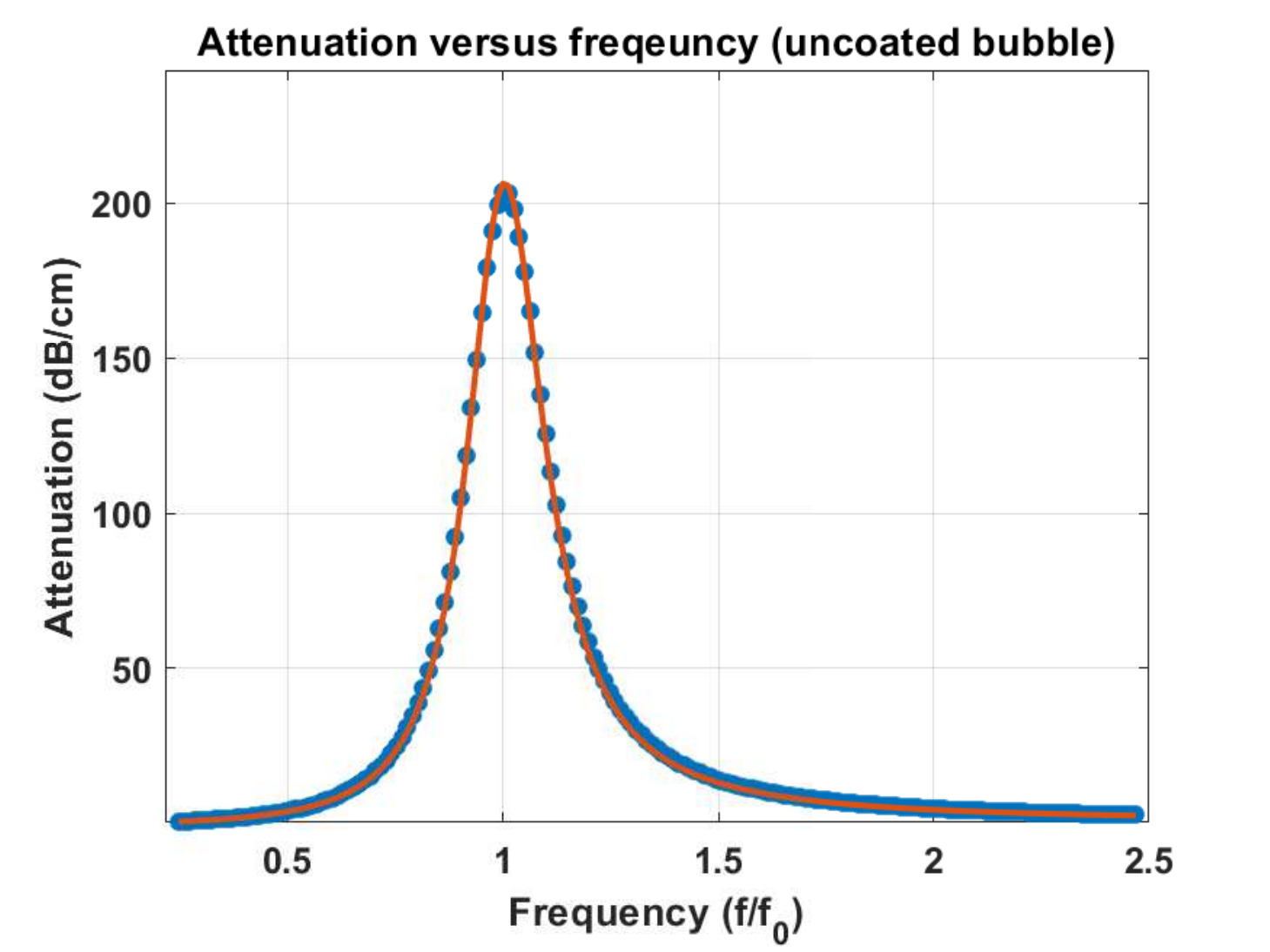} \includegraphics[scale=0.55]{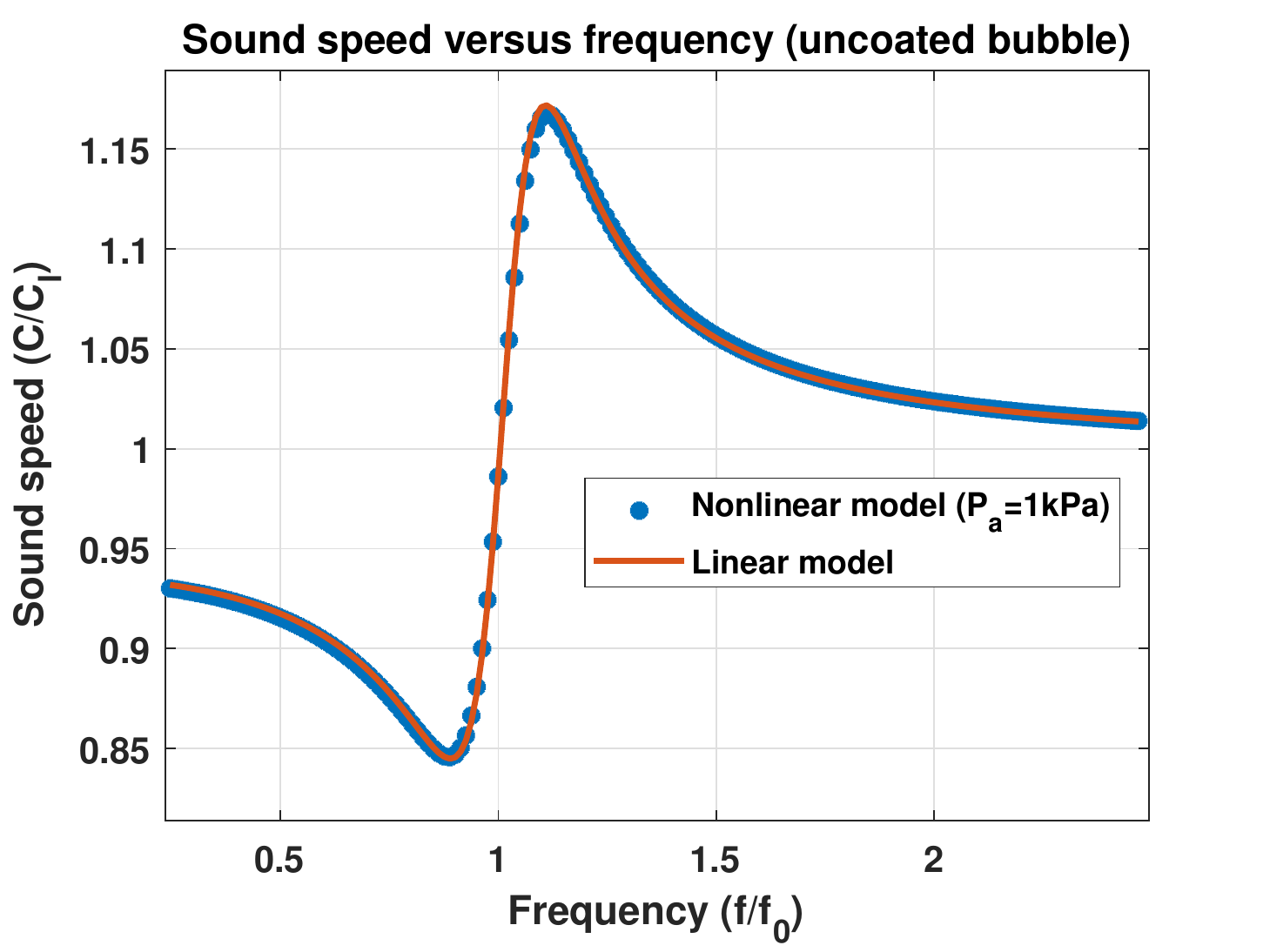}\\
		(a) \hspace{8 cm} (b)\\
		\includegraphics[scale=0.55]{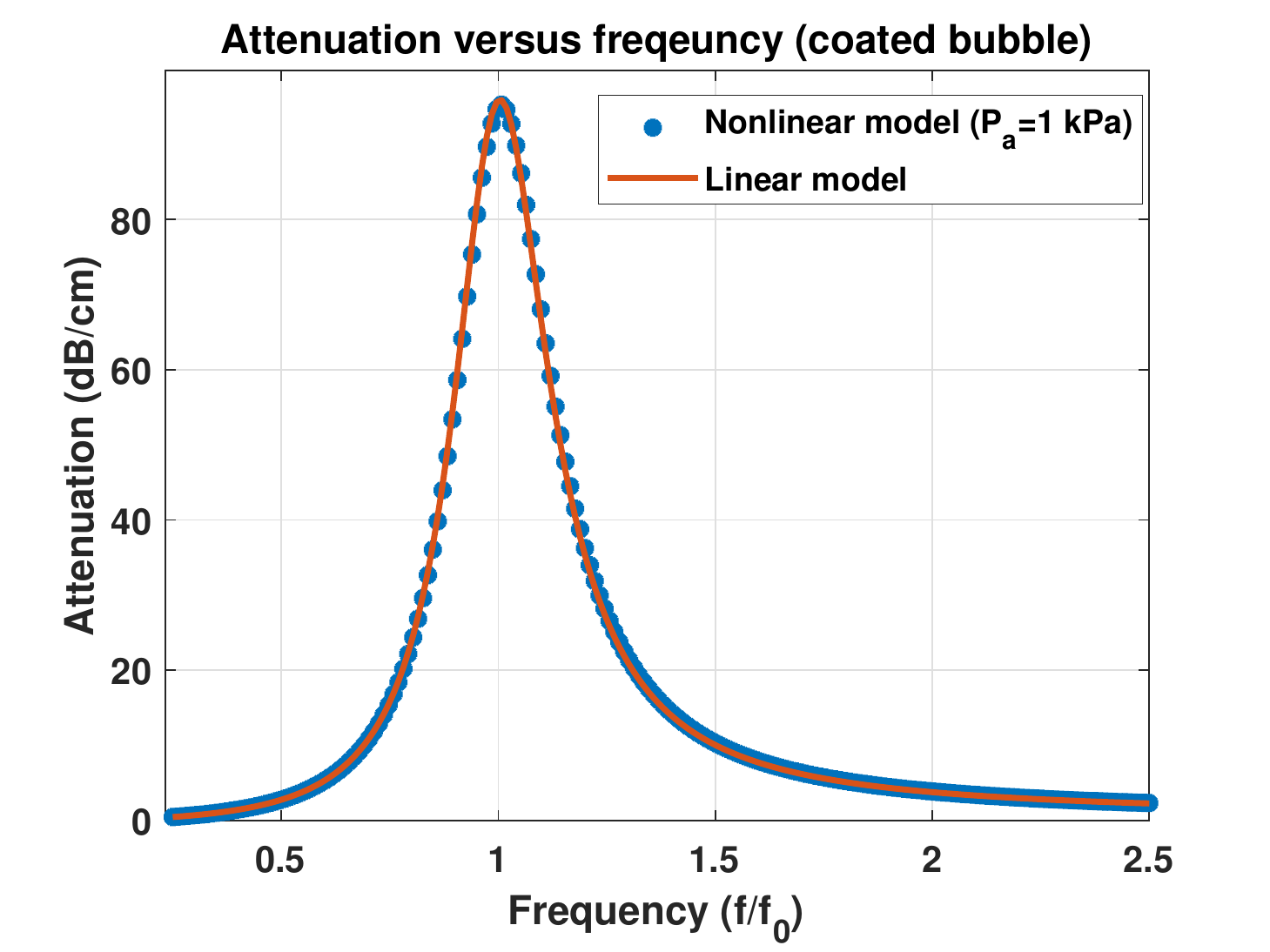}  \includegraphics[scale=0.55]{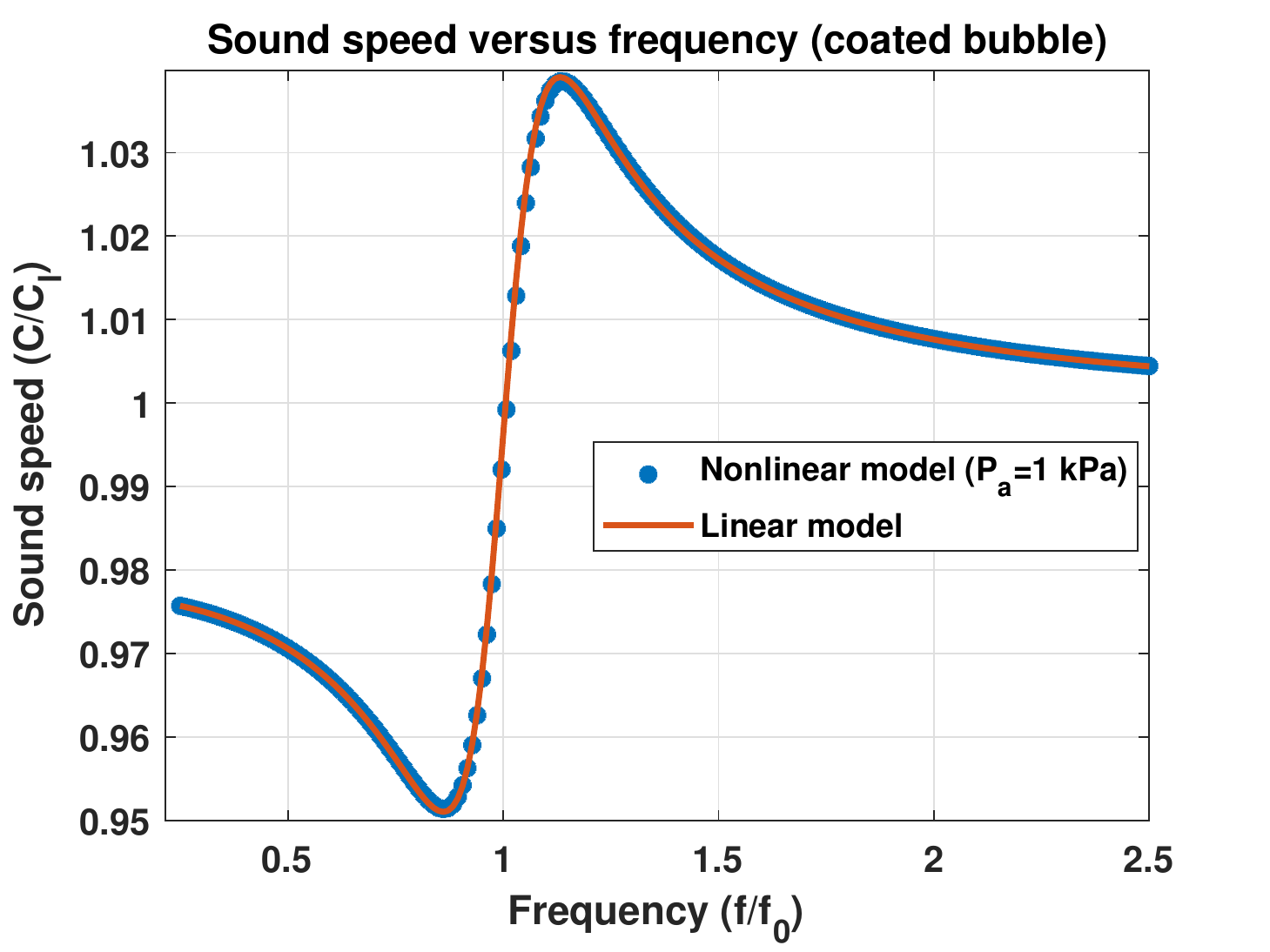}\\
		(c) \hspace{8 cm} (d)\\
		\includegraphics[scale=0.55]{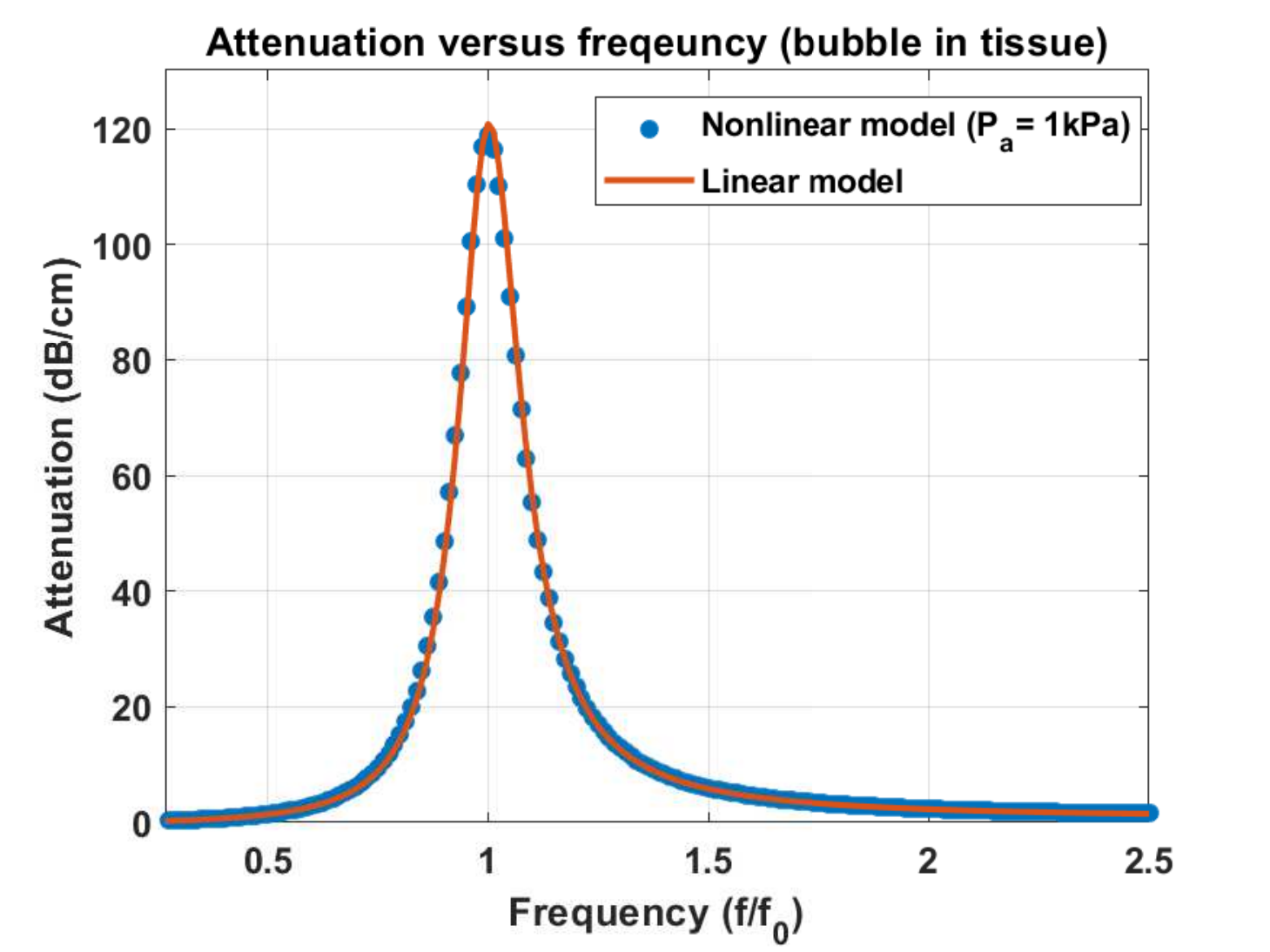}  \includegraphics[scale=0.55]{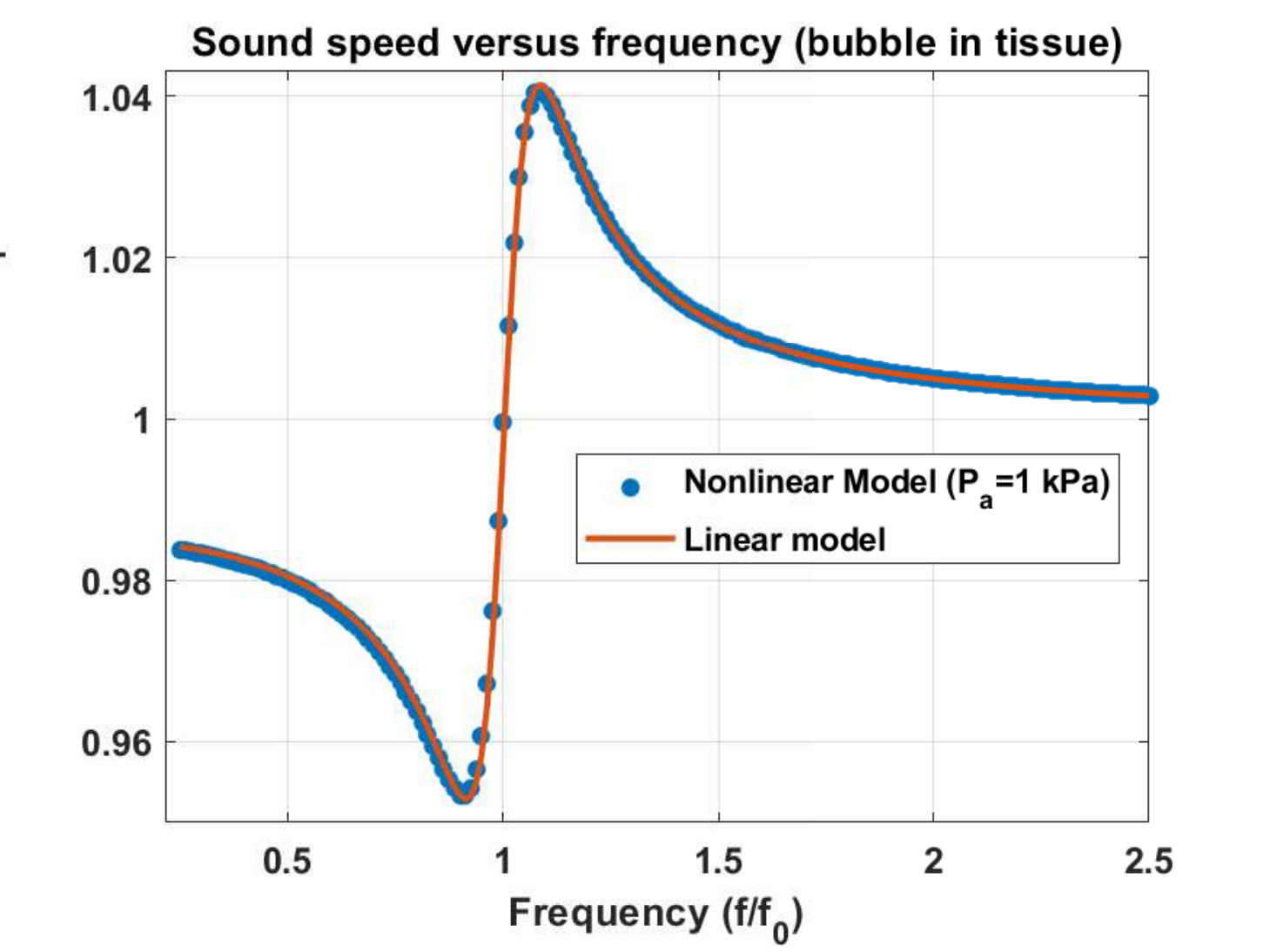}\\
		(e) \hspace{8 cm} (f)\\  
	\end{center}
	\caption{ Case of a bubbly medium with MBs with $R_0$= 2 $\mu m$ and {}{$\beta_0$}=10$^{-5}$. Attenuation calculated using the linear model and nonlinear model (left) and  sound speed calculated using the linear model and the nonlinear model at (P= 1kPa) (Right) for: uncoated bubbles in water (a and b), coated bubbles in water (c and d) and uncoated bubbles in tissue ($\rho$=1060 kg/m$^3$, $C_l$=1540 m/s, $\mu_s$= 0.00287 Pa.s, $G$=0.5 MPa, $\sigma$=0.056 N/m \cite{39}) (e and f).}
	\label{fig:6}
\end{figure*}
\subsection{Validation of the model at linear regimes against the linear models}
\label{subsection:D1}
The relevant models (Eq.\ref{eq:A1}  (uncoated bubble), Eq.\ref{eq:A2}  (coated bubble) and Eq. \ref{eq:A3} (bubble in tissue or sediment)) for large amplitude MB oscillations were coupled with the ordinary differential equations describing the thermal damping effects (Eq.\ref{eq:A8} and Eq.\ref{eq:A9}). The new set of differential equations were solved to calculate the MB radial oscillations. Constants of the linear models were calculated from Eq. \ref{eq:B8} (uncoated bubble), Eq.\ref{eq:B11} (coated bubble) and Eq.\ref{eq:B14} (bubble in tissue or sediment). Attenuation and sound speed were then calculated for each case using Eq.\ref{eq:B17}.\\ 
Since the linear model is only valid for narrowband pulses with small pressure
amplitudes, pulses of \textit{1kPa} amplitude with 60 cycles were chosen at
each frequency, and the last 20 cycles of the bubble oscillations were used (to eliminate the transient behavior) for the integration using Eqs.\ref{eq:3}
and \ref{eq:4}. For the linearized model, the initial MB radius is 2 $\mathrm{\mu}$m;
the gas inside the MB is\textit{ air} and the thermal properties are chosen
from \cite{49} (Table 1) and {}{$\beta_0$} was set to 10$^{-5}$. {}{Figures \ref{fig:6}a-f} compare the
attenuation and sound speed predictions between the linear model and the
non-linear model given by {}{Eqs}.\ref{eq:10} and \ref{eq:11}.\\ Model predictions are in excellent
agreement with the linear model for small amplitude radial oscillations ($R_{max}/R_0<1.01$) in Fig. \ref{fig:6}. The simple model given by {}{Eqs}.\ref{eq:10} and \ref{eq:11} predicted the attenuation and sound speed of the medium for the uncoated bubble, the coated bubble and the bubble in tissue. The model only requires as input the radial oscillations of the bubbles and reduces the complexity of deriving the linear terms in each cases. Fig. \ref{fig:6} also shows that the the effect of encapsulating shell (added viscosity and stiffness) reduced the bubble expansion ratio which translated to smaller changes in attenuation and sound speed when compared to the uncoated counterpart.
\begin{figure*}
	\begin{center}
		\includegraphics[scale=0.48]{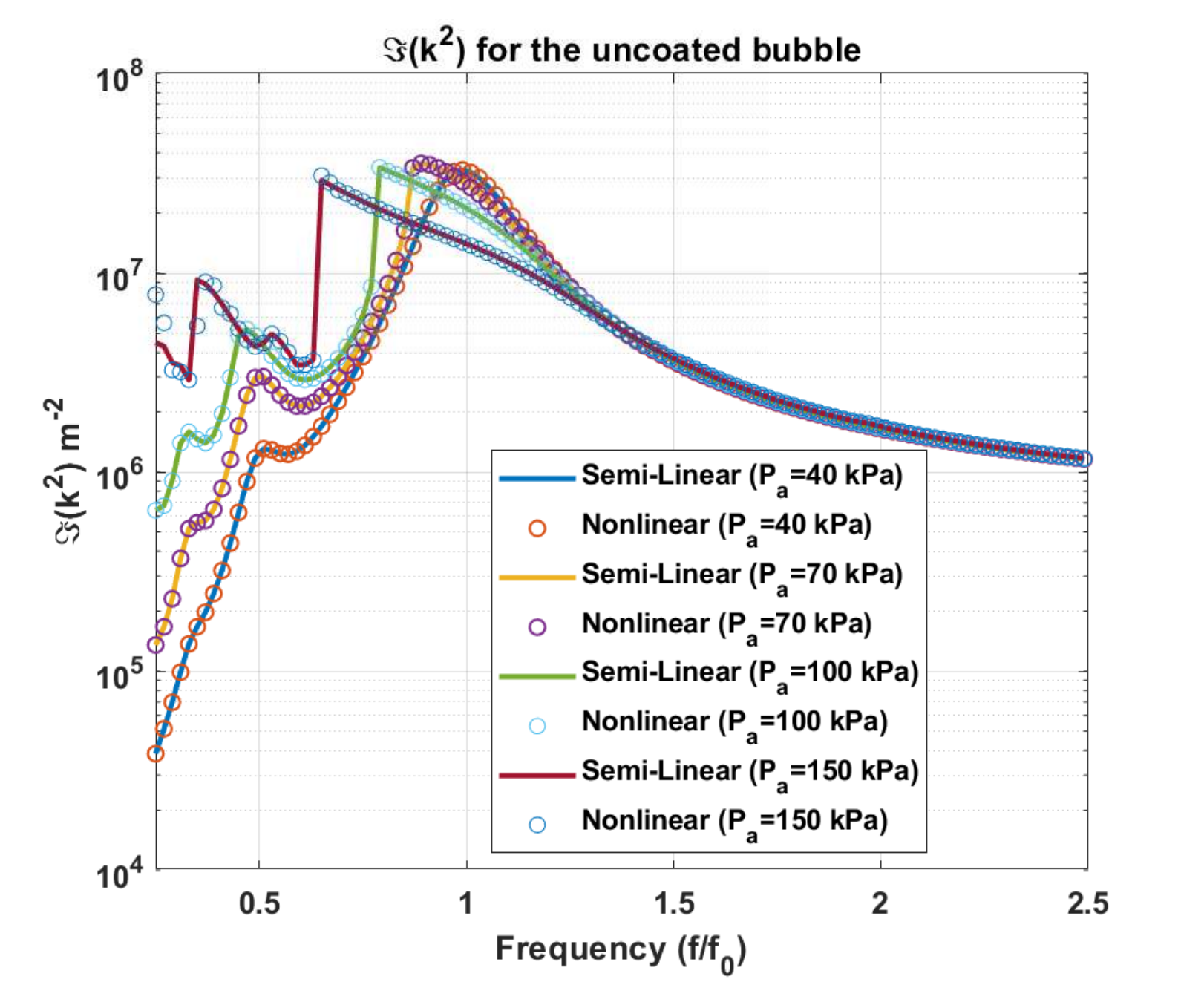}  \includegraphics[scale=0.48]{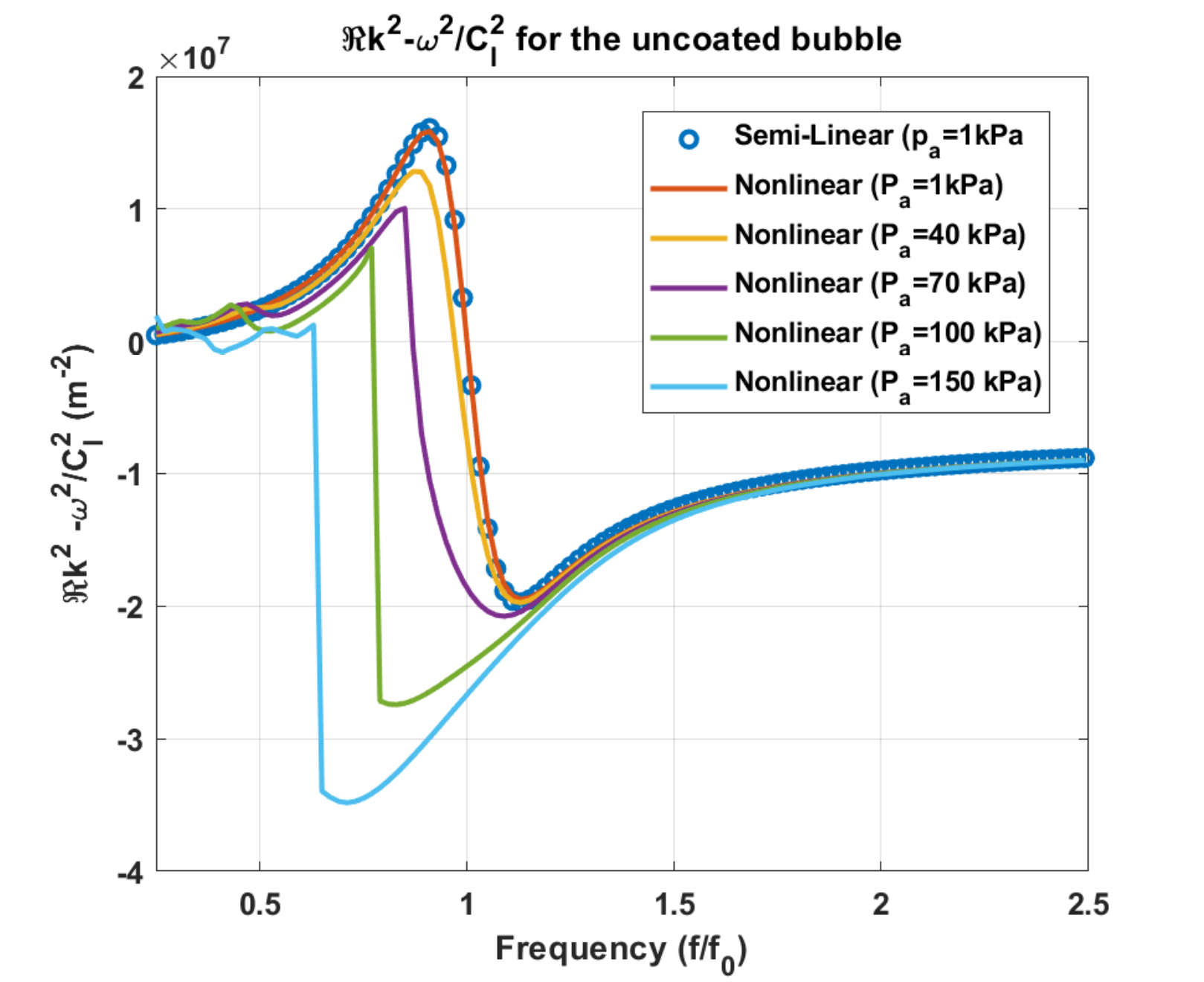}\\
		(a) \hspace{8 cm} (b)\\
		\includegraphics[scale=0.4]{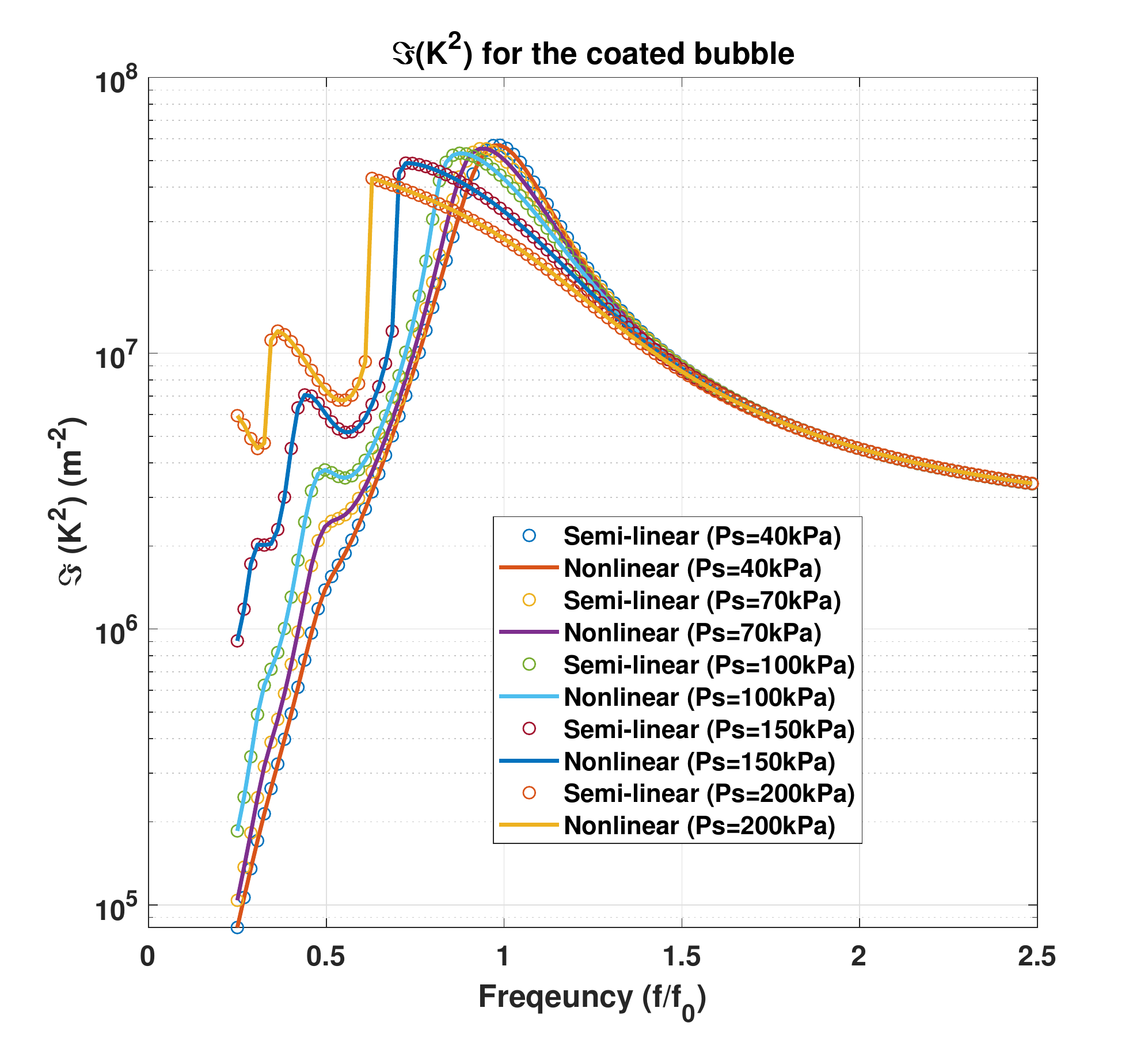}  \includegraphics[scale=0.4]{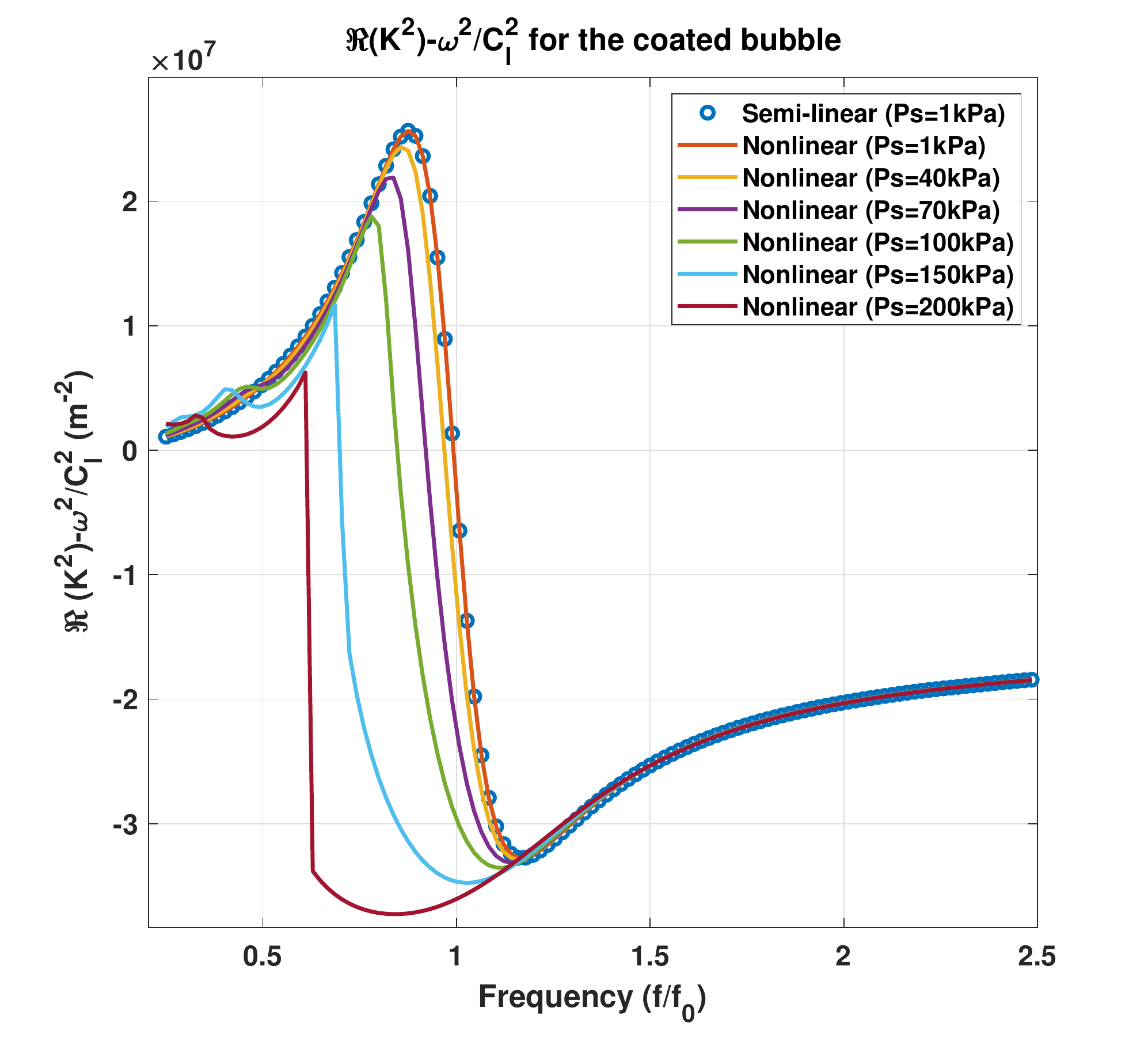}\\
		(c) \hspace{8 cm} (d)\\
		\includegraphics[scale=0.4]{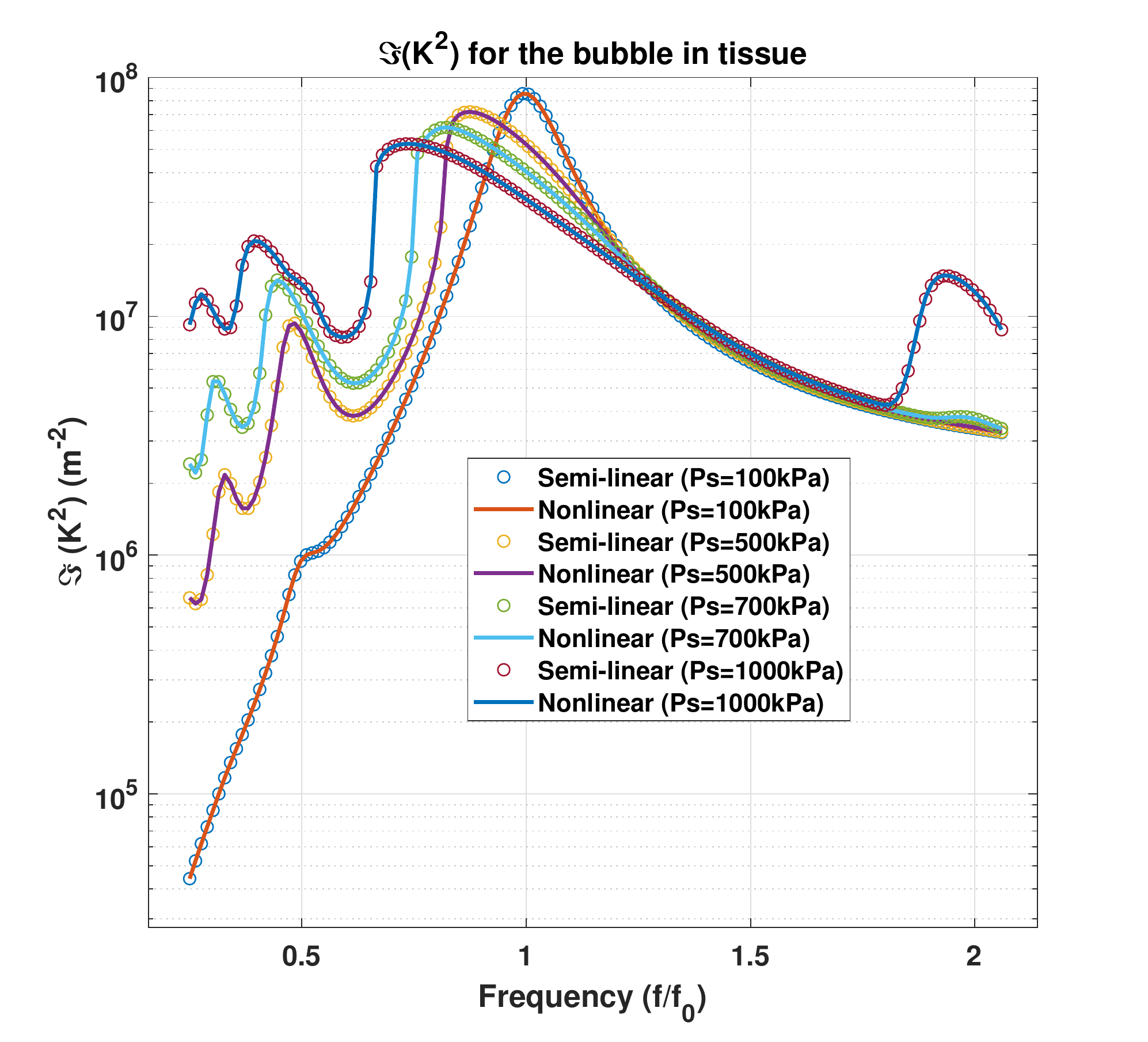}  \includegraphics[scale=0.4]{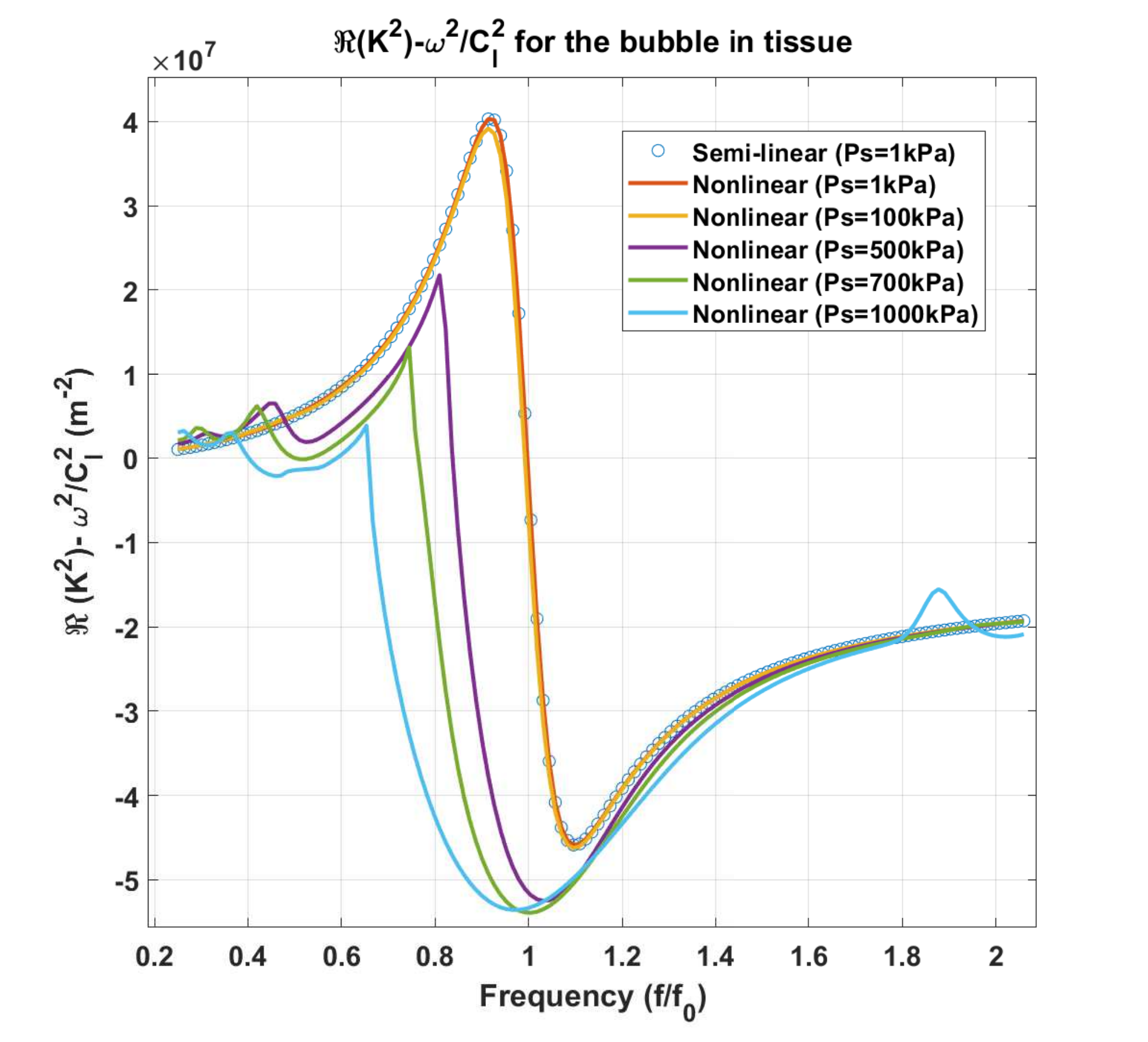}\\
		(e) \hspace{8 cm} (f)\\
	\end{center}
	\caption{ Case of a bubbly medium with MBs with $R_0$= 2 $\mu$m and {}{$\beta_0$}=10$^{-5}$ sonicated at various pressures. Left:  $\left\langle{}{\Im}\left(k^2\right)\right\rangle{}$ calculated using the nonlinear model (Eqs.\ref{eq:11}) and Louisnard model (Eq.\ref{eq:C8})  and  Right: $\left\langle{}{\Re}\left(k^2\right)\right\rangle{}$ calculated using the nonlinear model (Eqs.\ref{eq:10}) and the Louisnard model (Eq.\ref{eq:C9}) (Louisnard model employs the linear model for the real part; thus it is pressure independent) for: uncoated bubbles in water (a and b), coated bubbles in water (c $\&$ d) and uncoated bubbles in tissue ($\rho$=1060 kg/m$^3$, $C_l$=1540 m/s, $\mu_s$= 0.00287 Pa.s, $G$=0.5 MPa, $\sigma$=0.056 N/m \cite{39}) (e and f).}
	\label{fig:7}
\end{figure*}
\subsection{Validation of the model at higher pressures against the semi-linear Louisnard model}
\label{subsection:D2}
\begin{figure*}
	\begin{center}
		\includegraphics[scale=0.55]{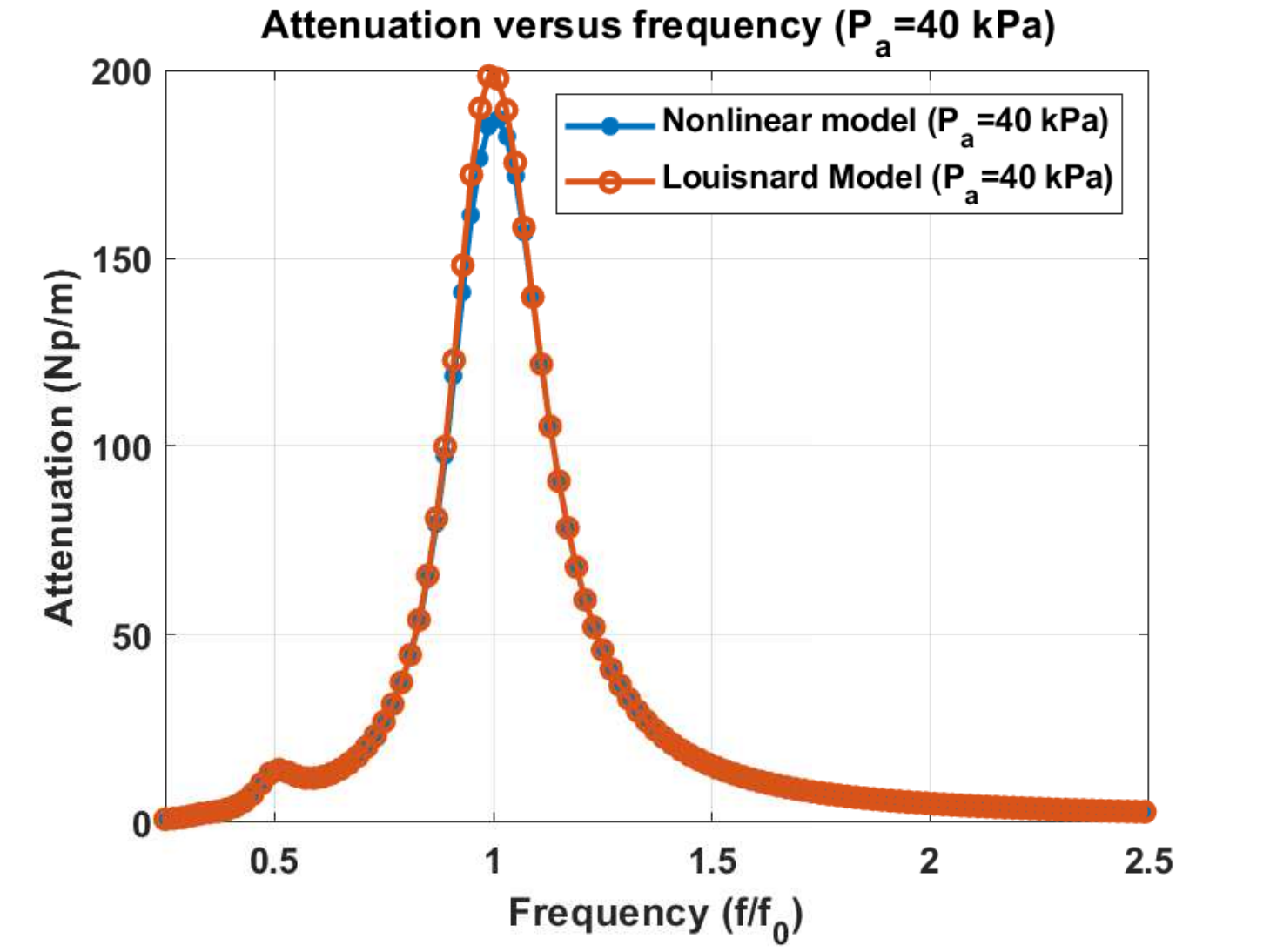}  \includegraphics[scale=0.55]{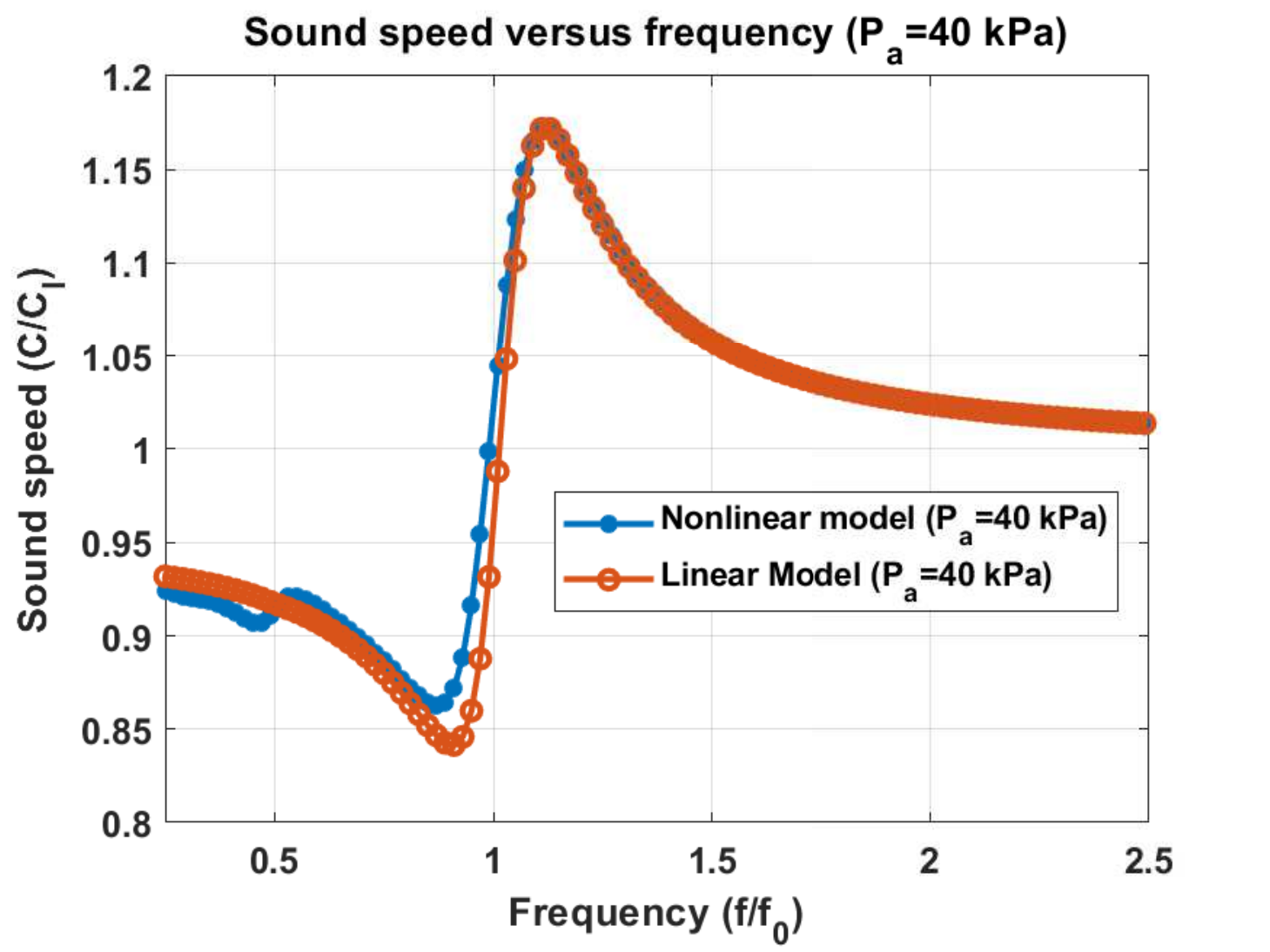}\\
		(a) \hspace{8 cm} (b)\\
		\includegraphics[scale=0.55]{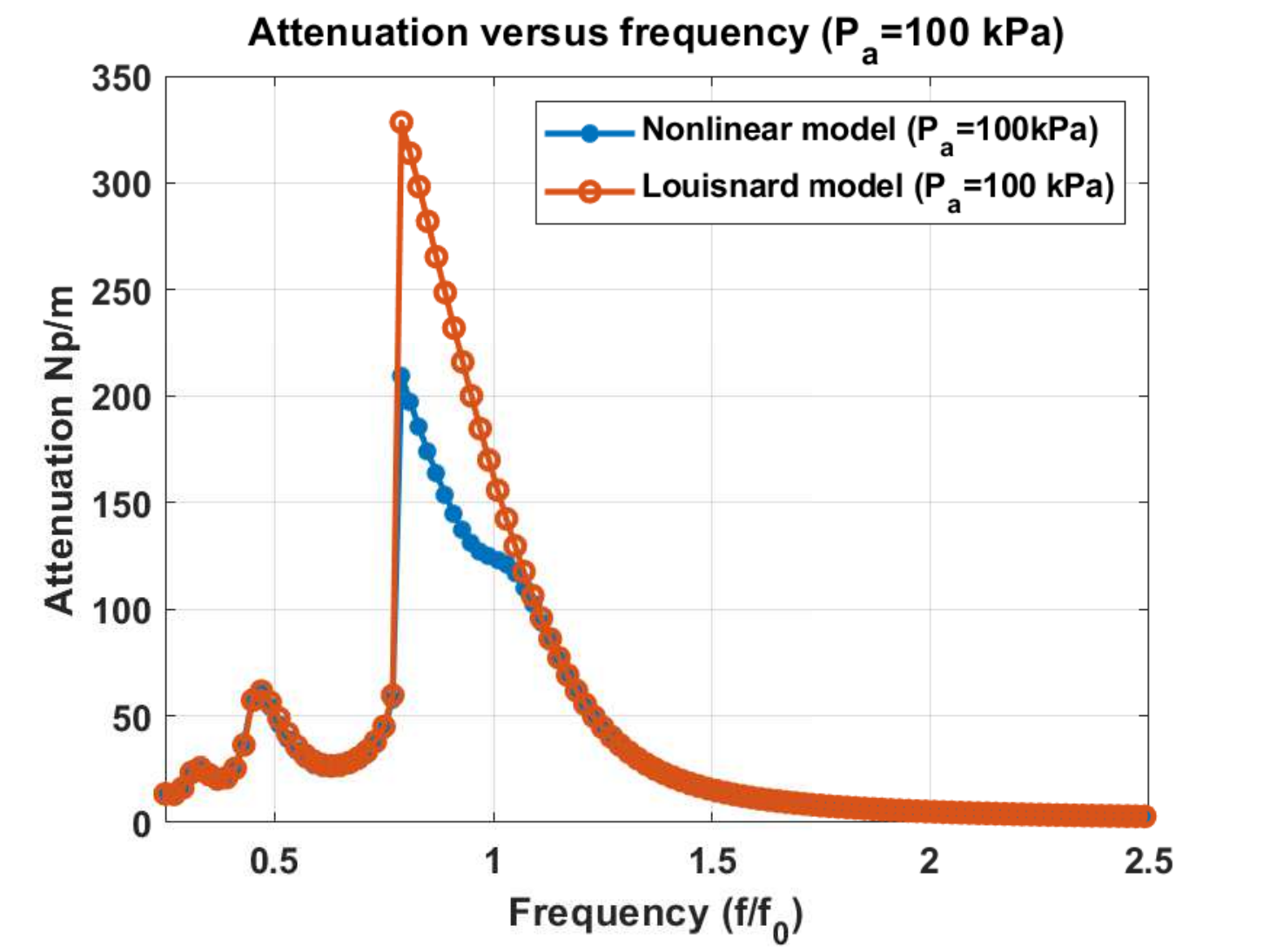}  \includegraphics[scale=0.55]{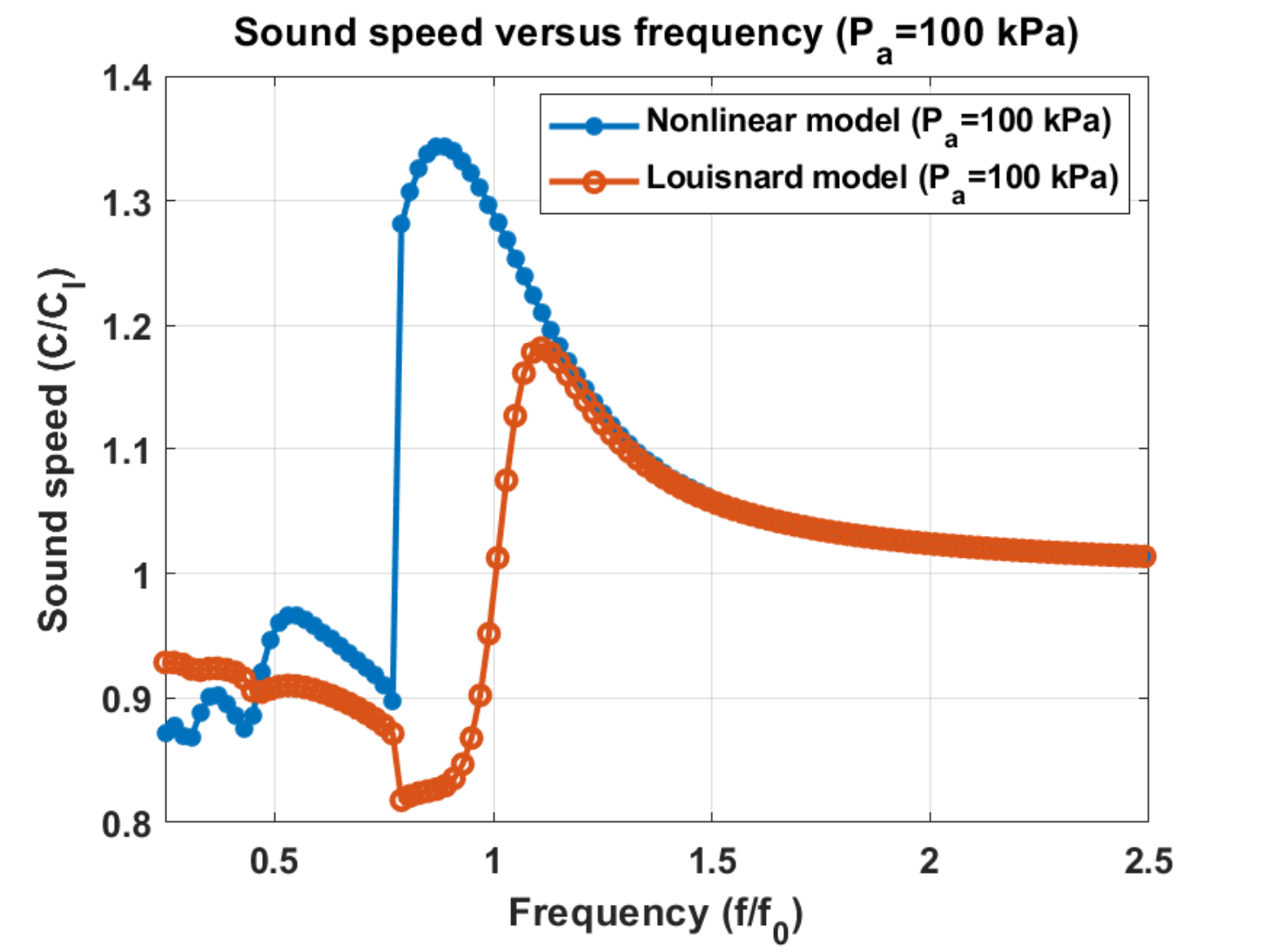}\\
		(c) \hspace{8cm} (d)\\
		\includegraphics[scale=0.55]{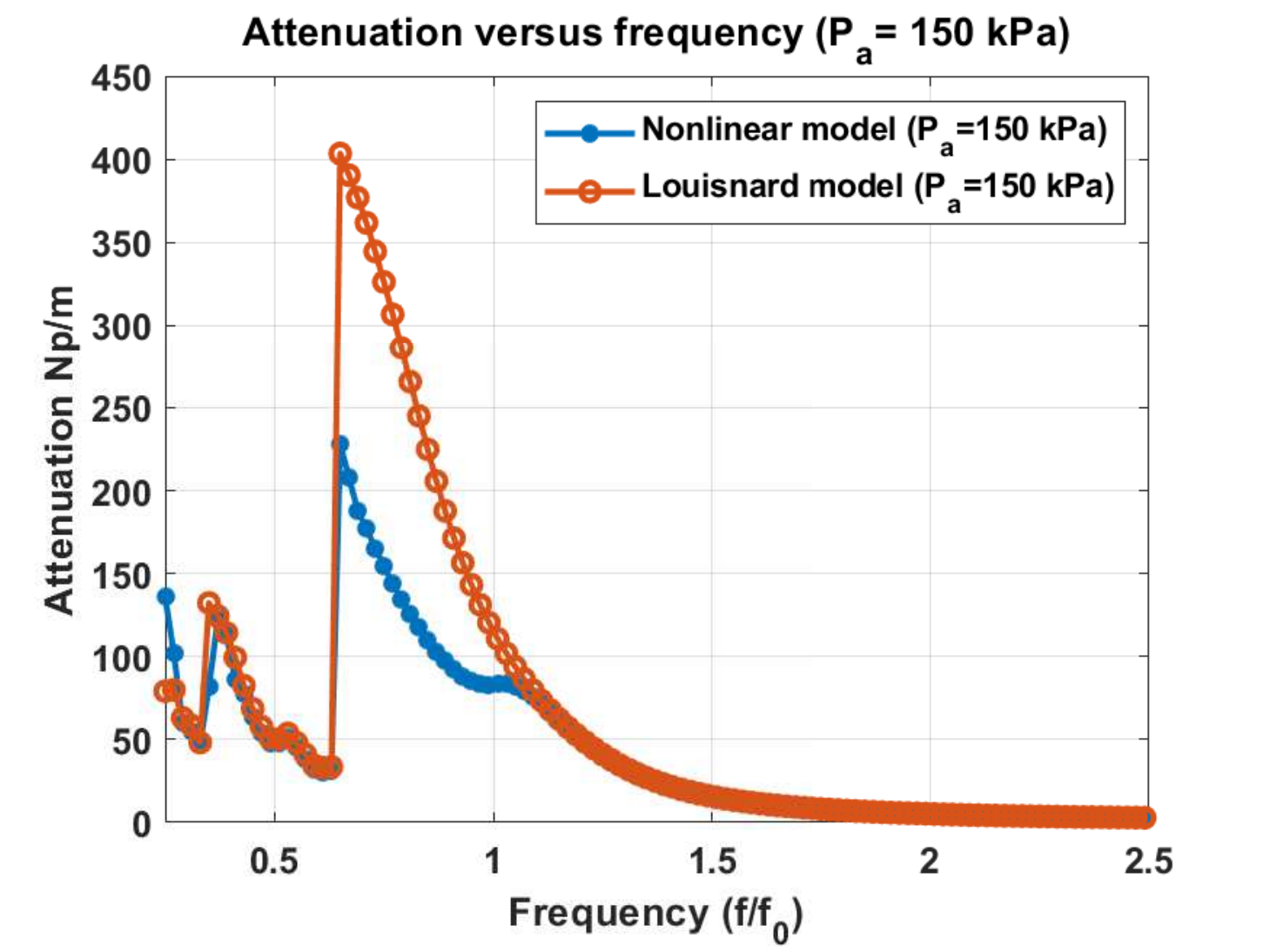}  \includegraphics[scale=0.55]{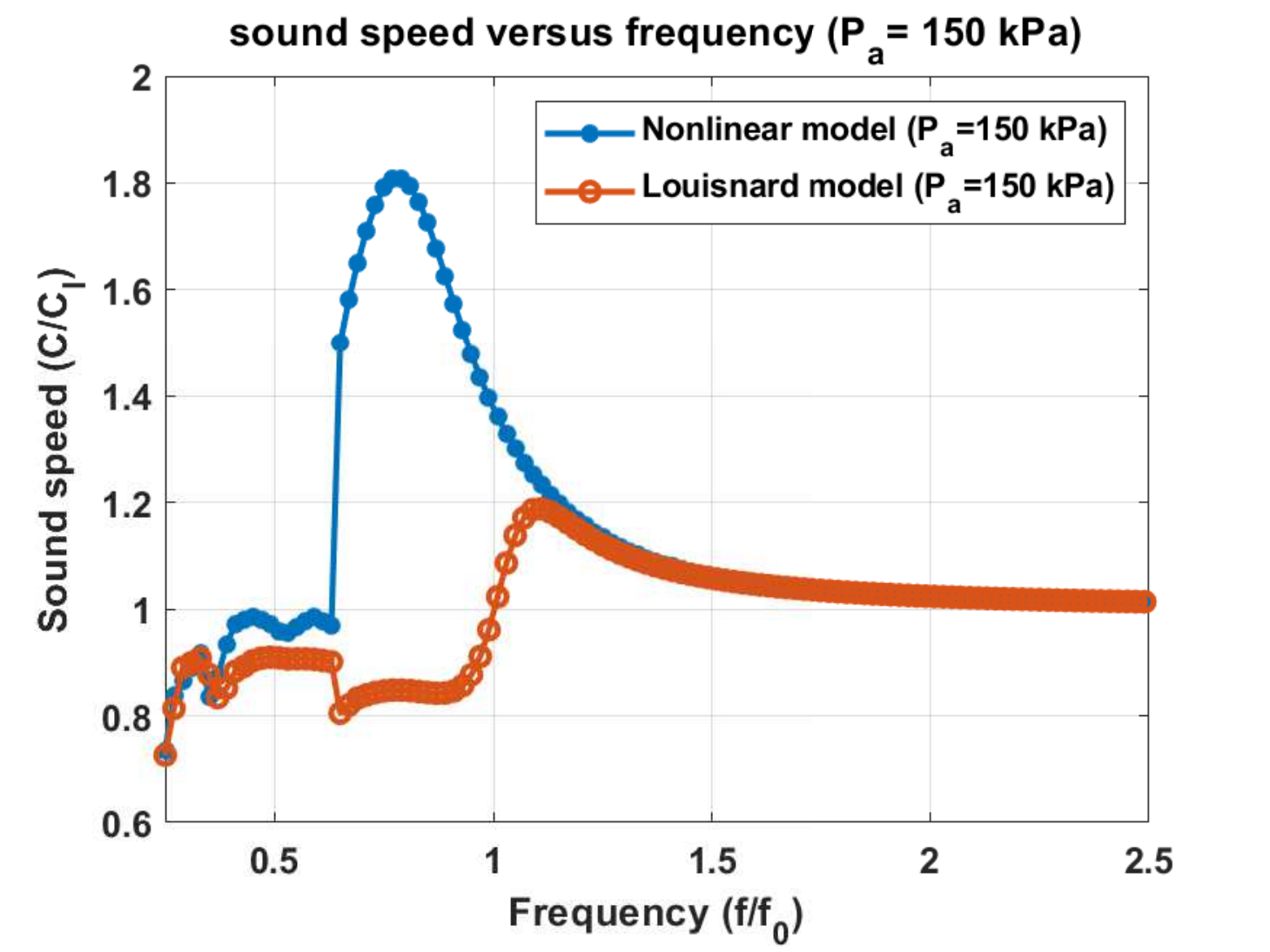}\\
		(e) \hspace{8 cm} (f)\\
	\end{center}
	\caption{ Comparison between the predictions the Louisnard $\&$ the nonlinear model for sound speed and attenuation. Case of a bubbly medium with uncoated MBs with $R_0$= {2 $\mu$ m} and {}{$\beta_0$}=10$^{-5}$. a) attenuation at $P_a$=40 {kPa}, b) sound speed at $P_a$=40 {kPa}, c) attenuation at $P_a$=100 {kPa}  d) sound speed at $P_a$=100 {kPa}, e) attenuation at $P_a$= 150 {kPa} and f) sound speed at $P_a$=150 {kPa}.}
	\label{fig:8}
\end{figure*}
As the pressure increases, assumptions (e.g. small amplitude MB oscillations) on which the linear model is based on are no longer valid. To investigate the effect of pressure, the radial oscillations of the MBs were simulated for exposures of  various acoustic pressure amplitudes. For the uncoated bubble $P_a$= 40, 70, 100, 150 kPa, for the coated bubble $P_a$= 40, 70, 100, 150, 200 kPa and for the bubble in tissue $P_a$= 100, 500, 700, 1000 kPa were chosen.  The power dissipation expressions for nonlinear damping effects which are given by Eq.\ref{eq:C5} (for uncoated bubble), Eq.\ref{eq:C6} for coated bubble and Eq.\ref{eq:C7} for the bubble in tissue were used to calculate the total dissipated power. The imaginary and real part of the wave number were then calculated using Eqs.\ref{eq:11} and \ref{eq:10} in case of the nonlinear model and Eq.\ref{eq:C8} and Eq.\ref{eq:C9} in case of the Louisnard model \cite{18}. The predictions of the two models are illustrated in Fig. \ref{fig:7}.\\ The left column of Fig. \ref{fig:7} shows that the $\left\langle{}{\Im}\left(k^2\right)\right\rangle{}$ calculated by Eq. \ref{eq:11} is in excellent agreement with the Louisnard model (Eq. \ref{eq:C8}) for all the acoustic pressures and the bubble models that investigated. The simple approach introduced here, only needs the radial oscillations of the bubble as input and reduces the complexity of the Louisnard model where the equations for different dissipation mechanisms must be derived for each bubble case. Variations of $\left\langle{}{\Im}\left(k^2\right)\right\rangle{}$ with pressure shows the importance of the considerations of the pressure effects as the linear model fails to predict phenomena like the resonance shift (e.g. \cite{14}), changes in the amplitude of the $\left\langle{}{\Im}\left(k^2\right)\right\rangle{}$ with pressure and the generation of SuH (e.g. \cite{sojahrood3}) and subharmonic (SH) resonances (e.g. \cite{54,55}).\\ As an instance in case of the uncoated bubble in Fig. \ref{fig:7}a $\left\langle{}{\Im}\left(k^2\right)\right\rangle{}\approxeq$ 8.5$\times$10$^{8}$ m$^{-2}$ at pressure dependent resonance( $\frac{f}{f_r}\approxeq 0.98$) when $P_a$=40 kPa. However, as pressure increases to $P_a$=150 kPa resonance shifts to $f/f_r\approxeq$ 0.64 and $\left\langle{}{\Im}\left(k^2\right)\right\rangle{}\approxeq$ 7.3$\times$10$^{8}$ m$^{-2}$. Moreover, a SuH occurs at ${f}/{f_r}\approxeq$ 0.34 with $\left\langle{}{\Im}\left(k^2\right)\right\rangle{}\approxeq$ 2.4$\times$10$^{8}$ m$^{-2}$. When $P_a=40 kPa$ and at $\frac{f}{f_r}\approxeq$ 0.34, $\left\langle{}{\Im}\left(k^2\right)\right\rangle{}\approxeq$ 3.9$\times$10$^{8}$ m$^{-2}$. Thus, the pressure increase has a significant influence on the resonances of the system and the magnitude of the $\left\langle{}{\Im}\left(k^2\right)\right\rangle{}$.\\
The Louisnard model uses the linear assumptions (Eq.\ref{eq:C9}) to calculate the $\left\langle{}{\Re}\left(k^2\right)\right\rangle{}$. The predictions of the nonlinear model Eq.\ref{eq:10} for $\left\langle{}{\Re}\left(k^2\right)\right\rangle{}-({\omega}/{C_l})^2$, are compared with the predictions of Eq.\ref{eq:C9} in the right hand column of Fig. \ref{fig:7} and for 3 bubble cases (uncoated, coated and bubble in tissue). We have subtracted the constant $({\omega}/{C_l})^2$ from $\left\langle{}{\Re}\left(k^2\right)\right\rangle{}$ to better highlight the pressure dependent changes. In each case, pressure increase leads to significant changes in 
$\left\langle{}{\Re}\left(k^2\right)\right\rangle{}$, and predictions of Eq. \ref{eq:3} significantly deviate from the linear values (Eq. \ref{eq:C9}). 
As an instance for the uncoated bubble (Fig. 2b) the linear model predicts a maximum for $\left\langle{}{\Re}\left(k^2\right)\right\rangle{}-({\omega}/{C_l})^2\approxeq$ 4.1$\times$10$^{7}$ m$^{-2}$ at ${f/f_r}\approxeq$ 0.9 and a minimum for  $\left\langle{}{\Re}\left(k^2\right)\right\rangle{}-({\omega}/{C_l})^2\approxeq$ -5$\times$10$^{7}$ m$^{-2}$ at $\frac{f}{f_r}\approxeq 1.12$. However, when $P_a=100 kPa$ the maximum of $\left\langle{}{\Re}\left(k^2\right)\right\rangle{}-({\omega}/{C_l})^2\approxeq$ 1.9$\times$10$^{7}$ m$^{-2}$ at ${f}/{f_r}\approxeq$ 0.761 and the minimum is $\left\langle{}{\Re}\left(k^2\right)\right\rangle{}-({\omega}/{C_l})^2\approxeq$ -8.5$\times$10$^{7}$ m$^{-2}$ at ${f}/{f_r}\approxeq$ 0.773.\\
The nonlinear model incorporates the pressure-dependent changes in $\left\langle{}{\Re}\left(k^2\right)\right\rangle{}$ and thus can be used to predict the changes of the $\left\langle{}{\Re}\left(k^2\right)\right\rangle{}$ with pressure. To our best knowledge this is the first time that the frequency-pressure dependence of the $\left\langle{}{\Re}\left(k^2\right)\right\rangle{}$ in a bubbly medium has been calculated. The ability of the nonlinear model to calculate both the $\left\langle{}{\Im}\left(k^2\right)\right\rangle{}$ and $\left\langle{}{\Re}\left(k^2\right)\right\rangle{}$ with pressure changes increase the accuracy of the predictions of the medium attenuation and sound speed changes.\\
	\begin{figure*}
	\includegraphics[scale=0.55]{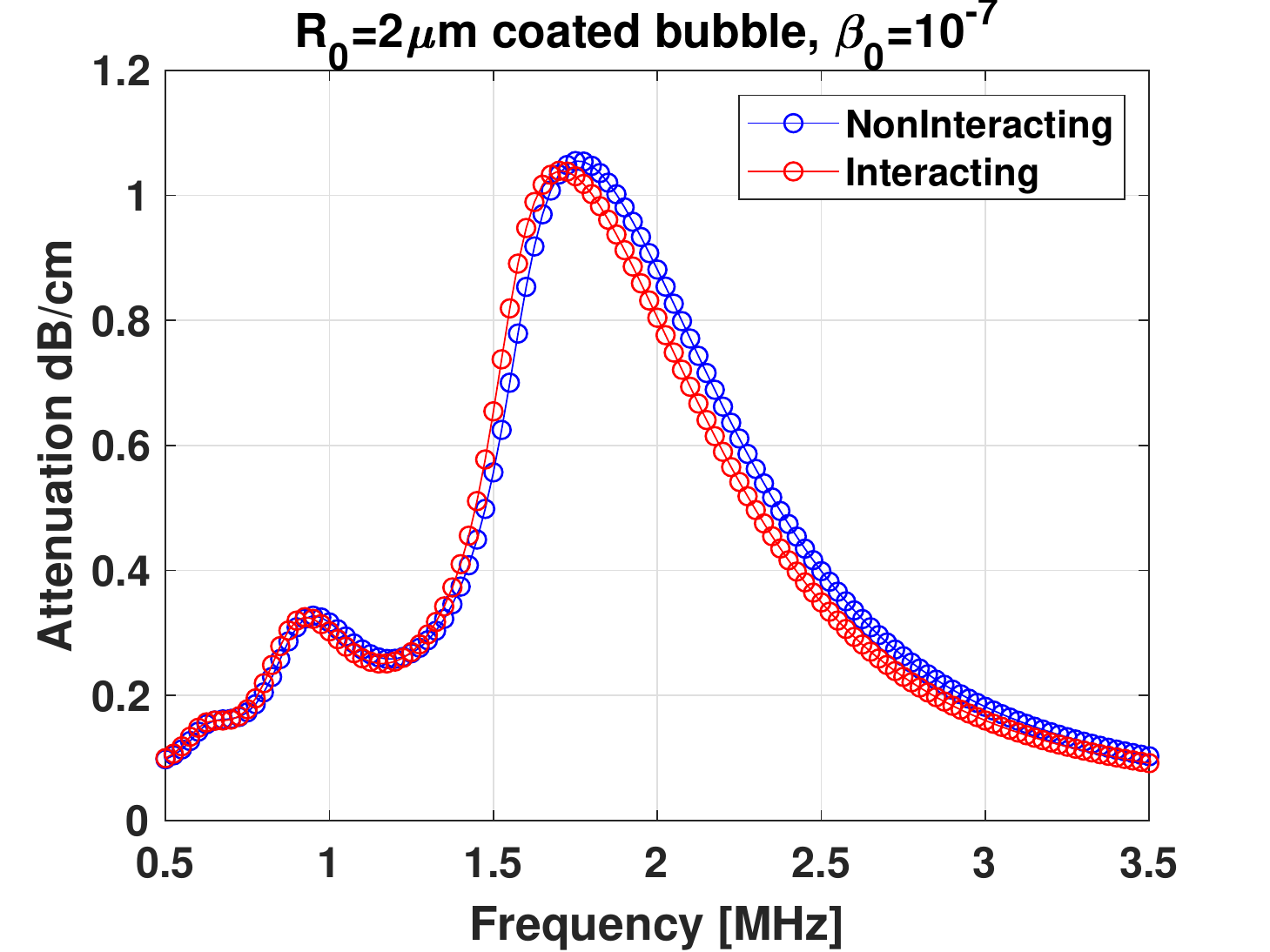} \includegraphics[scale=0.55]{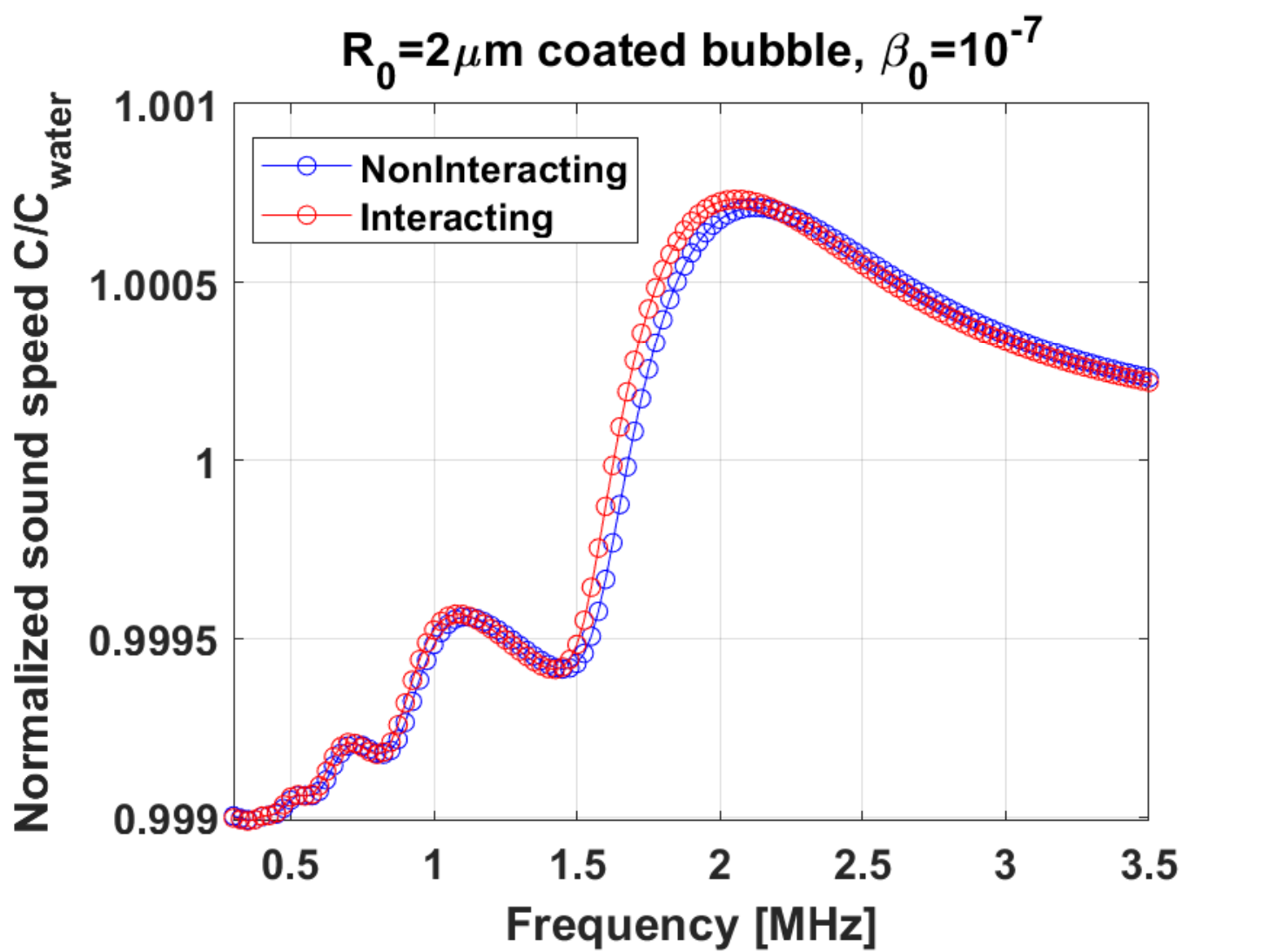} \\
	(a) \hspace{8 cm} (b)\\
		\includegraphics[scale=0.55]{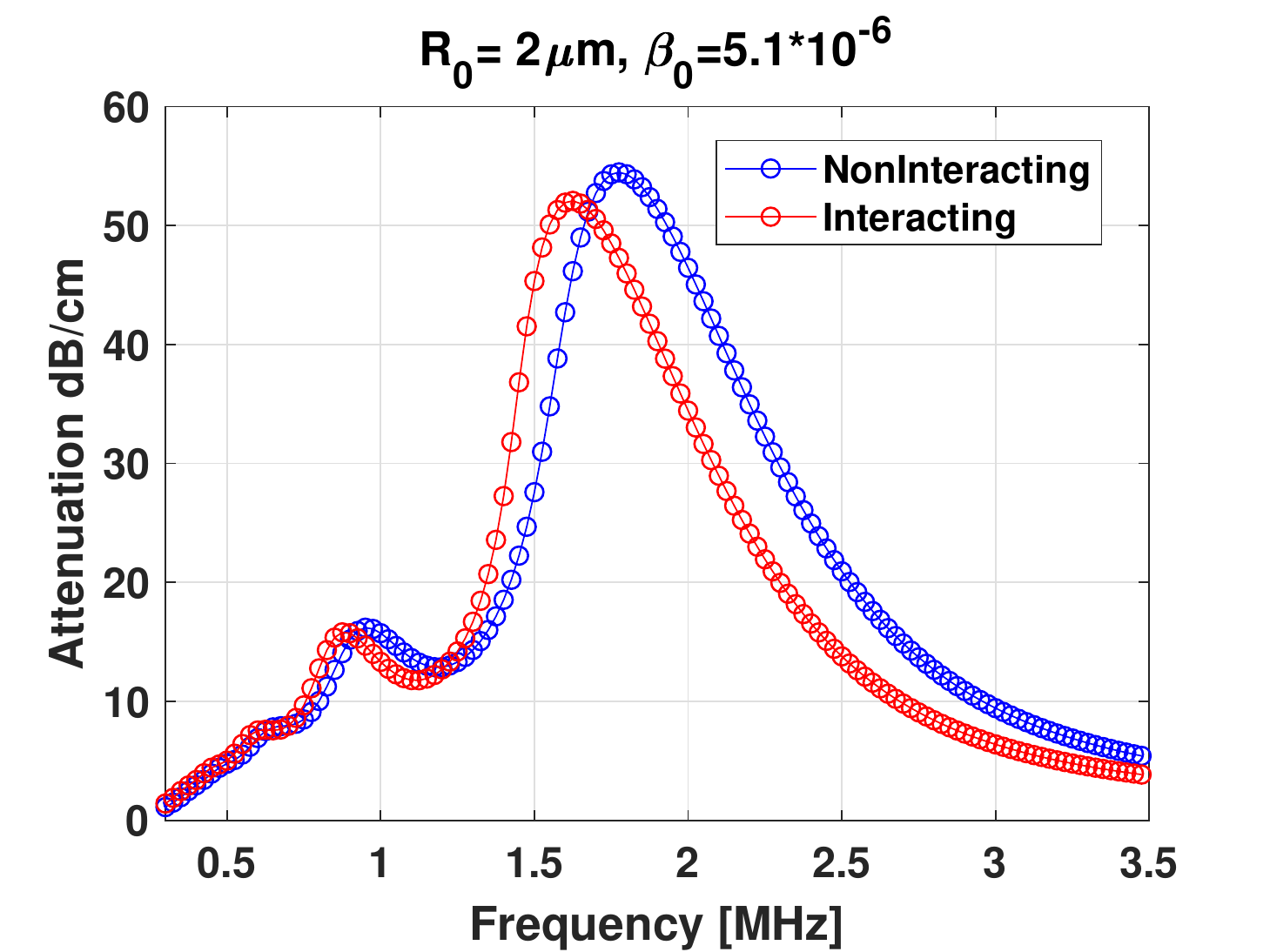} \includegraphics[scale=0.55]{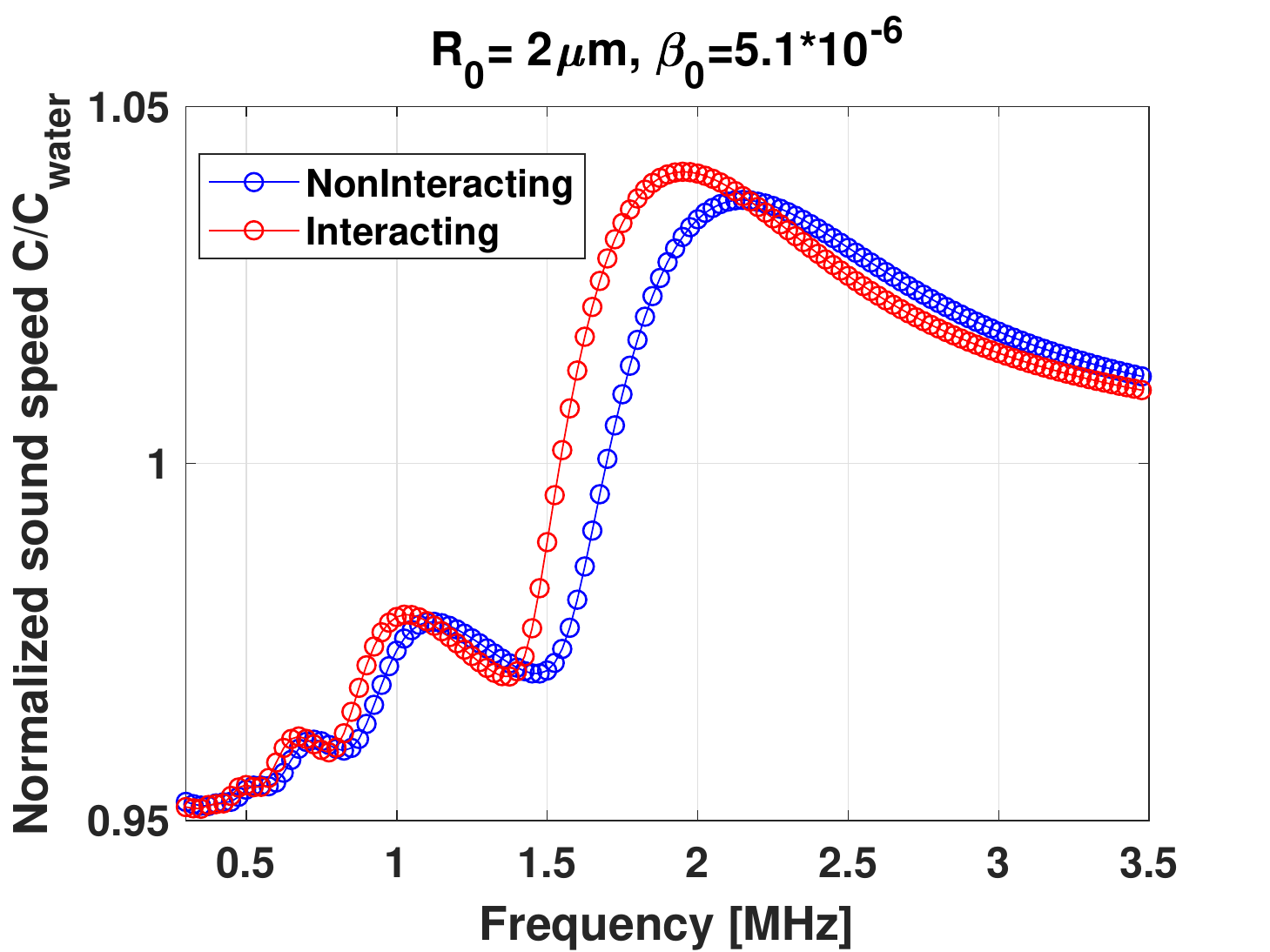} \\
	(c) \hspace{8 cm} (d)\\
	\includegraphics[scale=0.55]{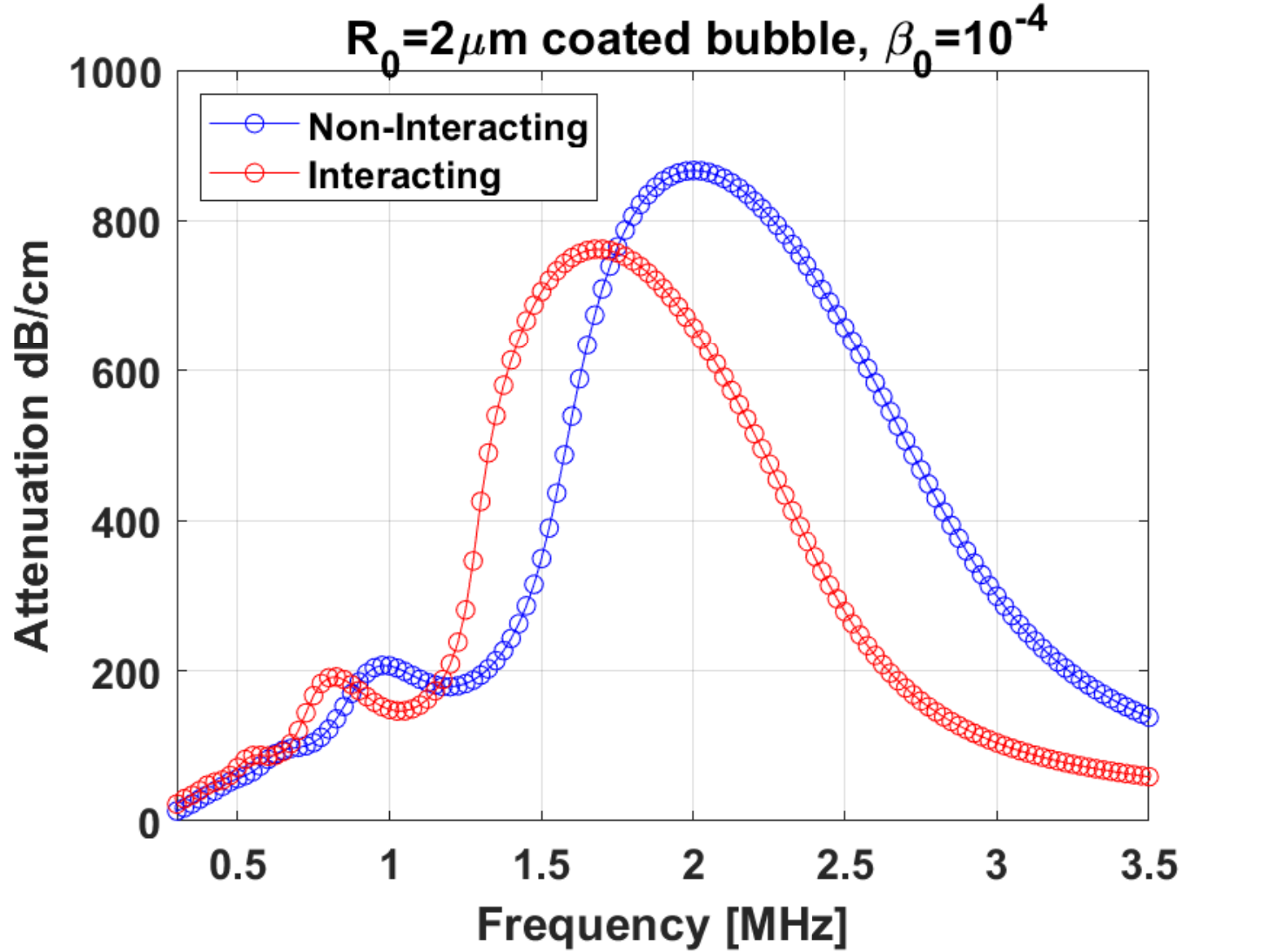} \includegraphics[scale=0.55]{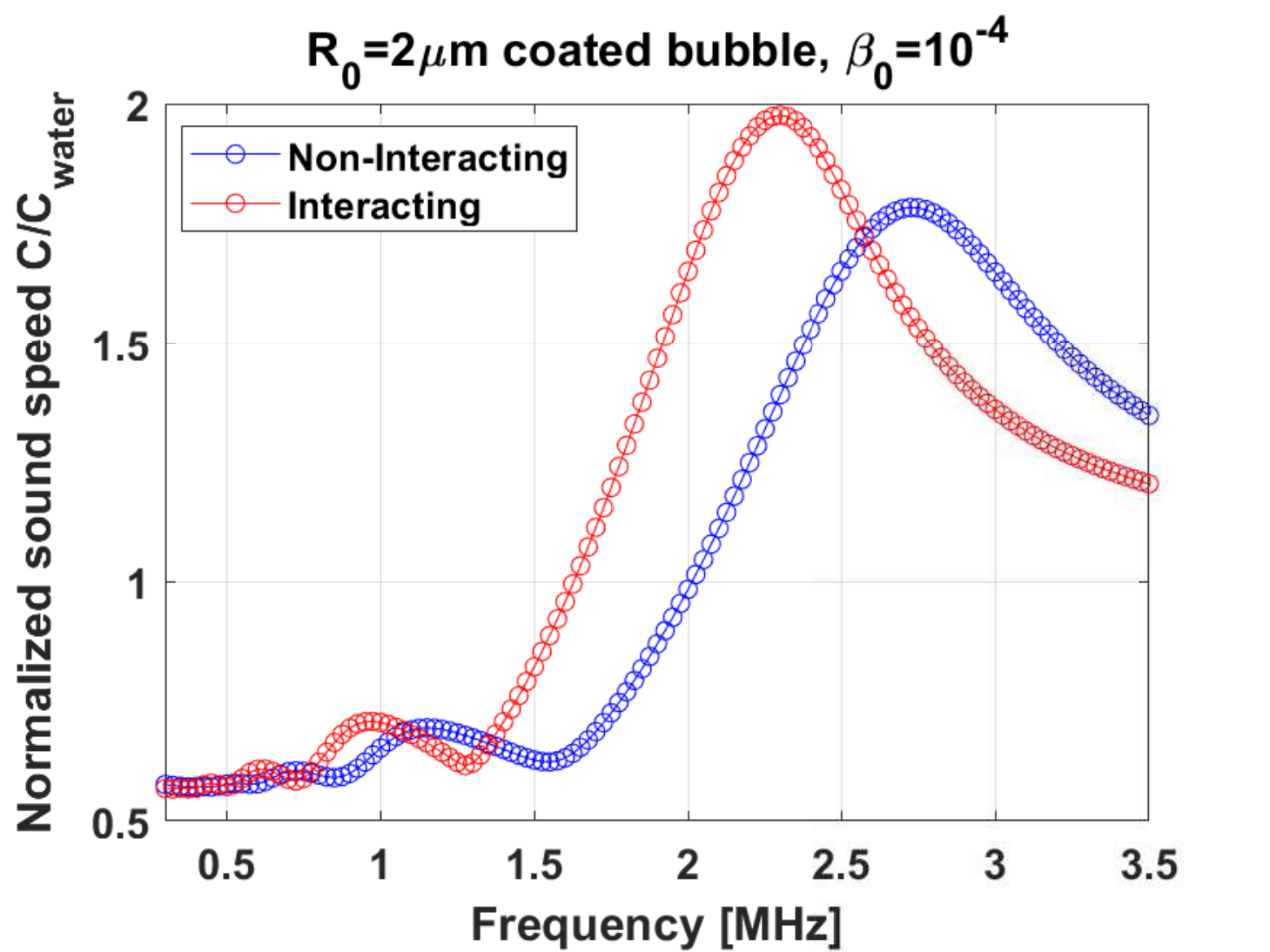} \\
	(e) \hspace{8 cm} (f)
	\caption{Influence of bubble-bubble interaction on the pressure dependent sound speed and attenuation at 100kPa for a coated bubble with $R_0=2\mu m$ in Eq. A2: a-b) {}{ $\beta_0$}=10$^{-7}$, {}{c-d) $\beta_0=5.1\times10^{-6}$ and e-f) $ \beta_0=10^{-4}$.} In each case 20 bubbles are considered and randomly distributed in a cube. The side lengths of the cube were chosen to replicate the {}{$\beta_0$} in each case. The side length can be calculated as $d=(20\times4\pi R_0^3/3{}{\beta_0})^{1/3}$. {}{The minimum distance between neighboring MBs was chosen to be 10$\mu$m.}}
\end{figure*}
\subsection{Importance of the accurate calculation of $\left\langle{}{\Re}\left(k^2\right)\right\rangle{}$ in pressure dependent attenuation and sound speed estimation}
\label{subsection:D3}
Fig. \ref{fig:8} compares the attenuation and sound speed that are calculated using the nonlinear model and the Louisnard model. The values are calculated for the uncoated bubble in Figs. \ref{fig:8}a-b and at $P_a$=40 kPa, $P_a$= 100 kPa and $P_a$= 150 kPa. At 40 kPa (Figs. \ref{fig:8}a-b), the Louisnard model fails to capture the sound speed fluctuation around ${f}/{f_r}\approxeq$ 0.5 due to the occurrence of 2nd order superharmonic (SuH) regime. Moreover, the Louisnard model over-estimates the attenuation at the resonance frequency by about 10 $\%$. The deviation in the predicted values between the two models increases with increasing pressure. At $P_a$= 100 kPa (Figs. \ref{fig:8}c-d), Louisnard model overestimates the attenuation by about 40 $\%$. Moreover, Louisnard model can not capture the the shift in the maximum sound speed to lower frequencies as well as the $\approx$ 15$\%$ increase in its magnitude. At 150 kPa (Figs. \ref{fig:8}e-f) the Louisnard model overestimates the attenuation peak by 77 $\%$ and underestimates the sound speed peak by about 52 $\%$. The nonlinear model predicts a shift in the frequency of the sound speed peak by about 42 $\%$. Once again, the frequency at which the attenuation peaks (${f}/{f_r}$=0.65) corresponds to the frequency at which ${C}/{C_l}$=1.\\ This, shows that pressure dependent effects of $\left\langle{}{\Re}\left(k^2\right)\right\rangle{}-({\omega}/{C_l})^2$  can not be neglected and must be included in the calculation of sound speed and attenuation. The proposed nonlinear model has the advantage of calculating both of the pressure dependent  $\left\langle{}{\Re}\left(k^2\right)\right\rangle{}$  and  $\left\langle{}{\Im}\left(k^2\right)\right\rangle{}$.   \\
As the pressure increases, the resonance frequency of the bubbles decreases \cite{14}, which is observed as the peak of $\Im{(k^2)}$ in Fig. \ref{fig:7} and attenuation curve in Fig. \ref{fig:8} shift towards lower frequencies; this corresponds to the frequencies at which the sound speed in the bubbly medium is equal to the sound speed in the absence of the bubbles. This is seen in Fig. \ref{fig:8} where the frequency in which attenuation peaks corresponds to the frequency in which ${C}/{C_l}$=1 in the blue curves that can only be captured by the nonlinear proposed model. At pressure dependent resonances, the oscillations are in phase with the driving acoustic pressure similar to the case of linear resonance(when $f=f_r$ and at $P_a\approxeq <$ 1 kPa ${C}/{C_l}$=1 page 290 \cite{56}).  As the pressure increases, the maximum sound speed of the bubbly medium increases and occurs at a lower frequency, which depends on the driving acoustic pressure amplitude. The abrupt increases in the sound speed and attenuation at particular frequencies in Figs. \ref{fig:8}c-d and in $\left\langle{}{\Im}\left(k^2\right)\right\rangle{}$ in Figs. \ref{fig:7}a, \ref{fig:7}c and \ref{fig:7}d are due to the pressure dependent resonance frequency which is described in detail in \cite{14}. We have previously shown that when MBs are sonicated with their pressure dependent resonance frequency, the radial oscillation amplitude of the MBs undergo a saddle node bifurcation (rapid increase in amplitude) as soon as the pressure increases above a threshold \cite{14} and the maximum stable scattered pressure increases considerably.
\begin{figure*}
	\begin{center}
		\includegraphics[scale=0.55]{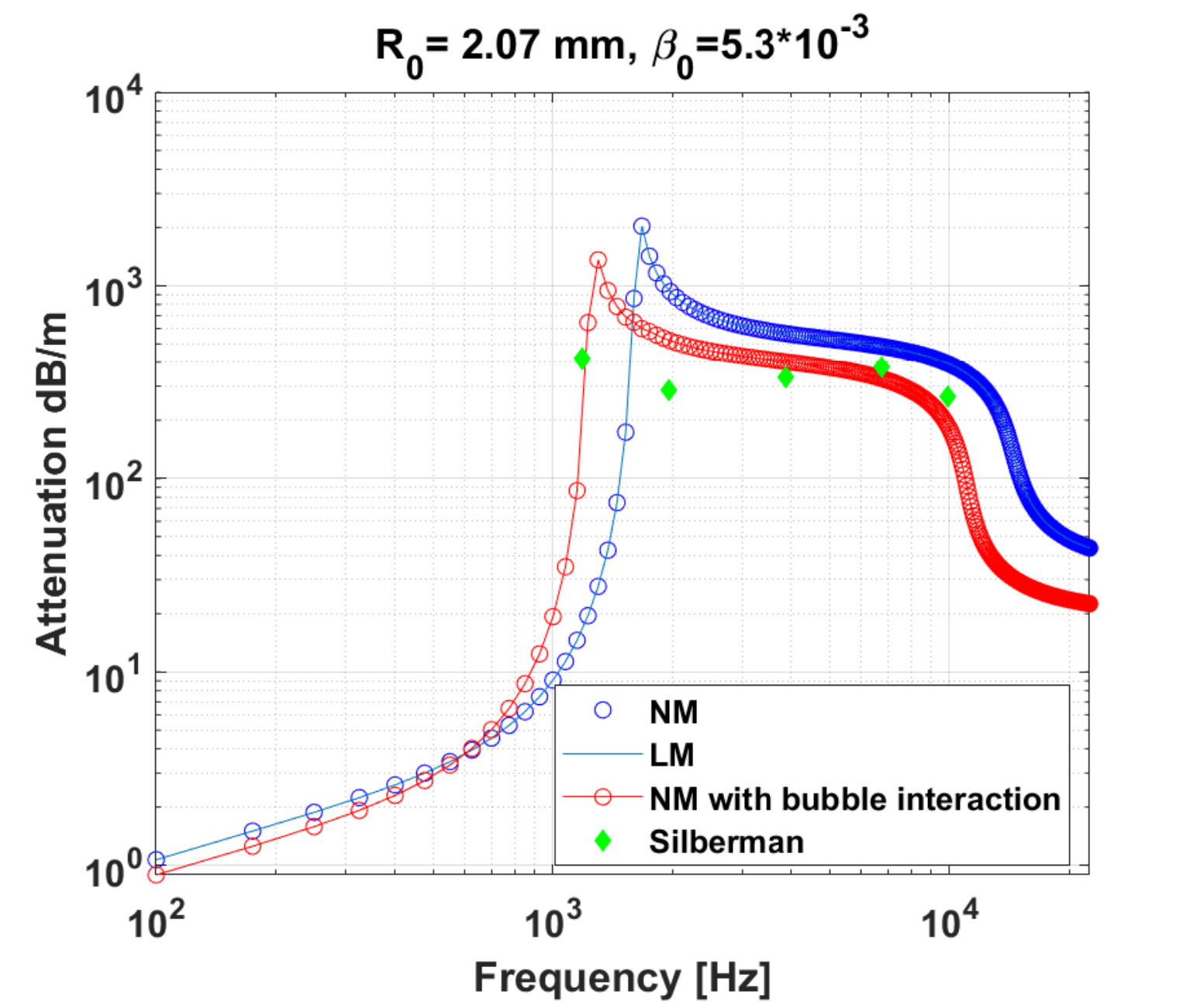}  \includegraphics[scale=0.55]{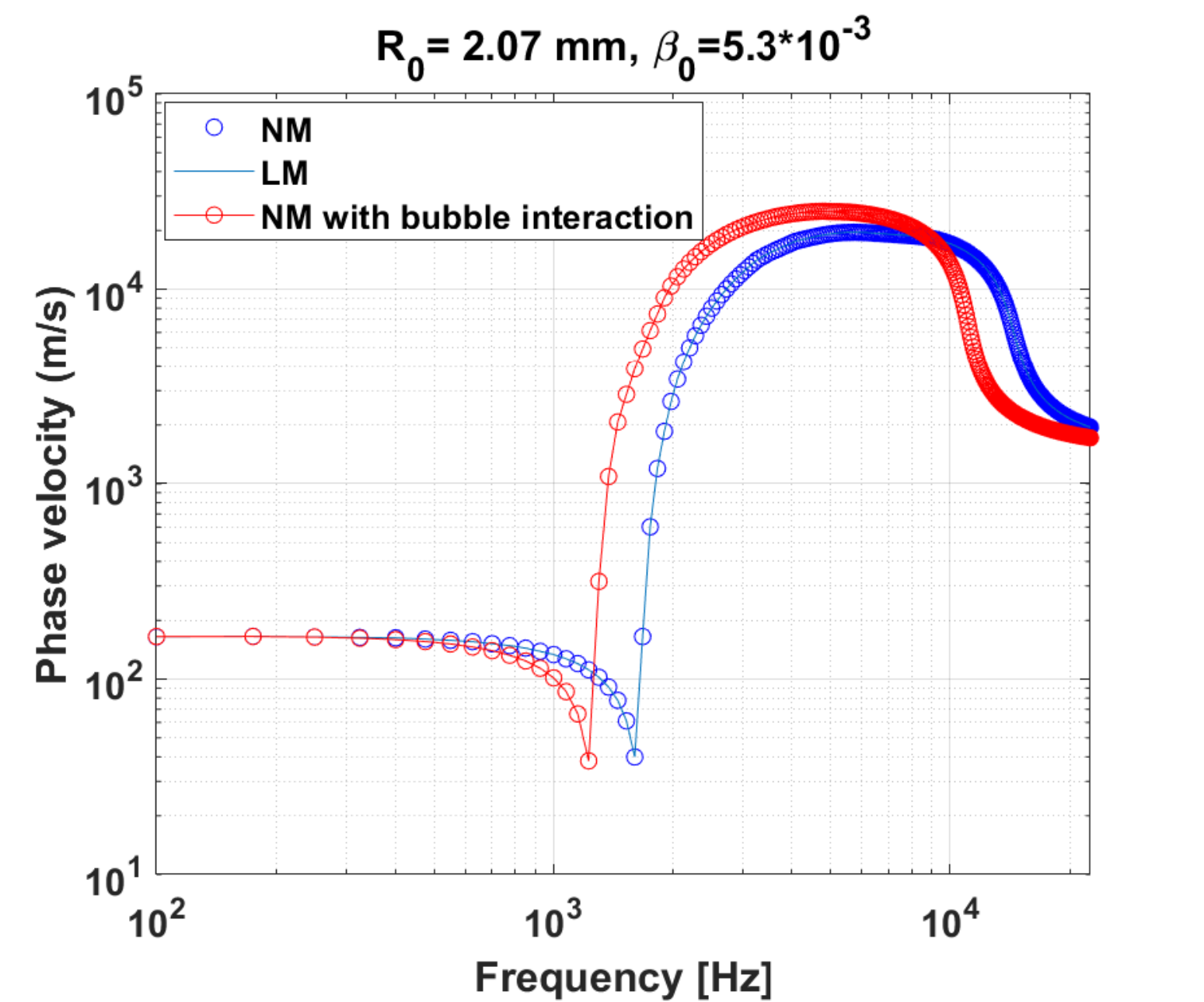}\\
		(a) \hspace{8 cm} (b)\\
	\end{center}
	\caption{Case of a bubbly medium with {}{$\beta_0=5.3\times10^{-3}$} and $R_0$=2.07 mm \cite{silberman} (a- attenuation and b-sound speed curves). Blue circles are constructed by solving the nonlinear model (NM) without bubble-bubble interaction. Blue solid line is constructed by the linear Commander and Prosperetti model \cite{17}. The red line-circle is constructed by solving the nonlinear model and incorporating bubble-bubble interaction. Green diamonds are experimentally measured values by Silberman \cite{silberman}. For the simulations, similar to \cite{trujio} pressure amplitude of 10Pa is used.}
\end{figure*}
\section{{Attenuation and sound speed changes at higher void fractions}}
{Some applications of MBs (e.g. pre-clinical drug delivery applications) employ high concentration of MBs. At higher void fractions, bubble-bubble interactions become significant. The resonance frequency and maximum oscillation amplitude has been shown to decrease with MB-MB interaction \cite{guedra,55,Qin1,Qin2}. In order to shed light on the pressure dependent changes of the attenuation and sound speed at higher void fractions we have performed numerical simulations on a monodisperse population of MBs  with and without considering MB-MB interactions.  Figure 9 shows the results of the numerical simulations for a case of MBs with $R_0$=2$\mu$m at 100kPa of driving acoustic pressure {}{at void fractions of 10$^{-7}$ (very dilute suspension), 5.1$\times10^{-6}$ (similar to the void fractions in experiments in Figs. 4-5) and 10$^{-4}$ (highly concentrated suspension)}. Radial oscillations in the absence of interaction at each frequency were computed using Eq. A2. In the presence of  MB-MB interactions radial oscillations were calculated by considering the pressure radiated by each MB in the location of other MBs by adding the term  $-\rho\sum_{j\neq i}^{20}{R_j}/{d_{ij}}(R_j\ddot{R_j}+2\dot{R_j}^2)$ to the right side of Eq. A2. An approach similar to Eq. \ref{eq:14} and \cite{55} is used for the solution of the large number of the coupled ordinary differential equations. In case of interacting bubbles, for simplicity we considered a case of randomly distributed 20 MBs in a cube with a side length of $d$. The side length can simply be calculated from the void fraction ($d=(20\times4\pi R_0^3/3{}{\beta_0})^{1/3}$)). For a sinusoidal acoustic excitation, the attenuation and sound speed can be calculated using Eqs. \ref{eq:E1} and \ref{eq:E2} as follows:
	 	  \begin{equation}
	  \langle\Re(k^2)\rangle=\frac{{\omega{}}^2}{C_l^2}+\frac{2{\rho_l{}}}{T{P_a}}\frac{N}{20}\sum_{i=1}^{20}\int_0^T{\sin(2\pi ft)}\frac{{\partial{}}^2{\beta{}}_i}{\partial{}t^2}dt
	  \label{eq:E1} 
	  \end{equation}
	  \begin{equation}
	  \langle\Im(k^2)\rangle =-\frac{2{\rho_l{}}}{T{P_a}}\frac{N}{20}\sum_{i=1}^{20}\int_0^T {\cos(2\pi ft)}\frac{{\partial{}}^2{\beta{}}_i}{\partial{}t^2}dt
	  \label{eq:E2} 
	  \end{equation}
	  where $N$ can be calculated from $N$=$\beta$/(4/3$\pi$$R_0^3$).\\
	At the lower void fraction of {}{$\beta_0$}=1$\times$10$^{-7}$ (Fig. 9a-b), MB-MB interactions are negligible. As concentration increases, the attenuation and sound speed peak do not linearly scale. {}{At the void fraction of {}{$\beta_0$}=5.1$\times$10$^{-6}$, MB-MB interactions lead to a $\approx$5.6$\%$ and $\approx$8.5$\%$ decrease in the peak attenuation and the frequency of the peak respectively (Fig. 9c). Sound speed peak increases by 3$\%$, while the frequency of the sound speed peak decreased by 9.3$\%$ (Fig. 9d). When in a solution, each MB receives a sum of the pressures of the multiple neighboring bubbles, plus the incident acoustic pressure field. Since the scattered pressure by each MB is not negligible, the interaction effects can not be neglected. Moreover, these effects are stronger near the MB resonance frequency. The effective pressure amplitude (pressure of the sound filed plus the pressure radiated by the other bubbles) at the location of one of the bubbles in Fig. 9c-d was calculated.  Compared to the case in the absence of interaction, the maximum pressure amplitude felt by the bubble increased by 14$\%$ (114kPa) suggesting a strong MB-MB interaction which can not be neglected.}
 At the higher void fraction of {}{$\beta_0$}=1$\times$10$^{-4}$, MB-MB interactions lead to a 13.6$\%$ and $\approx$16.2$\%$ decrease in the peak attenuation and the frequency of the peak respectively (Fig. 9e). Sound speed peak increases by 10.6$\%$, while the frequency of the sound speed peak decreases by 16.3$\%$ (Fig. 9f). The results of the numerical simulations emphasize the influence of the interactions within MBs. Analysis of the experimental results at large void fractions must include these effects.}\\
{Bubble-bubble interactions are even important in the linear regime of the oscillations. This is emphasized in Commander and Prosperetti \cite{17} as one of the reasons behind the disagreement between the linear model and the experiments. For a large enough void fraction, the distance between bubbles decreases such that the average pressure field exciting a bubble is smaller than, or comparable with, the pressure wave scattered by a neighboring bubble, and thus the linear model fails \cite{17}. Since bubble-bubble interactions were neglected, further analysis by Trujillo \cite{trujio,trujio2}, also did not achieve good agreement with experiential results near resonance (attenuation was over-estimated by at least an order of magnitude).\\  Here, we briefly investigate this important effect, and extend the experimental verification of our nonlinear model to higher void fractions (in case of bigger uncoated bubbles). We considered the experimental results by Silberman \cite{silberman} (green diamonds) in Figure 10 for air bubbles with $R_0$=2.07mm and {}{$\beta_0=5.3\times10^{-3}$}. Similar to Trujillo \cite{trujio} we used an excitation pressure of 10Pa to solve equation A1 at each given frequency in Fig. 10. Equation A1, was solved with and without bubble interactions effects. In case of interaction, we considered 20 randomly distributed bubbles in a cube with a side length $d$. {}{The minimum distance between bubbles was set to be 10mm}. Attenuation and sound speed were then calculated from Eq. \ref{eq:E1} and \ref{eq:E2}. Figure 10 shows that the linear model and the nonlinear model in the absence of bubble-bubble interactions overestimate the near resonance attenuation by about and order of magnitude (blue line in Fig. 10a). However, when bubble-bubble interactions are considered, there is a significant improvement in the the prediction of the attenuation curve (red line in Fig. 10a). Similar to the case of the coated MB in Fig. 9d, bubble-bubble interaction increases the sound speed peak and reduces the frequency of the peak (Fig. 10b).\\
Comparing Fig. 9 and Fig. 10 we see that for comparable void fractions attenuation of MBs are orders of magnitude higher than mm sized bubbles. Attenuation of a population of MBs with ${}{\beta_0}=1\times10^{-4}$ is about 800 dB/cm (Fig. 9c) while this is only about 10dB/cm for bubbles with $R_0=2.07$mm at ${}{\beta_0}=5.4\times10^{-3}$ (Fig. 10a). For the sound speed changes, one can see the exact opposite relationship (Fig. 9d and Fig. 10b).}   
\section{{}{Attenuation and sound speed changes at higher pressures}}
	\begin{figure*}
	\includegraphics[scale=0.55]{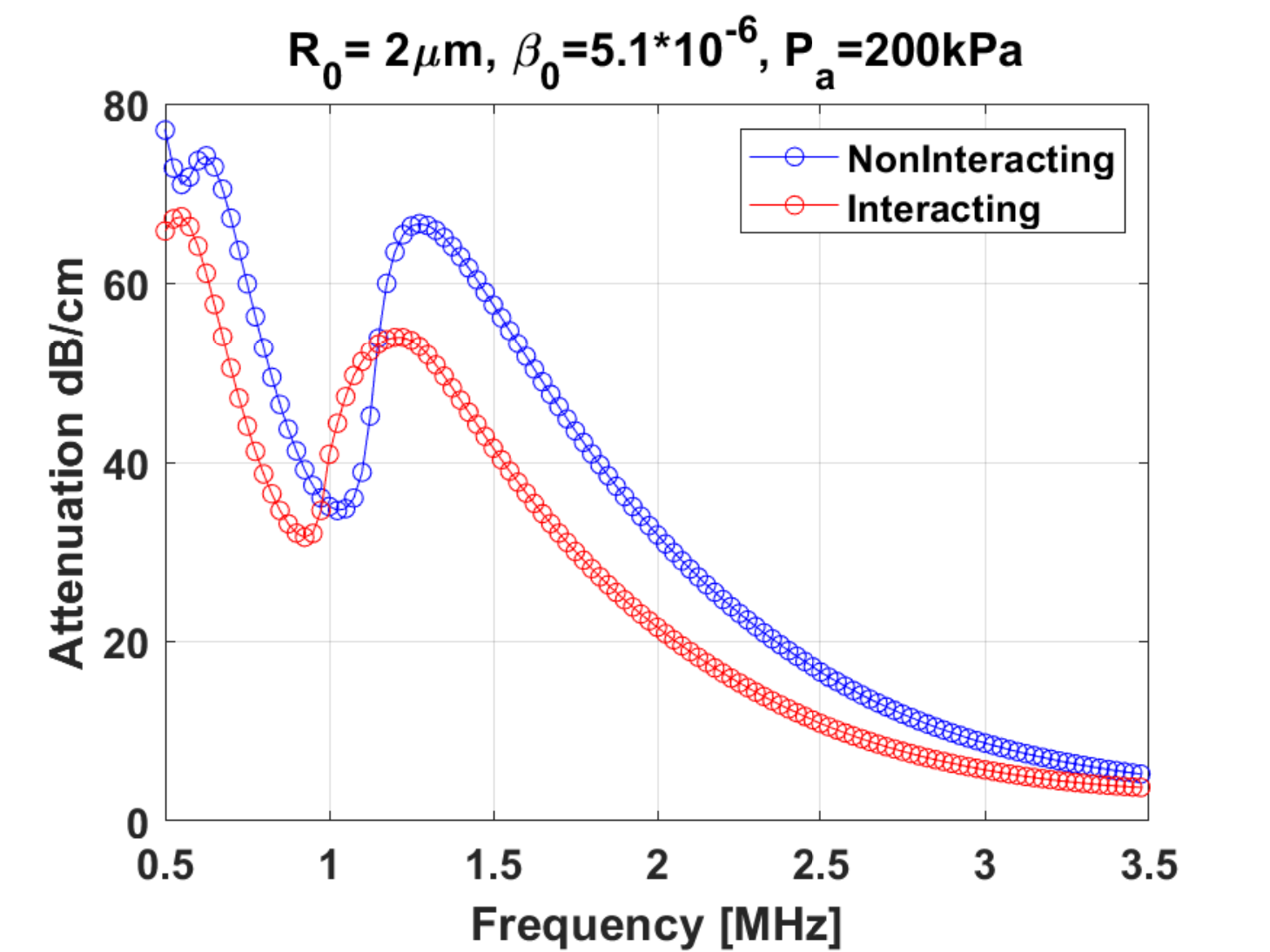} \includegraphics[scale=0.55]{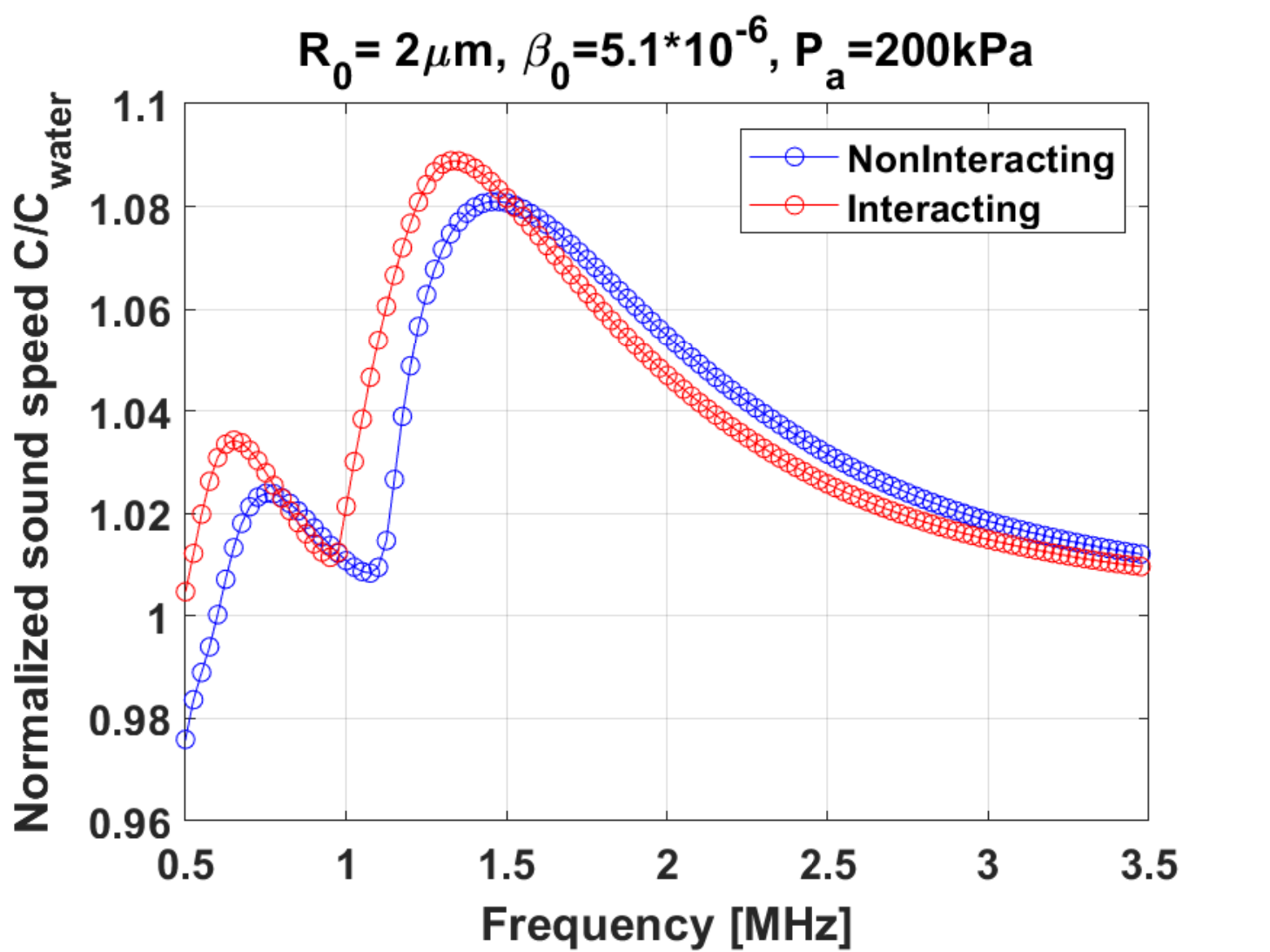} \\
	(a) \hspace{8 cm} (b)\\
	\includegraphics[scale=0.55]{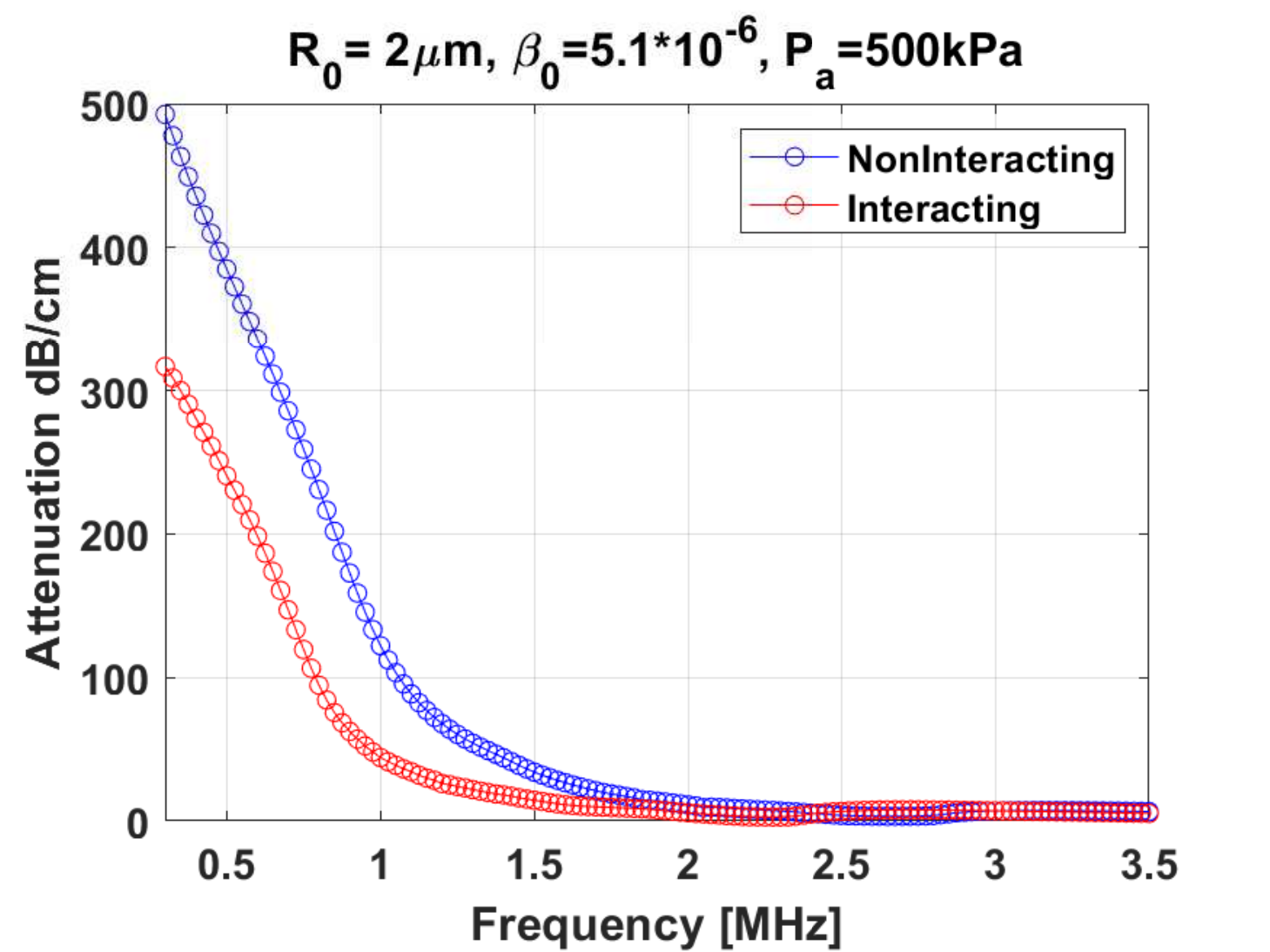} \includegraphics[scale=0.55]{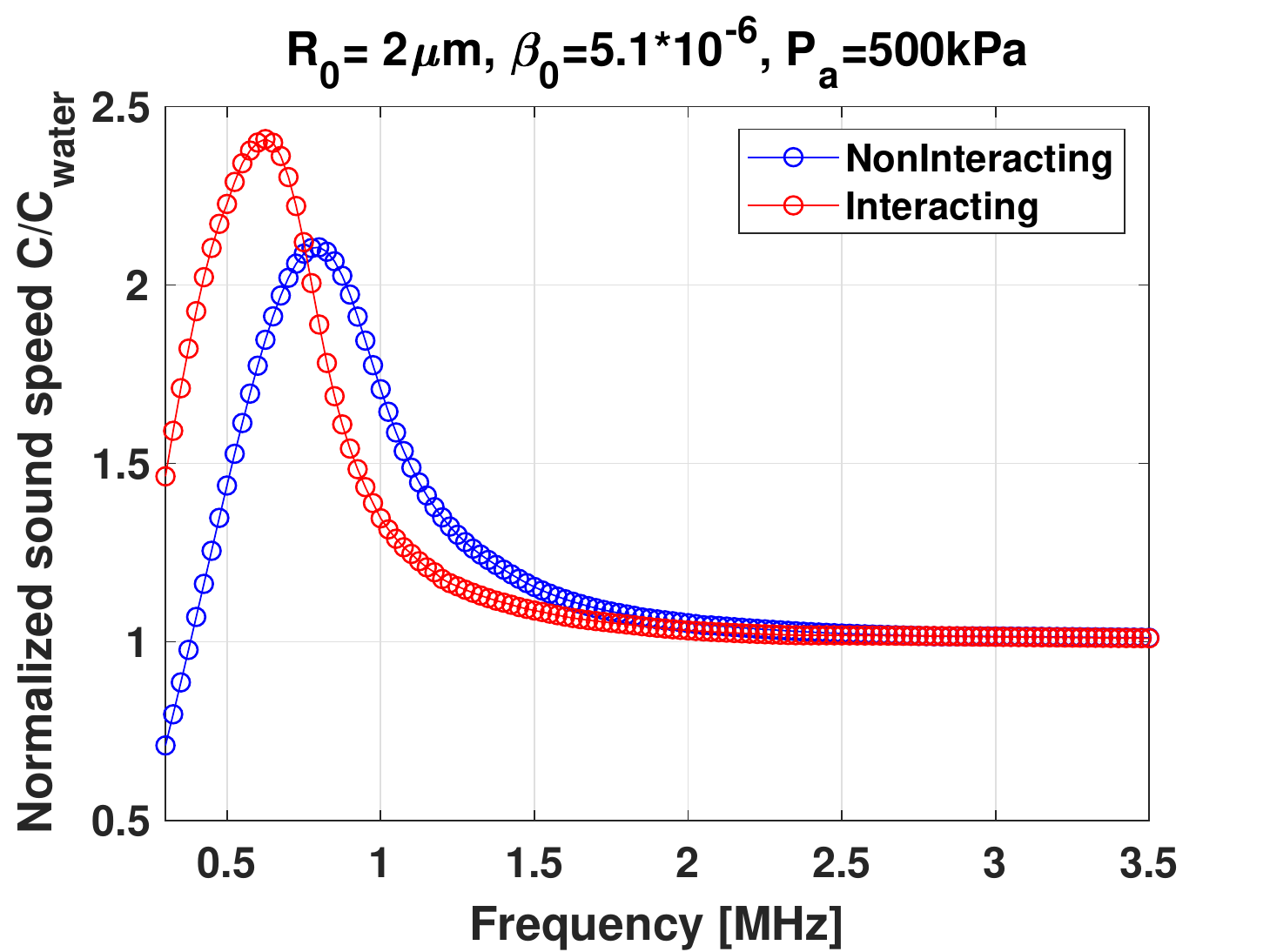} \\
	(c) \hspace{8 cm} (d)\\
	\includegraphics[scale=0.55]{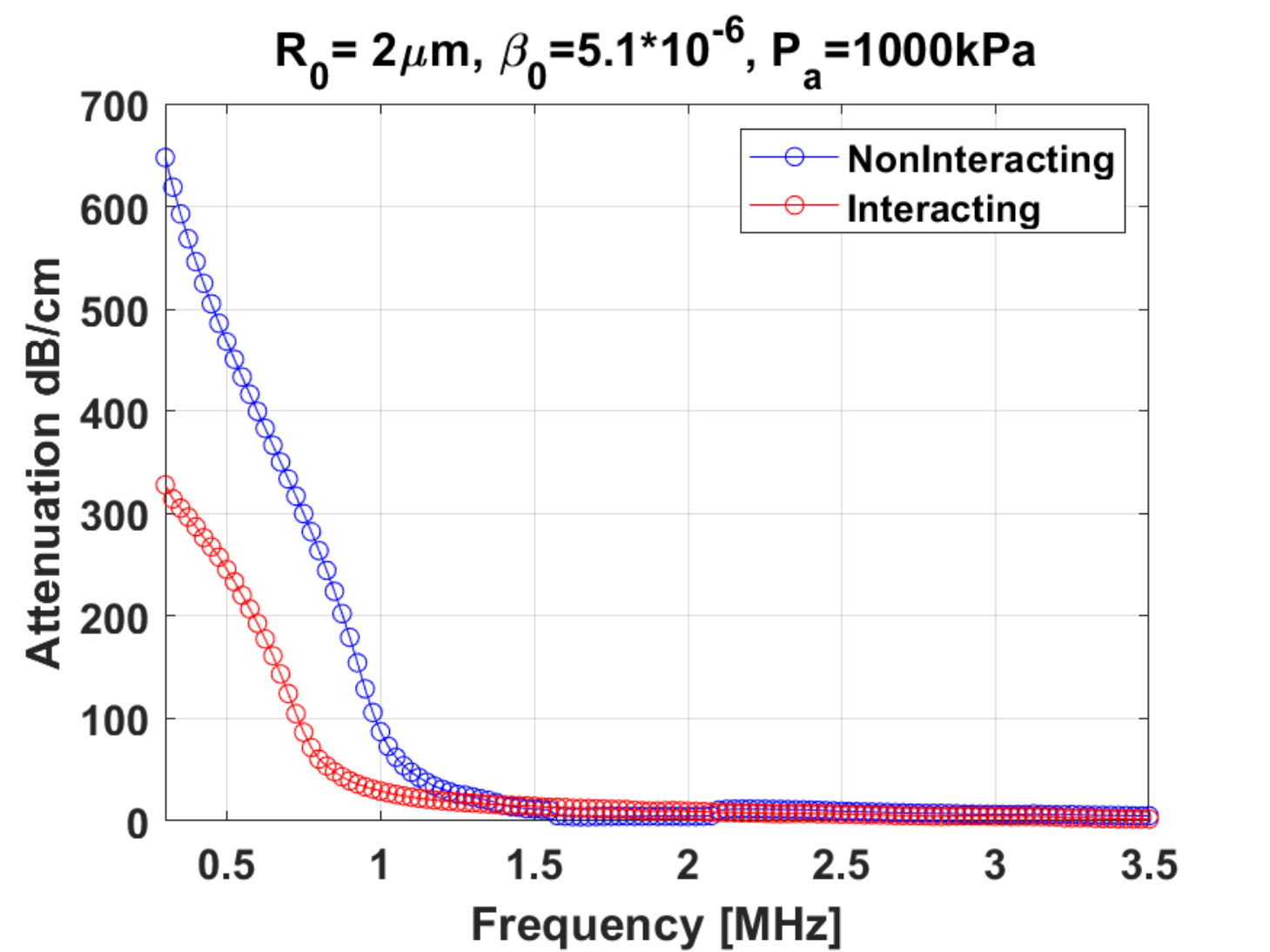} \includegraphics[scale=0.55]{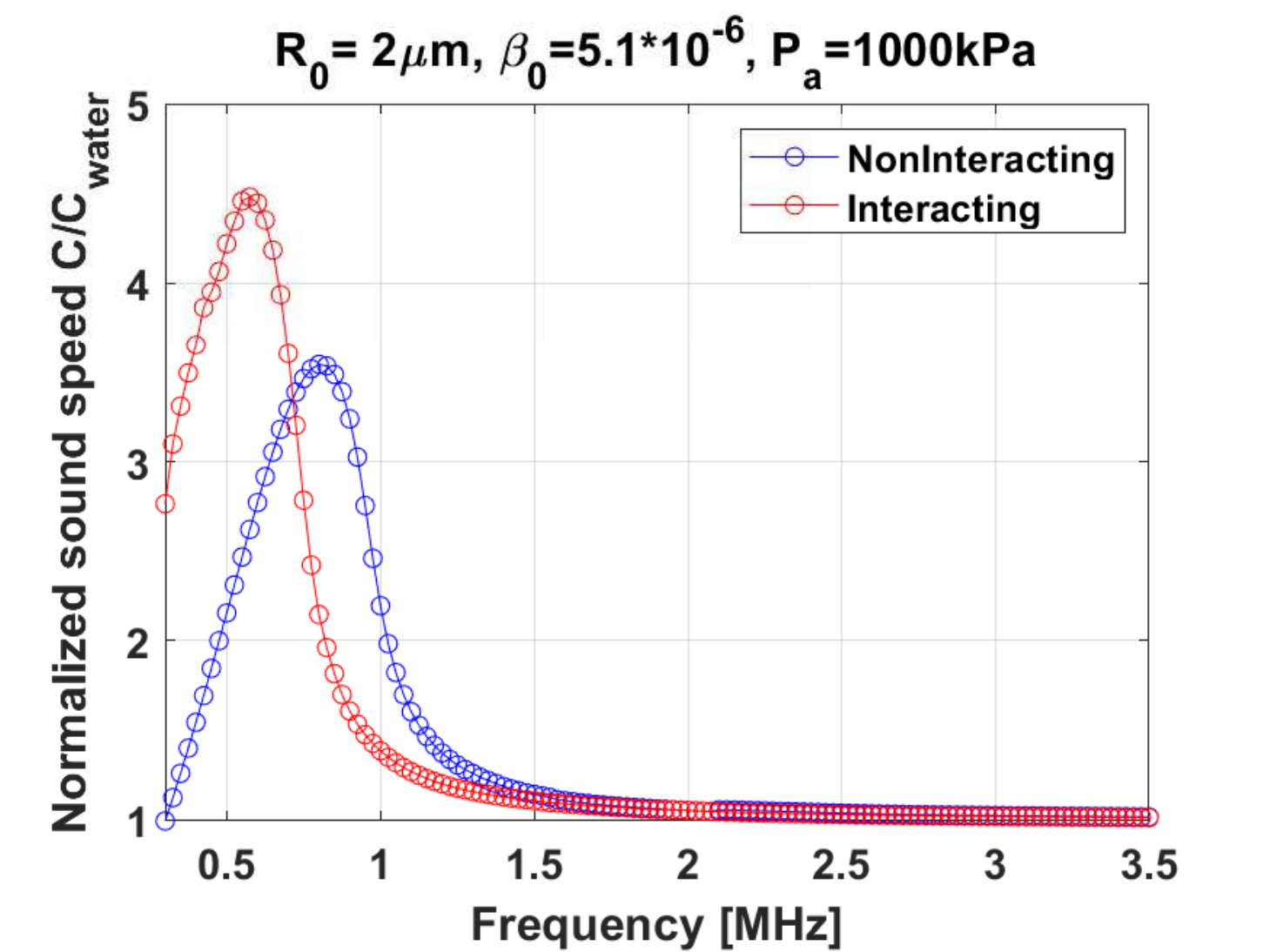} \\
	(e) \hspace{8 cm} (f)
	\caption{Influence of increasing the pressure amplitude on the sound speed and attenuation at $\beta_0=5.1\times10^{-6}$ for a coated bubble with $R_0=2\mu m$ in Eq. A2 when $P_a$ is: a-b) 200kPa  c-d)500kPa  and e-f)1000kPa . In each case 20 bubbles are considered and randomly distributed in a cube. The dimensions of the cube were chosen to replicate the {$\beta_0$} in each case. The dimension can be calculated as $d=(20\times4\pi R_0^3/3{\beta_0})^{1/3}$. The minimum distance between neighboring MBs was chosen to be 50$\mu$m to eliminate the possibility of MBs collisions at higher pressures .}
\end{figure*}
\vspace{-5pt}
{}{Many applications employ pressures above the Blake threshold of the MBs. To investigate the frequency dependent attenuation at higher pressures, in this section we considered a coated MB with $R_0$=2$\mu$m with $\beta_0$=5.1$\times$10$^{-6}$. It is assumed that MB integrity is maintained for all exposures. Figure 11a-b shows the attenuation and sound speed when $P_a$ is 200kPa. The fundamental frequency of the attenuation peak further decreases (30$\%$ compared to when $P_a$=100kPa in Fig. 9c) when bubble-bubble interaction is considered. The attenuation of the 2nd order superharmonic (SuH) resonance frequency exceeds that of the main resonance ($\approx$27$\%$) in Fig. 11a. In the vicinity of the 2nd order SuH frequency (550kHz), the sound speed peak which was below the medium sound speed at 100kPa (0.95$C_{water}$) becomes larger ($\approx$1.04$C_{water}$). Bubble-bubble interactions reduce the attenuation maxima and frequencies of the attenuation peaks while increasing the maximum sound speed (Figs.11a-f).\\ The attenuation peak increase with pressure in the studied frequency range (0.5-3.5MHz) and the influence of bubble-bubble interactions becomes stronger with increasing pressure. The maximum sound speed increases with increasing pressure and at 500kPa(625kHz) and 1MPa(575kHz) becomes approximately 2.5 and 4.5 times the $C_{water}$. It is interesting to note that at a $P_a$ of 100kPa, and at the same frequencies, the sound speed was below that of $C_{water}$.  The results indicate strong nonlinear changes in the attenuation and sound speed even at the simulated low void fraction. We will classify these changes during major non-linear regimes of the oscillations (e.g. \cite{pof1}) in future studies.}         
\section{{}{Importance of pressure dependent measurements in shell characterization of \textit{lipid coated MBs} undergoing buckling and rupture}}
	\begin{figure*}
	\includegraphics[scale=0.5]{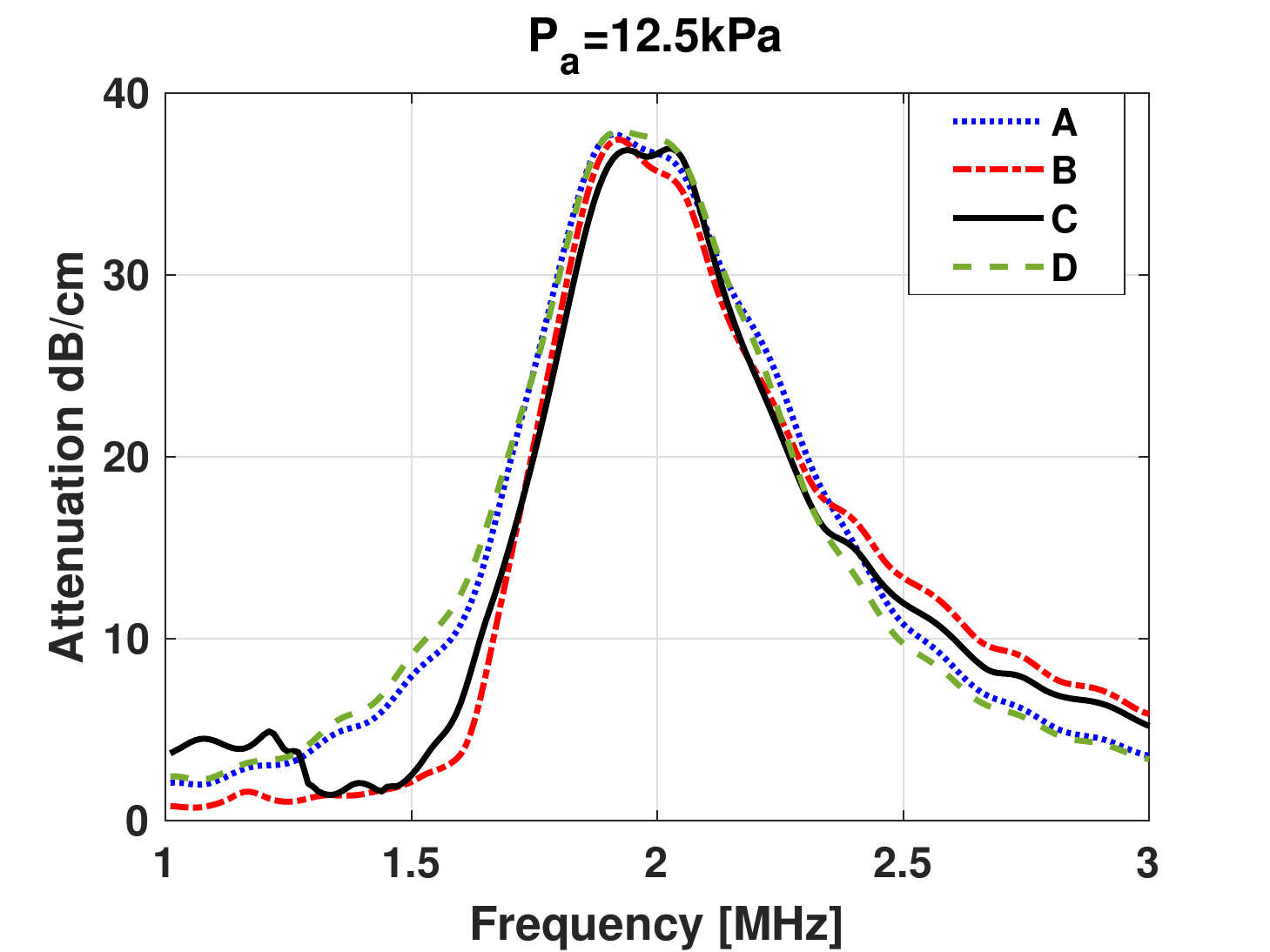} \includegraphics[scale=0.5]{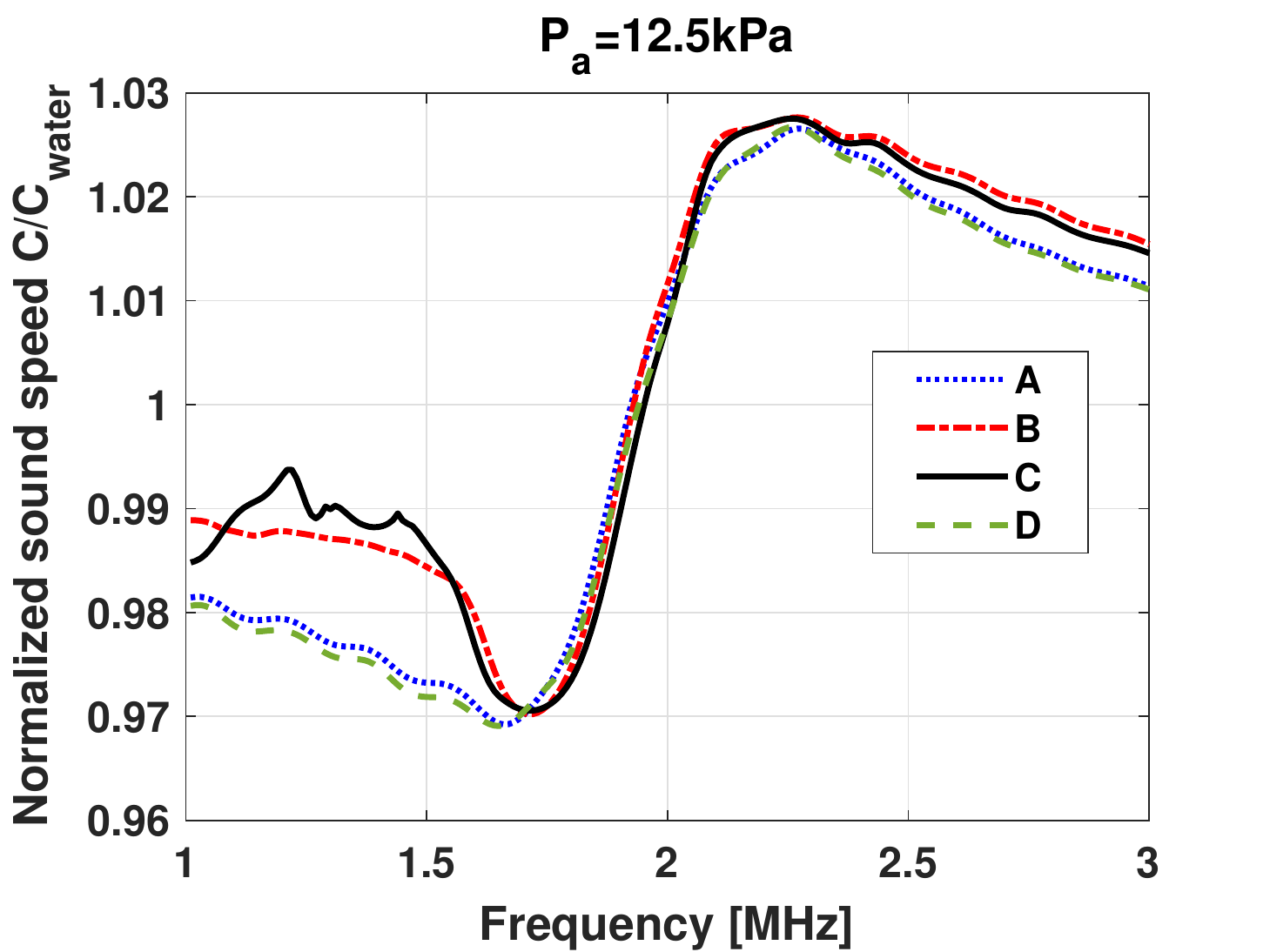} \\
	(a) \hspace{8 cm} (b)\\
	\includegraphics[scale=0.5]{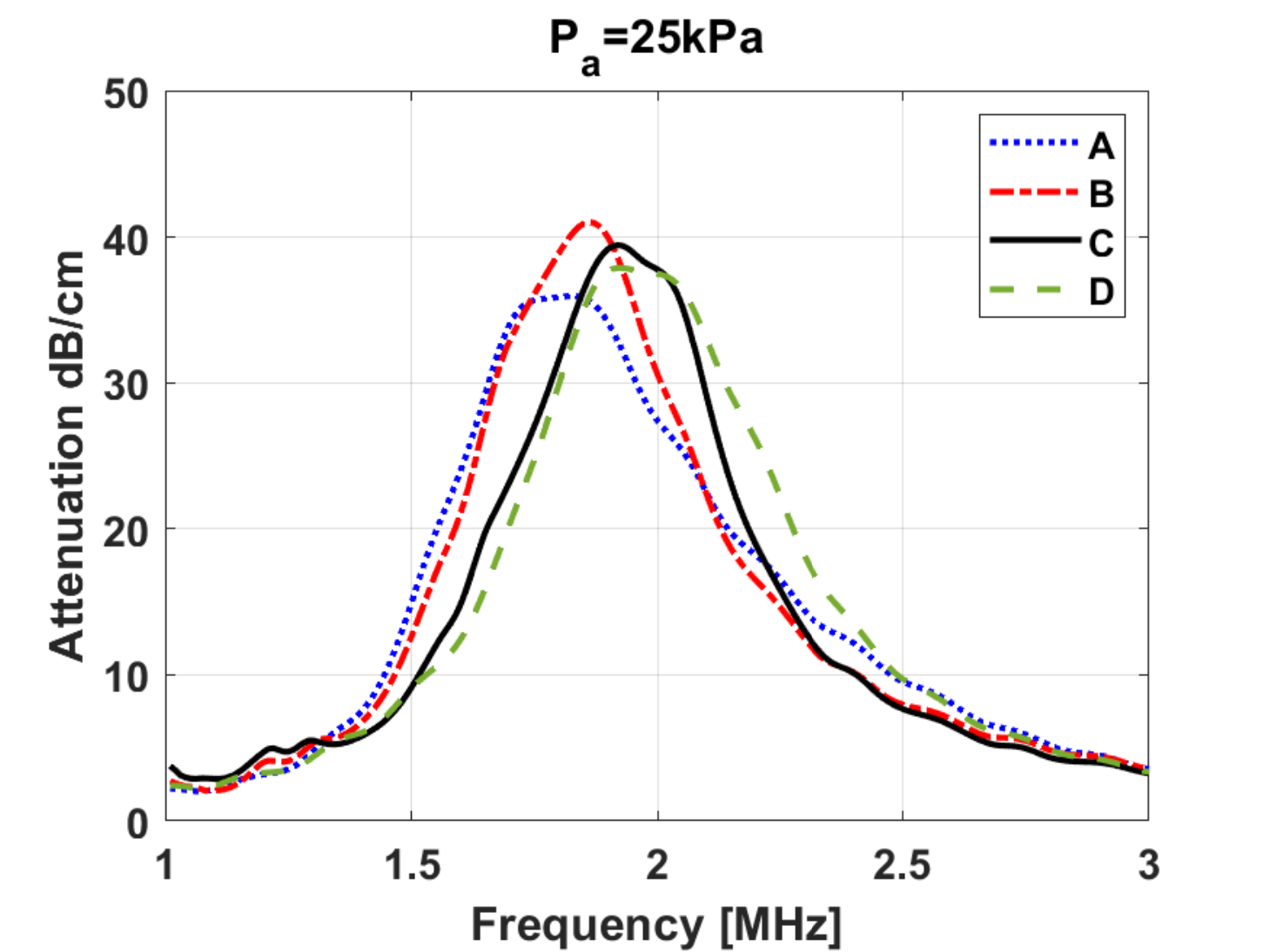} \includegraphics[scale=0.5]{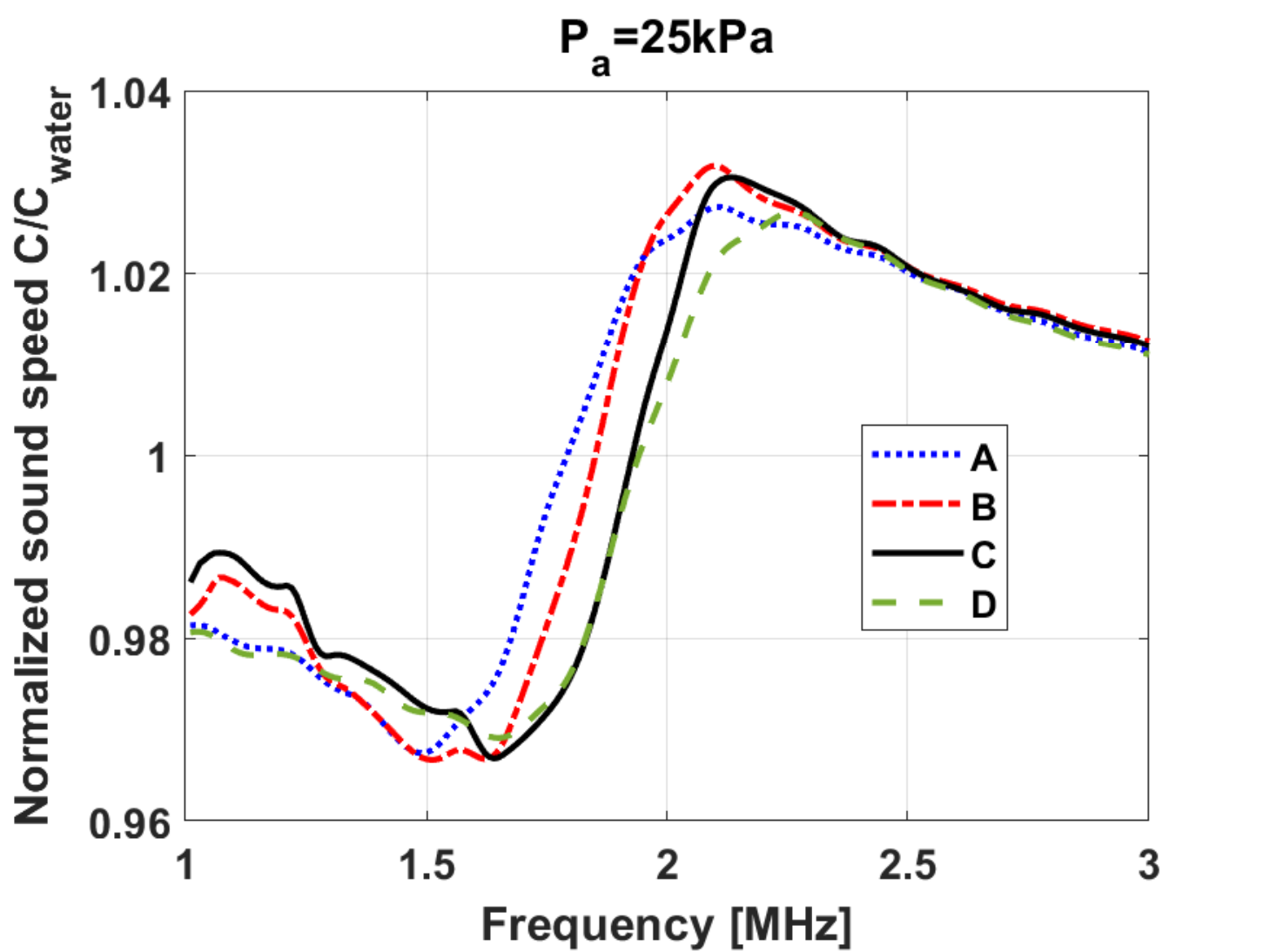} \\
	(c) \hspace{8 cm} (d)\\
	\includegraphics[scale=0.5]{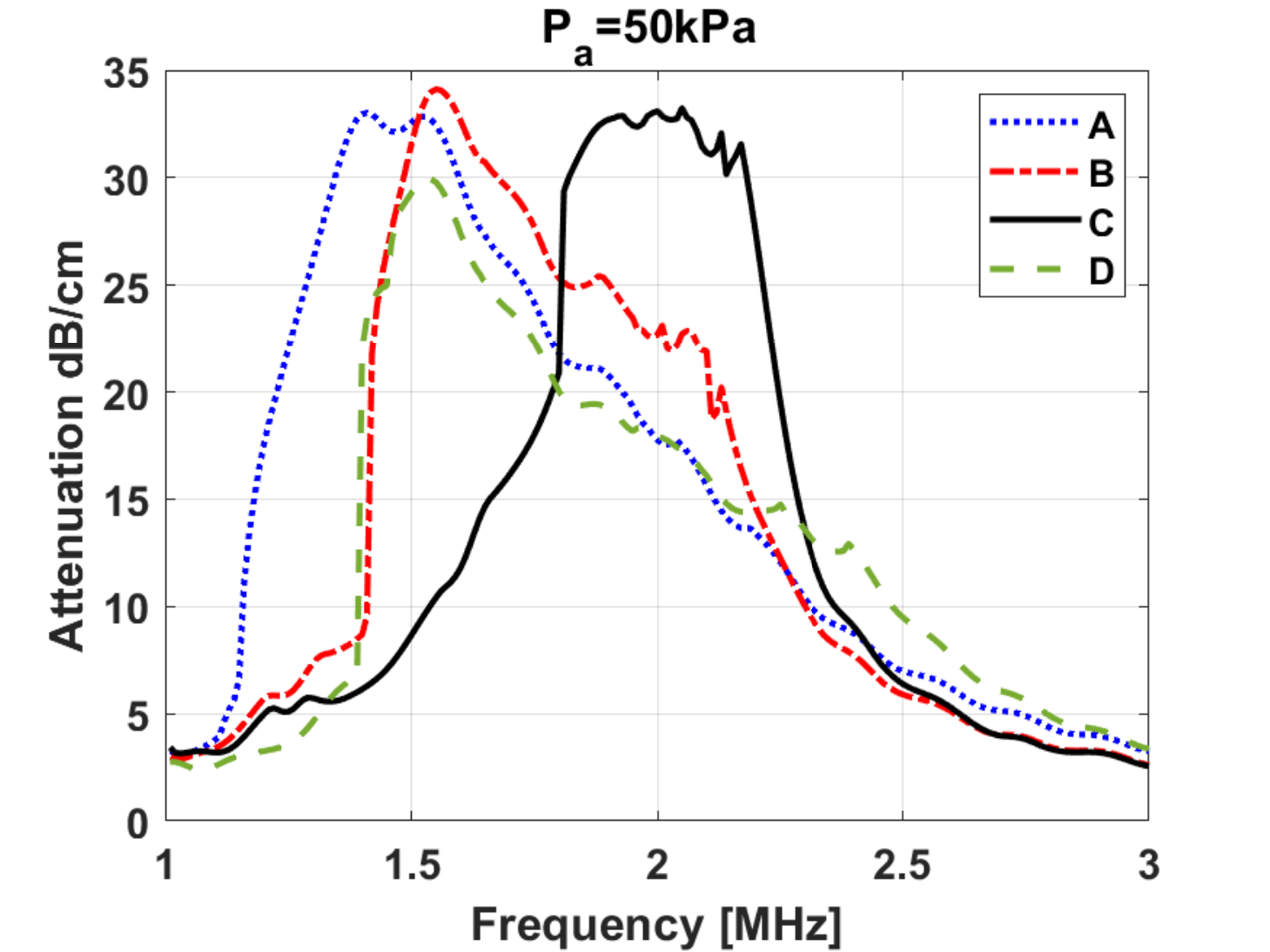} \includegraphics[scale=0.5]{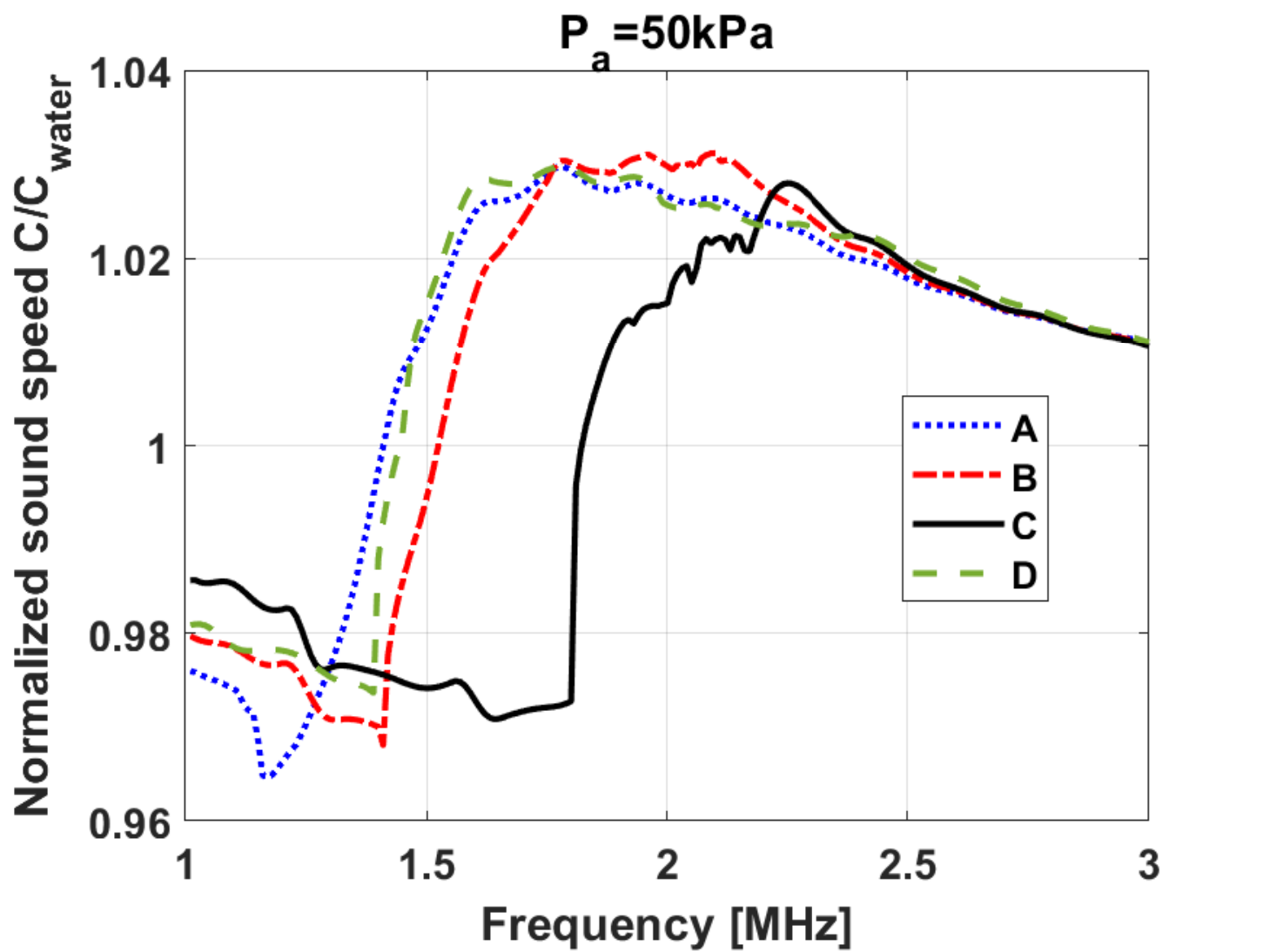} \\
	(e) \hspace{8 cm} (f)\\
		\includegraphics[scale=0.5]{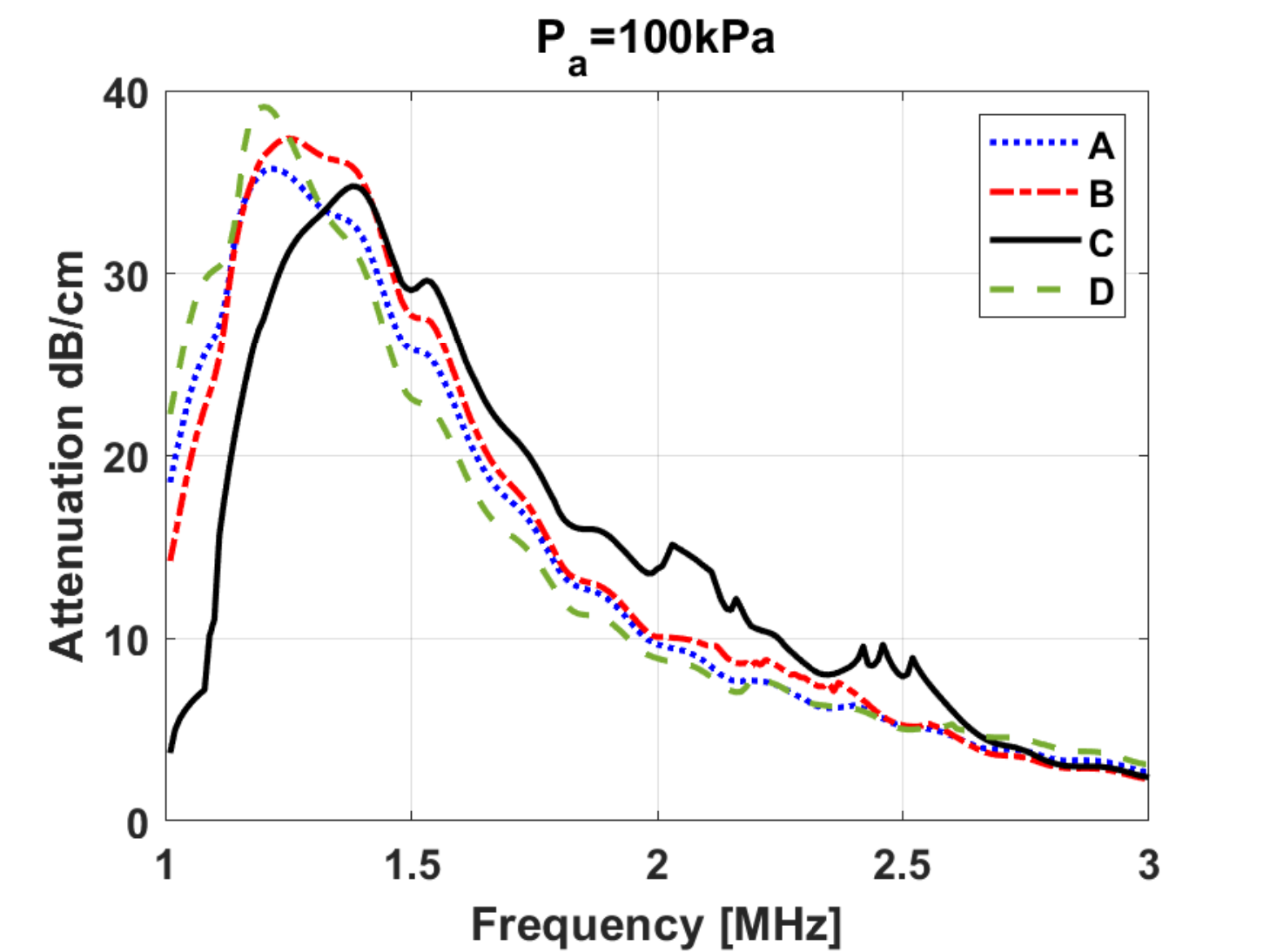} \includegraphics[scale=0.45]{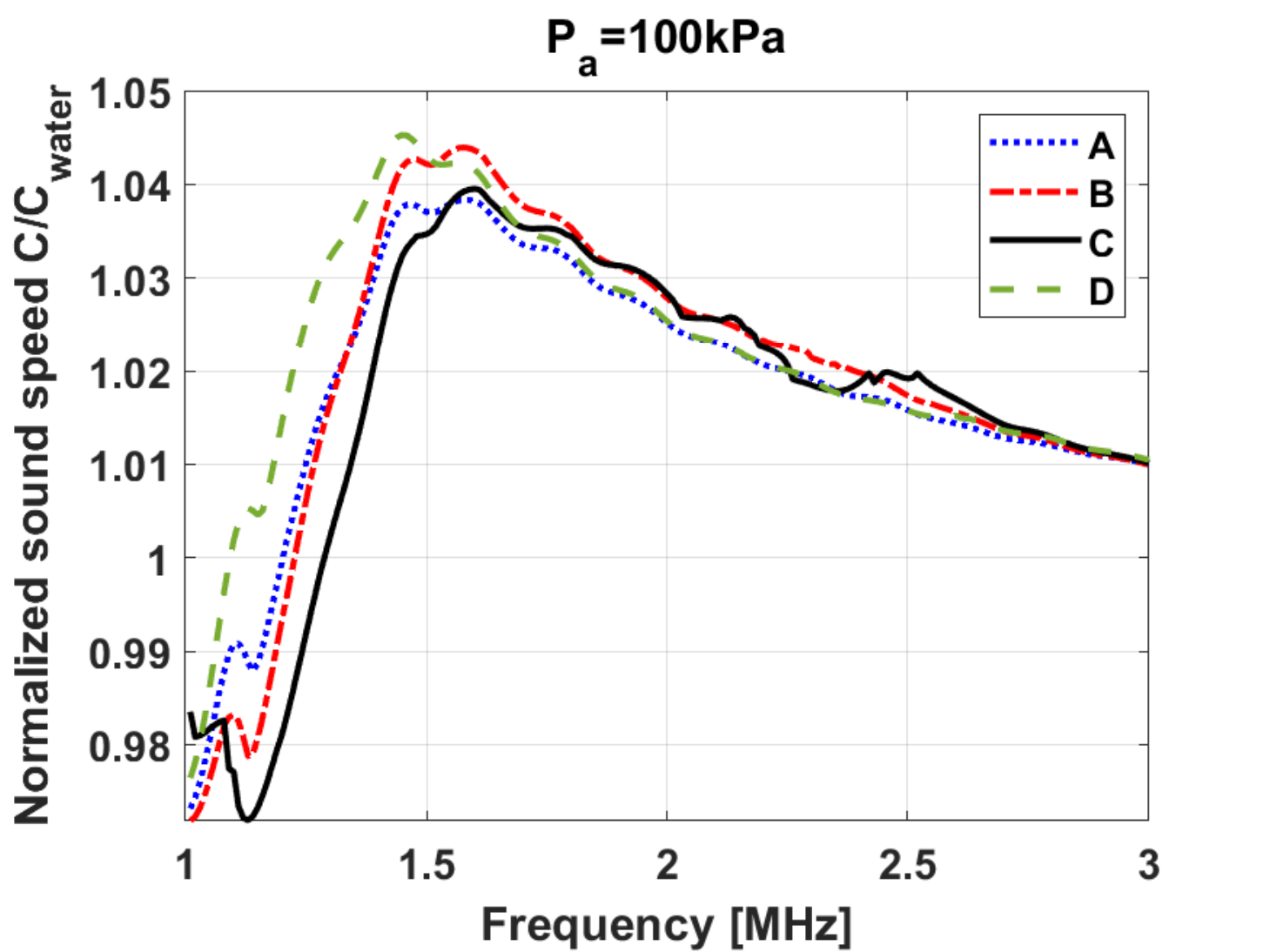} \\
	(g) \hspace{8 cm} (h)
	\caption{Influence of increasing the pressure amplitude on the sound speed and attenuation of a lipid coated MB with $R_0$=2.7$\mu$m, $\beta_0=5.1\times10^{-6}$ and different sets of shell parameters A,B,C and D: a-b)12.5kPa  c-d)25kPa  and e-f)50kPa, g-h)100kPa. For each group, the shell parameters are given in table \ref{table:2}. Radial oscillations are calculated using Eq. \ref{eq:12} and bubble-bubble interactions and transducer response are neglected for simplicity. The duration of the sonicating pulse is 3$\mu$s.}
\end{figure*}
{}{Here we demonstrate one of the applications of the introduced nonlinear model which is in the accurate characterization of the shell parameters of lipid coated MBs. The nonlinear behavior of the lipid coating including bucking and rupture \cite{16} intensifies the nonlinear changes of the resonance frequency.\\
The linear resonance frequency of the Marmottant model\cite{xia} is given by:
\begin{equation}
f_r=\frac{1}{2\pi R_0}\sqrt{\frac{1}{\rho}\left(3kP_0+(3k-1)\frac{2\sigma_0}{R_0}+\frac{4\chi}{R_0}\right)}
\label{eq:G1}
\end{equation}
This equation has limited sensitivity to the $\sigma_0$ value. For a MB with $R_0$=2.7$\mu$m and $\chi$ of 1N/m, eq. \ref{eq:G1} predicts only 80kHz changes in the resonance frequency for $\sigma_0$ between 0 and 0.072N/m. More importantly, the attenuation measurements are often performed using pressures of $\approx >$20kPa \cite{43,44,45,46,faez}. Thus, these MBs are already in their nonlinear regime, and since the resonance frequency of the lipid coated MBs shifts with increasing pressure (as small as 5kPa) thus, these methods may underestimate the shell elasticity. Moreover, using the linear model to fit the pressure dependent shell parameters \cite{17} casts doubts on the accuracy of the claims such as stiffness softening and shear thinning with increasing pressure \cite{17}.\\     
An accurate fit requires interrogation of the frequency dependent attenuation at multiple pressures in increasing steps such as what was done in this study or in \cite{26}. At the lower pressures (e.g. $\approx$$<$50kPa) and in the absence of the shell rupture, the changes in the resonance frequency and attenuation peak are majorly affected by the $\sigma_0$, $\chi$ and $k_s$ (Fig. 12). At pressures where the rupture occurs, the resonance frequency undergoes a sudden decrease.  The magnitude of the shift in the resonance frequency and the attenuation are largely affected by the $R_r$, $\chi$ and $k_s$. At lower pressures (e.g. 12.5kPa) several parameters of $\sigma_0$, $\chi$, $R_r$ and $k_s$ can provide a good fit to the attenuation and sound speed data (Fig. 12a-b). However, as the pressure increases (Figs. 12c-h), the predictions of each group diverge and only one group provides the best fit to the experimental curves at all excitation pressures. These behaviors can not be captured by the linear model.} 
{}{\begin{table*}
	\begin{tabular}{ |p{2cm}||p{3.5cm}|p{2cm}|p{2cm}|p{2cm}|}
		\hline
		\multicolumn{5}{|c|}{Shell properties} \\
		\hline
		Shell  & {$\chi$(N/m)} &{$\sigma_0$(N/m)} &{$k_s$ ($\times$10$^{-9}$kg/s)}& {$R_r$ ($\times$$R_0$)}\\
		\hline
		A    & 0.5 & 0.03&  10&1.1\\
		\hline
		B   & 1& 0.016&  7&1.1\\
		\hline
		C& 1.8 & 0.01&  6.5&1.1\\
		\hline
		D& 0.4 & 0.063&  10&1.1\\
		\hline
	\end{tabular}
	\caption{Shell properties of the lipid coated bubble in Fig. 12}
	.
	\label{table:2}
\end{table*}}

\end{document}